\documentclass[3p,11pt]{elsarticle}
\usepackage[utf8]{inputenc}
\usepackage[colorlinks,citecolor=red,urlcolor=red]{hyperref}
\usepackage[usenames,dvipsnames]{xcolor}
\usepackage{amsmath,amsthm,amssymb,amsfonts,graphicx,subfigure,moreverb} 
\usepackage[english]{babel}
\usepackage[ruled,linesnumbered]{algorithm2e} 
\usepackage{myshortcuts} 
\usepackage{siunitx,tabularx,etoolbox,xcolor,booktabs} 
\usepackage{mathtools}
\usepackage{booktabs}
\usepackage{grffile} 
\usepackage{comment} 

\providecommand*{\mrm}[1]{\mathrm{#1}}
\newcommand{\diff}{\,\mrm{dx}}

\newcommand{\R}{{\mathbb R}}
\newcommand{\C}{{\mathbb C}}

\newcommand{\numax}{{\nu_{\rm max}}}

\usepackage{pgfplots}
\usetikzlibrary{external}
\usetikzlibrary{decorations.pathreplacing}
\usetikzlibrary{positioning,arrows}

\newcommand{\dis}{\displaystyle}
\newcommand{\om}{\omega}
\newcommand{\Om}{\Omega}

\newcommand{\eps}{\epsilon} 
\newcommand{\pseudo}{\delta} 
\newcommand{\disG}{l} 

\newcommand{\n}[1]{{\bf #1}}

\newcommand{\fr}[2]{\displaystyle\frac{#1}{#2}}

\newcommand{\mx}[1]{\displaystyle\mathbb{#1}} 
\newcommand{\mc}[1]{\mathcal{#1}} 
\newcommand{\mr}[1]{\mathrm{#1}} 

\newcommand{\dpar}[2]{\displaystyle\frac{\partial #1}{\partial #2}}

\newcommand{\dtot}[2]{\displaystyle\frac{d #1}{d #2}}

\newcommand{\fem}{{\gamma}} 
\newcommand{\dtn}{a} 
\newcommand{\pml}{\ell} 
\newcommand{\PML}{{P\!M\!L}}
\newcommand{\OmOneD}{\Om} 

\newcommand{\Honedtn}{H^{1}(\OmOneD_{\dtn})}

\usepackage[noend]{algpseudocode}

\makeatletter
\def\BState{\State\hskip-\ALG@thistlm}
\makeatother

\providecommand{\myceil}[1]{\left \lceil #1 \right \rceil } 

\iftrue 
  \usepackage{tikz}
  \usepackage{pgfplots}
  \pgfplotsset{
    compat=newest,
    tick label style={font=\scriptsize},
    label style={font=\scriptsize},
    legend style={font=\scriptsize}
  }
  \usetikzlibrary{decorations.pathreplacing}
  \iffalse 
     \usepgfplotslibrary{external}
  \else
     \renewcommand{\tikzsetnextfilename}[1]{}
  \fi
\else
  \usepackage{tikzexternal}
  \tikzsetexternalprefix{gfx_tmp/}
  
\fi

\newtheorem{theorem}{Theorem}

\newtheorem{definition}[theorem]{Definition}

\newtheorem{remark}[theorem]{\textit{Remark}}

\journal{JOURNAL}

\begin{document}

\begin{frontmatter}

\title{On spurious solutions encountered in Helmholtz scattering resonance computations in $\R^d$ with applications to nano-photonics and acoustics}
\author{Juan C. Ara\'ujo C.\fnref{umit}}
\fntext[umit]{Department of Computing Science, Umeå University, MIT-Huset, 90187 Umeå, Sweden}

\author{Christian Engström\fnref{lnu}}
\fntext[lnu]{Department of Mathematics, Linnaeus University, Hus B, 35195  Växjö, Sweden}

\begin{abstract}
In this paper, we consider a sorting scheme for potentially spurious scattering resonant pairs in one- and two-dimensional electromagnetic problems and in three-dimensional acoustic problems. The novel sorting scheme is based on a Lippmann-Schwinger type of volume integral equation and can, therefore, be applied to structures with graded materials as well as to configurations including piece-wise constant material properties. For TM/TE polarized electromagnetic waves and for acoustic waves, we compute first approximations of scattering resonances with finite elements.
Then, we apply the novel sorting scheme to the computed eigenpairs and use it to mark potentially spurious solutions in electromagnetic and acoustic scattering resonances computations at a low computational cost. 
Several test cases with Drude-Lorentz dielectric resonators as well as with graded material properties are considered.
\end{abstract}

\end{frontmatter}

\section{Introduction}\label{sec:intro}

The most common approach to approximate scattering resonances is to truncate the domain with a perfectly matched layer (PML) and discretize the differential equations with a finite element method. This results in approximations of the  true resonances but in practice also a large number of solutions that are unrelated to the true resonances.
One approach to lessen the problem with nonphysical solutions in finite element (FE) computations and decrease the number of \emph{spurious solutions} is to design an appropriate $hp$-refinement strategy \cite{araujo+campos+engstrom+roman+2020}. However, for large problems \emph{spurious eigenvalues} remains a major problem in resonance computations \cite{ELBUSAIDY2020100161} and in quasimodal expansions \cite{PhysRevA.89.023829,MR3743646}.

The identification of spurious pairs from resonance computations with PML has been attempted using a sensitivity approach \cite{gopal08,ket12,ELBUSAIDY2020100161}. The approach is based on the observation that spurious eigenvalues are sensitive to parameter perturbations, while well-converged resonances are not. However, it has been observed that this approach is able to identify approximations to true resonances only in small regions of the complex plane and only if the finite element space is large. Additionally, the method is for a fixed discretization expensive as it requires the computation of all eigenpairs in a spectral window several times, from where re-meshing and re-assembling of the FE must be performed.

The origin of nonphysical eigenvalues in scattering resonance computations is \emph{spectral instability}, which is common for non-normal operators \cite{trefethen07,davies97}.  
Spectral instability is known to be less problematic with volume integral equations compared with formulations based on differential operators. Therefore, we proposed in \cite{araujo+engstrom+2017}
to use a Lippmann-Schwinger type of integral equation for marking potentially spurious solutions in a one-dimensional setting.  However, the direct application of the ideas presented in \cite{araujo+engstrom+2017} to higher dimensions turns out to be more challenging and computationally demanding.

In this paper, we develop a scheme for marking potentially spurious solutions
in higher dimensions when the computational domain is truncated by a PML or a Dirichlet-to-Neumann (DtN) map. In particular, we show that the developed scheme is computationally cheap and provides valuable information on the location of potentially spurious eigenvalues. For example, this information is  useful in quasimodal expansions of the field and when adaptive finite element methods are employed.

In the new sorting strategy, a particular approximation of a Lippmann-Schwinger based residual is computed for each eigenpair. The sorting strategy is computationally efficient since (i) the computation re-uses the pre-computed FE environment (ii) the testing has low memory requirements, and (iii) the algorithm is fully parallelizable. The method is successfully applied to metal-dielectric nanostructures in $\R^2$, where the metal is modeled by a sum of Drude-Lorentz terms, and to an acoustic benchmark problem in $\R^3$. Several additional benchmarks confirm the efficiency of the proposed sorting scheme in $\R^d$ with $d=1,2,3$.

\section{Electromagnetic and acoustic scattering resonances}

Assume that $\eps (x,\omega)=\eps (x_1,x_2,\omega)$ is independent of $x_3$ and consider {electromagnetic waves} propagating in the $(x_1,x_2)$-plane. The $x_3$-independent electromagnetic field  $(\n E,\n H)$ is then decomposed into transverse electric (TE) polarized waves 
$(E_1,E_2,0,0,0,H_3)$ and transverse magnetic (TM) polarized waves $(0,0,E_3,H_1,H_2,0)$ \cite{Cessenat1996}. This decomposition reduces Maxwell's equations to one scalar equation for $H_3$ and one scalar equation for $E_3$. 
The TM-polarized waves and the TE-polarized waves satisfy formally
\begin{equation}\label{eq:TMTE}
-\Delta E_3-\omega^2\eps E_3=0 \quad \text{and}\quad -\nabla\cdot\left (\frac{1}{\eps} \nabla H_3\right )-\omega^2 H_3=0,
\end{equation}
respectively. For the scattering resonance problems, $E_3$ and $H_3$ are assumed to be locally $L_2$-integrable functions 
that satisfy an outgoing condition \cite{MR1350074,MR1037774}.

Let the physical domain $\Om_\dtn\subset\mx R^d$ be an open ball of radius $\dtn$ with boundary $\Gamma_a$, 
and let $\Omega_r:=\text{supp}\, (\eps-1)\subset \Om_\dtn$ be the bounded domain defining the resonators. Hence, we assume that the relative permittivity $\eps$ for $x\in \R^d\setminus\Omega_r$ is one. Furthermore, let $\Om_r:=\cup_{i=1}^{N}\Om_i$ denote the union of disjoint resonators $\Omega_1,\,\Omega_2,...,\Omega_N$
as shown in Figure \ref{fig:domain_pml}. A \emph{scattering resonance} was in \cite{gopal08} formally defined as a complex number $\omega$ for which the Lippmann-Schwinger equation
\begin{equation}
T(\om)u:= u - K(\om) u=0
\label{eq:lippSchw}
\end{equation}
has a non-zero solution $u$. 

The integral operator $K$ in \eqref{eq:lippSchw} is for TM/TE waves given by
\begin{equation}
\begin{array}{l}
T\!M\!: \,  K(\om) u (x) := \om^2\int_{\Om_r} \Phi (x,y) \left(\eps(y)-1\right) u(y)\,dy \\[2mm]
T\!E: \,\,  K(\om) u (x) := \nabla\cdot 
\int_{\Om_r} \Phi (x,y) \left(\tfrac{1}{\eps(y)}-1\right) \nabla u(y)\,dy
\end{array},\,\,
\Phi (x,y):=\left\{
\begin{array}{ll}
\frac{i}{2\om} e^{i\om|x-y|},  & d=1\\[2mm]
\tfrac{i}{4}H_0^{(1)}(\om|x-y|), & d=2.
\end{array}
\right.
\label{eq:lippSchw_green}
\end{equation}
Here, $\Phi (x,y)$ is known as the outgoing Green function in free space for the Helmholtz equation \cite{MR1350074,MR1037774}. 
Notice that while we are interested in $x\in\Om_\dtn$, the integration in \eqref{eq:lippSchw} is only performed over $\Om_r$, since the integration over the air region $\Om_0:=\Om_\dtn\setminus\Omega_r$ is zero. 

The scattering resonance problem \eqref{eq:lippSchw} is a highly non-linear eigenvalue problem, where the matrices after discretization are large and full. It is possible to accurately solve the non-linear eigenvalue problem $T(\omega)u=0$ in $\R$ using a standard laptop; see e.g. \cite{bra13,araujo+engstrom+2017}. However, accurate computations of eigenpairs of \eqref{eq:lippSchw} in higher dimensions would require huge computer resources; See \cite{MR2406839} and the discussion in Section \ref{sec:comp_details}.

\subsection*{Acoustic scattering resonances in $\R^3$}

Sound-soft materials are characterized by the speed of sound $c(x)$ and we assume that acoustic resonators are defined by $\Omega_r:=\text{supp}\, (c^{-2}-1)\subset \Om_\dtn$. Then, the acoustic pressure $u$ satisfies formally the Helmholtz equation
\begin{equation}
	-\Delta u-\frac{\omega^2}{c(x)^2} u=0,
\end{equation}
where the outgoing condition can be expressed as $u$ satisfying an expansion in spherical harmonics outside the open ball $\Om_\dtn$ \cite{ihl98}.

Moreover, $(\om, u)$ is a scattering resonance pair if \eqref{eq:lippSchw} holds with
\begin{equation}
 	K(\om) u (x) := \om^2\int_{\Om_r} \Phi (x,y) \left(\frac{1}{c(y)^2}-1\right) u(y)\,dy, \quad  \Phi (x,y)=\fr{e^{i\om|x-y|} }{4\pi|x-y|}.
\end{equation} 	
Note that the acoustic problem in $\R^d$, $d=1,2,3$ is analogous to TM-polarized electromagnetic waves with $\eps=1/c^2$.

\subsection{Alternative formulations}

Understanding the resonance behavior of structures in unbounded domains is important and many different approaches have been proposed. Graded  material properties are increasingly popular in applications \cite{Wang2016} and we will therefore not consider boundary integral equations. However, boundary integral equation based methods are a good alternative for cases with piecewise constant coefficients and not too complicated geometries \cite{MR3622414}. The most popular method to compute resonances in $\R^d$, $d>1$ is the finite element (FE) method with a perfectly matched layer (PML). In recent years, finite element methods based on Hardy space infinite elements (HIF) \cite{MR2485441} and DtN maps \cite{araujo+engstrom+jarlebring+2017} have also been proposed as strong alternatives to compute resonances in higher dimensions. For the DtN map, recent developments in computational linear 
algebra are a key to the high performance of the method \cite{jar14,jar15J}. Discretization with FE of the PML and HIF formulations result in sparse matrices, and a formulation in terms of a DtN result in sparse matrices except a small dense block corresponding to the DtN map. Moreover, the PML and HIF formulations result in a standard generalized eigenvalue problem if $\eps$ is $\omega$-independent and in the general case the non-linearity in $\omega$ is completely determined by $\eps (x,\omega)$. Hence, the PML and HIF formulations seem to have the most attractive properties of the considered methods. However, it is very important to also take into account the so called spectral instability. Then, the picture changes completely, as discussed in the next section.

\subsection{Spectral instability and pseudospectra}\label{sec:stability}

Let $A$ denote an unbounded closed linear operator in a Hilbert space with domain $\text{dom}\, A$, spectrum $\sigma (A)$, and resolvent set $\rho (A)$. Then $A$ exhibits high \emph{spectral instability} if for a very small $\pseudo>0$ there exist many $\omega^2\in\C$ and $u\in\text{dom}\, A$ such that
\begin{equation}\label{eq:si}
	\|(A-\omega^2)u\|\leq \pseudo\|u\|
\end{equation}
even though $\omega^2$ is not close to $\sigma (A)$ \cite{MR1912874}. This is closely related to the pseudospectrum $\sigma_{\pseudo} (A)$ which is defined as the union of $\sigma (A)$ and all  $\omega^2$ 
in the resolvent set $\rho(A)$
for which it exists an $u\in\text{dom}\, A$ such that \eqref{eq:si} holds. The generalization of those results to an operator function $T$ is straightforward and we will in some of the numerical computations rely on the following alternative characterization of the pseudospectrum:
\[
	\sigma_{\pseudo} (T)=\sigma(T)\cup\{\omega\in\rho (T)\, :\, \|T^{-1}(\omega)\|>\pseudo^{-1}\}.
\]
It is well known that PML and HIF based methods encounter high spectral instability \cite{MR2485441,araujo+engstrom+2017}. Methods based on a DtN map encounter medium spectral instability \cite{araujo+engstrom+jarlebring+2017, araujo+engstrom+2017} and integral equation based methods encounter low spectral instability \cite{araujo+engstrom+2017}. Hence, the Lippmann-Schwinger equation is {in our setting} the preferred method in terms of spectral stability. This will be further discussed in the paper.

\subsection {Domain and material properties}\label{sec:domain_permittivity}

In optics, the material properties of {non-magnetic} metals are characterized by the complex relative permittivity function $\eps$, which changes rapidly at optical frequencies $\omega$. The most common accurate material model is then the Drude-Lorentz model
\begin{equation}
\eps_{metal} (\om):=\eps_\infty + \sum_{j=0}^{N_p}\fr{f_j\om_p^2}{\om_j^2-\om^2-i\om\gamma_j},
\label{eq:drude_lorentz}
\end{equation}
where $\eps_\infty\geq 1$ and $f_j$, $\om_p,\,\om_j$, $\gamma_j$ are non-negative \cite{Cessenat1996}. Hence, the Maxwell eigenvalue and scattering resonance problems in $\omega$ are nonlinear for metal-dielectric nanostructures. Assume that the domain of the resonators can be written in the form $\Om_r:=\cup_{i=1}^N\Om_i$ and let $\chi_{\Om_m}$ denote the characteristic function of the subset $\Om_m$. For material properties that are piecewise constant in $\Om_\dtn$, we assume a permittivity function in the form
\begin{equation}
\eps(x,\omega):=\sum_{m=0}^{N_r}\eps_m(\omega)\chi_{\Om_m}(x),\quad x\in \Om_\dtn,\quad \omega\in \mc D,
\label{eq:permittivity_function}
\end{equation}
where the dependencies on {$\omega\in \mc D\subset \mx C$} in $\eps_m$ for $m=0,1,\dots$ are of Drude-Lorentz type \eqref{eq:drude_lorentz}. In addition, we will consider graded material properties, meaning that $\eps$ is a continuous function in $x$.
In linear acoustics, the speed of sound $c$ is assumed to be independent of the frequency.

\section{DtN and PML based methods}

In the next sections, we will describe two common approaches to compute scattering resonances and the restriction of resonance modes to a compact subset of $\R^d$. In the following, we use the notation
\begin{equation}
\begin{array}{rcl}
-\nabla\cdot \left( \rho \nabla u\right) - \om^2 \eta u &=& 0,
\end{array} \label{eq:master_eq}
\end{equation}
where $u:=E_3,\,\rho:=1,\,\eta:=\eps$ for the TM-case and $u:=H_3,\,\rho:=1/\eps,\,\eta:=1$ for the TE-case.

We define for {$u,v\in\Honedtn$} the forms
\begin{equation}\label{eq:common_forms}
	\mathfrak{a} (\omega)[u,v] :=\!\!\int_{\Om_\dtn}\!\!\rho \nabla u\cdot \nabla \overline{v}\diff,\qquad
	\mathfrak{b} (\omega)[u,v] :=\!\!\int_{\Om_\dtn}\!\!\eta u\overline{v}\diff,
\end{equation}
{where in \eqref{eq:common_forms}, $\rho$ and $\eta$ are functions of $\om\in\mc D$. Let $\mc Z$ denote the set of values $\om$ that are zeros or poles of $\eps$ and set $\mc D:=\mx C\setminus \mc Z$.}
\subsection{DtN based methods}

{Scattering} resonances $\omega$ and quasi-normal modes $u$ restricted to $\Om_\dtn$ can be determined from a problem that utilizes a Dirichlet-to-Neumann (DtN) map operator denoted $\mc G(\om)$ \cite{lenoir92,schenk2011optimization,araujo+engstrom+jarlebring+2017}. Below we present variational formulations for $\R^d$, $d=1,2$. Formally, $(\omega,u)$ is a scattering resonance pair if \eqref{eq:master_eq} holds in $\Omega_a$ and
\begin{equation}\label{eq:DtN}
	\frac{\partial u}{\partial n}=\mc G(\om)u\quad \text{on } \Gamma:=\partial\Omega_a,
\end{equation}
where $\partial u/\partial n$ is the normal derivative.

\subsubsection{DtN formulation in $\R$} 

In one space dimension the {scattering} resonance problem restricted to $\OmOneD_\dtn:=(-\dtn,\dtn)$ is formally: Find a non-zero $u$ and a complex $\omega$ such that
\begin{equation}\label{eq:Helmholtz}
-\left( \rho u'\right)' - \om^2 \eta u=0 \,\,\,\hbox{for}\,\, x\in\OmOneD_\dtn,
\end{equation}
where the {DtN-map} at $x=\pm \dtn$ is
\begin{equation}\label{eq:formalDtN} 
u'(-\dtn)=-i \omega \,u(-\dtn), \quad u'(\dtn)  =i \omega\,u(\dtn).
\end{equation}
Define for $u,v\in\Honedtn$ and $\omega\in \mc D\subset\C$ 
the forms $\mathfrak{a},\, \mathfrak{b}$ as in \eqref{eq:common_forms}, and
\begin{equation}\label{eq:dtn_form}
g_1(\om)[u,v] := i\omega (u(\dtn)\overline{v}(\dtn)+u(-\dtn)\overline{v}(-\dtn)).
\end{equation}

The nonlinear eigenvalue problem is then as follows: Find $u\in\Honedtn\backslash\{0\}$ and $\omega\in \mc D$ satisfying 
\begin{equation}
q_1(\omega)[u,v]:=\mathfrak{a} (\omega)[u,v]-\omega^2 \mathfrak{b} (\omega)[u,v] - g_1(\om)[u,v]=0,
\label{eq:dtn1d_var}
\end{equation}
for all $v\in \Honedtn$. Note that \eqref{eq:dtn1d_var} is a quadratic eigenvalue problem if $\eps$ is independent of $\omega$ and a rational eigenvalue problem for Drude-Lorentz type of materials \eqref{eq:drude_lorentz}.

\subsubsection{DtN formulation in $\R^2$}

In this subsection, we present a DtN formulation in polar coordinates $(r,\theta)$. 
Let $H^{(1)}_\nu(z)$ denote the Hankel function of first kind, then the DtN operator \eqref{eq:DtN} on the circle $\Gamma_a$ has the explicit form
\begin{equation}
\mc G(\om)u:=\frac{1}{2\pi}
\sum_{\nu=-\infty}^{\infty}
\om \frac{ H^{(1)\prime}_\nu(\om a)}{H^{(1)}_\nu(\om a)}
\,e^{i\nu\theta}
\int_0^{2\pi}\!\!u(a,\theta')\,e^{-i\nu\theta'}d\theta'
\label{eq:DtN2}
\end{equation}
and $\mc G(\om):H^{1/2}(\Gamma_a)\rightarrow H^{-1/2}(\Gamma_a)$ is bounded \cite{lenoir92}.  

The resonance problem restricted to $\Om_\dtn$ is formally to find non-trivial solutions $(\om,u)$ such that \eqref{eq:master_eq} and \eqref{eq:DtN} with \eqref{eq:DtN2} holds. The theory presented in \cite{lenoir92} can with minor changes be used in the present case to derive properties of a variational formulation of the problem.

{\bf Variational formulation}: Let $S$ denote the union of the set of zeros of $H^{(1)}_\nu(\om a),\,\nu\in \mx Z$, and $\mc G_\numax(\om)$ denote the operator \eqref{eq:DtN} truncated after $|\nu|=\numax$. The eigenvalues of the truncated version of \eqref{eq:master_eq}-\eqref{eq:DtN} are determined by the following variational problem:
Find $u\in H^1(\Om_\dtn)\setminus \{0\}$ and 
$\om \in\mc D:=\mx C\setminus \{\mc Z\cup\mx R^-\cup S\}$ such that for all $v\in H^1(\Om_\dtn)$
\begin{equation}
q (\om)[u,v]:=\mathfrak{a} (\omega)[u,v]-\omega^2 \mathfrak{b} (\omega)[u,v]
- g (\om)[u,v] =0,
\label{eq:dtn2d_var}
\end{equation}
where the forms 
$\mathfrak{a},\, \mathfrak{b}$ are defined as in \eqref{eq:common_forms}, and
\begin{equation}
g(\om)[u,v]:=(\mc G_\numax(\om) u,v)_{\Gamma_a}
=\dis \!\sum_{\nu=-\numax}^{\numax}\!\!\!\om a\frac{H^{(1)\prime}_\nu(\om a)}{H^{(1)}_\nu(\om a)}\,\hat u_\nu\, \bar {\hat v}_\nu,\,\, \hat \varphi_\nu=\tfrac{1}{\sqrt{2\pi}}\int_{0}^{2\pi} \!\!\varphi(a,\theta)\,e^{-i\nu\theta}d\theta.
\label{eq:dtn_1}
\end{equation} 

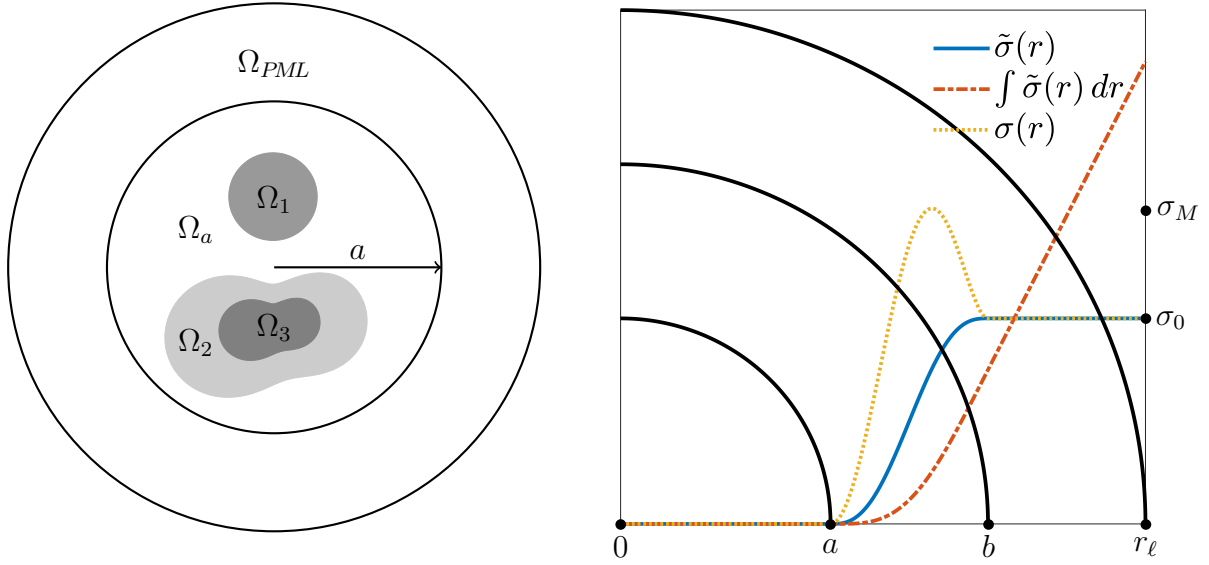
\begin{figure}
\centering
	\begin{tikzpicture}[thick,scale=1.0, every node/.style={scale=1.0}]	
		\tikzstyle{ann} = [fill=none,font=\large,inner sep=4pt]
		
		\draw( -4.10, -0.1) node { \includegraphics[scale=0.7,angle=0,origin=c]{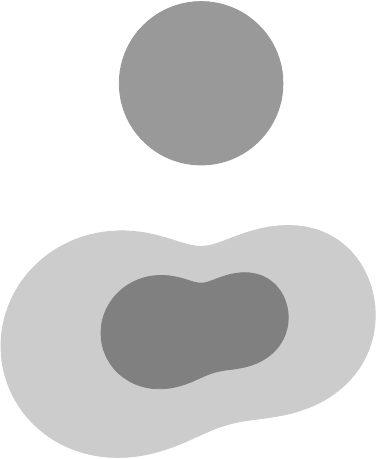} };
		
		\node[ann] at (-2.9,0.2) {$a$};
		\draw[arrows=->,line width=0.8pt](-4.0,0.0)--(-1.80,0.0);		
		
		\draw (-4.0,0) circle (2.2cm);
		\draw (-4.0,0) circle (3.5cm);
		
		\node[ann] at ( -4.00, 2.7) {$\Omega_{\PML}$};
		\node[ann] at ( -5.03, 0.5) {$\Om_\dtn$};
		
		\node[ann] at ( -4.00, 0.9 ) {$\Omega_1$};
		\node[ann] at ( -5.03,-1.00) {$\Omega_2$};
		\node[ann] at ( -4.00,-0.80) {$\Omega_3$};
		
		\draw( 4.00, 0.0) node { \includegraphics[scale=0.7] {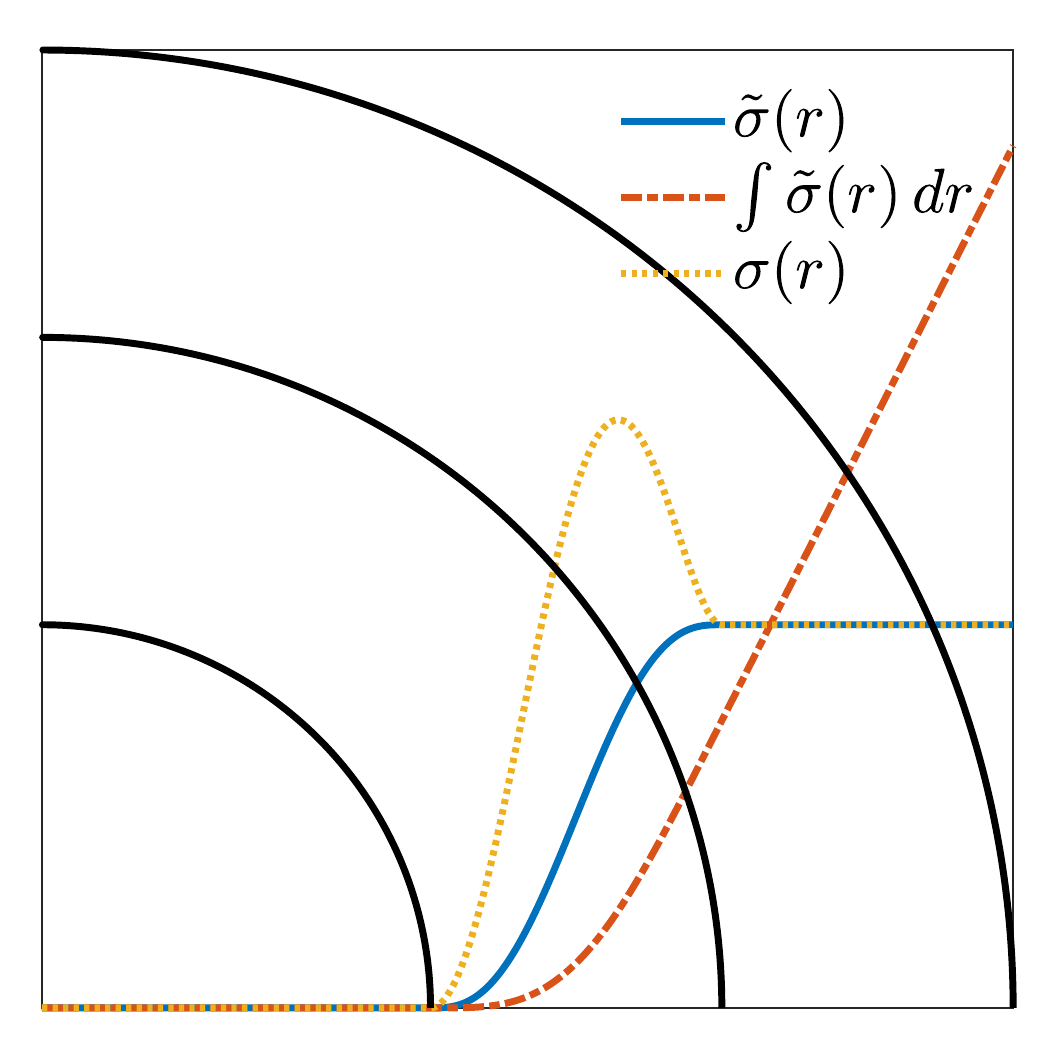} };
		
		\fill (0.56,-3.41,0.0) circle (2pt);
		\node[ann] at ( 0.56,-3.7) {$0$};
		
		\fill (3.32,-3.41,0.0) circle (2pt);
		\node[ann] at ( 3.32,-3.7) {$a$};
		
		\fill (5.40,-3.41,0.0) circle (2pt);
		\node[ann] at ( 5.40,-3.7) {$b$};
		
		\fill (7.47,-3.41,0.0) circle (2pt);
		\node[ann] at ( 7.47,-3.7) {$r_\ell$};
		
		\fill (7.47,-0.68,0.0) circle (2pt);
		\node[ann] at (7.80,-0.70) {$\sigma_0$};
		
		\fill (7.47,0.75,0.0) circle (2pt);
		\node[ann] at (7.90, 0.75) {$\sigma_M$};
	
	\end{tikzpicture}	
	\caption{\emph{Left) Arbitrary configuration of resonators. Right) 
	PML stretching function. }}\label{fig:domain_pml}
\end{figure}

\subsection{PML based methods}

In the previous section, a DtN-map was used to reduce the exterior Helmholtz problem to a bounded domain. In this section, we consider an alternative approach based on a complex coordinate stretching (the PML method), which results in a linear eigenvalue problem for non dispersive material coefficients \cite{kim09}. The method consists on attaching to $\Om_\dtn$ a buffer layer of thickness $\ell$, where outgoing solutions decay rapidly. The buffer domain is referred to as $\Om_{\PML}$ and the full computational domain $\Om:=\Om_\dtn\cup\Om_\PML$ is enlarged as shown in the Figure \ref{fig:domain_pml}.

\subsubsection{PML formulation in $\R$}\label{sec:pml1d} 

Let $\ell>0$, and $0<a<b<r_\ell$, with $r_\ell=b+\ell$. The action of the PML is defined through the stretch function
\begin{equation}
\tilde \sigma(r):=\left\{
\begin{array}{ll}
0,& \hbox{if}\,\, r < a \\
P(r),& \hbox{if}\,\, a \leq r \leq b \\
\sigma_0,& \hbox{if}\,\, r > b 
\end{array}
\right.
\label{eq:sig_pol}
\end{equation}
with $r=|x|$. The polynomial $P$ in \eqref{eq:sig_pol} is required to be increasing in $[a,b]$ and $\tilde\sigma(r)$ is sufficiently smooth $C^{2}(0,r_\pml)$. For this we introduce the fifth order polynomial $P$ satisfying: 
$P(a)=P'(a)=P''(a)=P'(b)=P''(b)=0$ and $P(b)=\sigma_0$.

The PML problem is restricted to $(-r_\pml,r_\pml)$ and the PML strength function has then the profile shown in Figure\,\ref{fig:domain_pml}. In the following sections, we consider the 
transformation rule
\begin{equation}
\dtot{}{\tilde x}=\frac{1}{\tilde\alpha( x)}\dtot{}{ x},
\,\,\,\hbox{with}\,\,\,\tilde\alpha(x)=1+i\tilde \sigma(x).
\label{eq:coordstr}
\end{equation} 

For finite element computations we restrict the domain to $\OmOneD_{\pml}:=(-r_\pml,r_\pml)$ 
and choose as in \cite{kim09} homogeneous Dirichlet boundary conditions. Formally, the \emph{finite PML problem} is then: Find the eigenpairs $(\om,u)$ such that
\begin{equation}\label{eq:trun_PML}
-\dtot{}{x} \left(\frac{\rho}{\tilde\alpha}\dtot{u}{x}\right)-\om^2\eta\tilde\alpha \,u=0,\,\, x\in \OmOneD_{\pml},\,\,\,
u(r_\pml)=0 \,\,\,\hbox{and}\,\,\, 	u(-r_\pml)=0. 
\end{equation}

Let $\OmOneD_\PML=(-r_\pml,-a)\cup (a,r_\pml)$ and denote by $\mathfrak{a},\, \mathfrak{b}$ the forms in \eqref{eq:common_forms}.
As variational formulation of \eqref{eq:trun_PML}, we consider:\\ 
Find $u\in H_0^1(\OmOneD_{\pml})\setminus \{0\}$ and $\om  \in\mc D:=\mx C\setminus \mc Z$ such that for all $v\in H_0^1(\OmOneD_{\pml})$
\begin{equation}
t_1(\om)[u,v]:= \mathfrak{a} (\omega)[u,v]-\omega^2 \mathfrak{b} (\omega)[u,v] +
\hat t_1(\om)[u,v] =0,
\label{eq:pml1d_var}
\end{equation}	
where $\hat t_1(\om)[u,v]=(\tfrac{1}{\tilde\alpha} u', v')_{\OmOneD_{\PML}}-\om^2(\tilde\alpha u,v)_{\OmOneD_{\PML}}$.

\subsubsection{PML formulation for $\R^d$}

Approximation of resonances using a radial PML was analyzed in \cite{kim09} and we will here only consider the PML problem truncated to the ball $\Omega \subset\R^d,\,d=2,3$. {The used complex stretching functions are, as in one-dimension, of the form \eqref{eq:sig_pol}}. Let $\mc Z$ denote the set of values $\om$ that are zeros or poles of $\eps$ and let $\Om:=\Om_\dtn\cup\Om_\PML$ denote a partition into the PML-region and the part of the domain $\Om_\dtn$ containing the resonators.

In the sequel we need the following definitions
\begin{equation}
\begin{array}{lllll}
\tilde \alpha(r)  \!\!\!& := 1+i\tilde \sigma(r), & \, &
\tilde r(r) \!\!\!& := (1+i\tilde \sigma)r=\tilde \alpha(r)\,r, \\[2mm]
\sigma(r) \!\!\!& := \tilde \sigma(r)+r\dpar{\tilde \sigma}{r}, & \, &
\alpha(r) \!\!\!& := \dpar{\tilde r}{r}=1+i\sigma(r),
\end{array}
\label{eq:sig_pol2}
\end{equation}
with the properties $\sigma(r) = \partial (r\tilde \sigma)/\partial r$ and 
$\alpha(r)=\tilde \alpha(r)=1+i\sigma_0\quad \hbox{for}\quad r>b$.

It is clear that the PML coefficients are designed such that for $r\leq a$, we obtain $\tilde\sigma(r)=0$ and $\alpha(r)=1$, which is the same as $\tilde r=r$. Hence, the PML operator restricted to $r\leq a$ corresponds to the original operator in problem \eqref{eq:master_eq}, where there is no PML effect. 

{\bf Variational formulation}: The eigenvalues of \eqref{eq:master_eq} are then determined by the following variational problem: 
Find $u\in H_0^1(\Om)\setminus \{0\}$ and $\om  \in\mc D:=\mx C\setminus \mc Z$ such that for all $v\in H_0^1(\Om)$
\begin{equation}
t(\om)[u,v]:= \mathfrak{a} (\omega)[u,v]-\omega^2 \mathfrak{b} (\omega)[u,v] 
+ \hat t (\om)[u,v] =0,
\label{eq:pml2d_var}    
\end{equation}	
where $\hat t (\om)[u,v]:=(\mc A\nabla u,\nabla v)_{\Om_\PML}-\om ^2(\mc B u,v)_{\Om_\PML}$, and the forms $\mathfrak{a},\, \mathfrak{b}$ are defined as in \eqref{eq:common_forms}.

As an example for $d=2$, direct transformation of \eqref{eq:pml2d_var} from polar to Cartesian coordinates results in
\begin{equation}
\mc A=\left(
\begin{array}{cc}
\tfrac{\tilde\alpha}{\alpha}\cos^2\theta+\tfrac{\alpha}{\tilde\alpha}\sin^2\theta &
\bigg(\tfrac{\tilde\alpha}{\alpha} -\tfrac{\alpha}{\tilde\alpha}\bigg)\sin\theta\cos\theta \\ 
\bigg(\tfrac{\tilde\alpha}{\alpha} -\tfrac{\alpha}{\tilde\alpha}\bigg)\sin\theta\cos\theta &
\tfrac{\tilde\alpha}{\alpha}\sin^2\theta+\tfrac{\alpha}{\tilde\alpha}\cos^2\theta
\end{array}
\right),\quad \mc B:=\alpha \tilde\alpha.
\label{eq:def_pml_r1}
\end{equation}
Note that even though $\mc A$, and $\mc B$ are defined in  
$\Om$, their action takes place only in $\Om_\PML$.

\section{Discretization of the Lippmann-Schwinger equation}

In this section we present a collocation method for the discretization of the Lippmann-Schwinger equation \eqref{eq:lippSchw}, which will be used to compute resonances in one-dimension and is the base for the numerical sorting algorithm in Section \ref{sec:sorting}.
Further computational details are given in Section \ref{sec:comp_details}.


\subsection{A Galerkin-Nystr\"om method} \label {sec:lipp_collocation}

We present a Galerkin-Nystr\"om discretization method for linear Fredholm integral equations of the second kind. 
In \cite{ikebe72}, this method is referred to as case (A) of the Galerkin methods, and convergence for the problem with sources is also discussed. In \cite[Sec.~3.2]{gopal08} the method was used for resonance computations for $d=2$.

Let $\{\varphi_j\}_{j=1}^N$ be piecewise polynomial functions with the property $\varphi_j(x_i)=\delta_{ji}$, $\{x_i\}_{i=1}^{N}\in \Om_\dtn$.
We introduce the representation $u^\fem=\sum^{N}_{j} \xi_j\varphi_j$, and with the use of \eqref{eq:lippSchw} we obtain the nonlinear eigenvalue problem: Find $\xi\in\mx C^{N}$ and $\om^\fem\in\mx C$ such that 
\begin{equation}
\begin{array}{ll}
& T^\fem(\om^\fem)\xi=(I-K(\om^\fem))\xi=0, \,\,\,\hbox{with}\\[3mm]
T\!M: &  K_{ij}(\om) := \om^2\int_{\Om_r} \Phi (x_i,y) \left(\eps(y)-1\right)\varphi_j (y)\,dy, \\[2mm]
T\!E: &  K_{ij}(\om) := \nabla\cdot 
\int_{\Om_r} \Phi (x_i,y) \left(\tfrac{1}{\eps(y)}-1\right) \nabla\varphi_j(y) \,dy. 
\end{array}	
\label{eq:lipp_collocation}
\end{equation}
The Nystr\"om method consists in choosing the collocation points $x_i$ as the nodes of a high order quadrature rule. By doing this, the convergence of the scheme is considerably improved. 
In Section \ref{sec:comp_details} we describe some of the implementation details of the Galerkin-Nystr\"om discretization.

The resulting nonlinear matrix eigenvalue problem \eqref{eq:lipp_collocation} is solved by using a contour integration based method \cite{Sakurai09,Beyn12,MR3423605}.

\begin{remark}\label{flexibility_ls}
The formulation in \eqref{eq:lippSchw} uses information of the exact solution of the problem at every discretization node $x_i$, through its fundamental solution $\Phi(x,y)$. From where a numerical scheme based on \eqref{eq:lipp_collocation} is flexible in the way that it can be posed in the smallest domain $\Om_r$, as well as in larger domains $\Om\supset\Om_r$, without taking any special treatment for boundary conditions.
\end{remark}

\section{FE discretization of the DtN and PML based formulations}\label{sec:fe_formulations}
In this section we discuss briefly the details involved in the assembly of the 
matrices corresponding to the discretization of
the formulations given in \eqref{eq:master_eq}, and \eqref{eq:lippSchw}.

\subsection{Discretization with the finite element method} \label {sec:FE_mesh}

Let the domain $\Omega\subset\mx R^d$ be covered with a regular and quasi uniform finite element mesh $\mc T(\Om)$ consisting of elements $\{K_j\}^{N_K}_{j=1}$. 
The mesh is designed such that the permittivity function $\eps$ is continuous in each $K_j$. 
Let $h_j$ be the length of the largest diagonal of the non-curved primitive $K_j$
and denote by $h$ the maximum mesh size $h:=\max_j{h_j}$.

Let $\mc P_p$ denote the space of polynomials on $\mx R^d$ of degree $\leq p$ in each space coordinate and define the $N$ dimensional finite element space
\begin{equation}
	S^{\fem}(\Omega):=\{u\in H^1(\Om_\dtn):\left. u\right|_{K_j} \in\mc P_{p}(K_j)\,\,\hbox{for}\,\,K_j\in\mc T\}.
\end{equation}
The computations of discrete resonance pairs $(u^\fem,\omega^\fem)$ are for $d=2,3$ performed in the approximated domain $\Omega^\fem$ using curvilinear elements \cite{Babuska92}. The meshes used are \emph{shape regular} in the sense of \cite[Sec.~4.3]{Schwab1998}, and consist of  quadrilateral/brick elements with curvilinear edges/surfaces that deviate slightly from their non-curved primitives.


\subsection{Assembly of the FE matrices}\label{sec:assem}  

In this section we refer to domains $\Om\in\R^d$.
Let $\{\varphi_1,\dots,\varphi_N\}$ be a basis of $S^{\fem}(\Omega)$. Then $u^\fem\in S^{\fem}(\Omega)$, and the entries in the finite element matrices are of the form
\begin{equation}
	u^\fem=\sum_{j=1}^N \xi_j\,\varphi_j,\quad A_{ij}= (\rho \nabla\varphi_j, \nabla\varphi_i)_{\Om_\dtn},\quad 
M_{ij}=(\eta \,\varphi_j, \varphi_i)_{\Om_\dtn}.
\label{eq:fem_ans}
\end{equation}

The matrix eigenvalue problem is then: Find the eigenpars $(\om ,{\xi})\in\mc D\times \C^N\setminus \{0\}$ such that
\begin{equation}
F (\om) \,{\xi}:=\left(A-\om^2 M + Q \right)\!(\om)\,{\xi}=0,
\label{eq:matrix_nep}
\end{equation}
where the corresponding matrix valued function is 
\begin{equation}
	Q_{ij}(\om):=\left\{
	\begin{array}{rl}
		-g_1 (\om)[\varphi_j,\varphi_i], & \text{DtN and }\, d=1\\
		-g (\om)[\varphi_j,\varphi_i], &   \text{DtN and }\, d=2
	\end{array} \right.\!,\,\,\text{or}\,
	\left\{
	\begin{array}{rl}
		\hat t_1 (\om)[\varphi_j,\varphi_i], & \text{PML and }\, d=1\\
		\hat t (\om)[\varphi_j,\varphi_i], &   \text{PML and }\, d=2,3
	\end{array} \right..
	\label{eq:discrete_Q}
\end{equation}
In the case where $\eps(\om,x)$ is given as piecewise smooth function of space, we write \eqref{eq:matrix_nep} as
\begin{equation}
F (\om) \,{\xi}:=\left(\sum_{m=0}^{N_r} \{\rho_m(\om)\tilde A_m-\om^2\eta_m(\om) 
\tilde M_m\} + Q(\om) \right){\xi}=0,
\label{eq:matrix_nep_piecewise}
\end{equation}
with matrices
$
		\tilde A^m_{ij}= (\nabla\varphi_j, \nabla\varphi_i)_{\Om_m}, \,\, 
		\tilde M^m_{ij}=(\varphi_j, \varphi_i)_{\Om_m},\,\,m=0,1,\ldots,N_r.
$

\begin{remark}\label{trunc_dtn}
{\bf Truncation of the DtN:} Let $\myceil{z}$ be the smallest integer greater than or equal to $z$. We use the rule $\numax=\myceil{a\,\om_M}$ according to the results in \cite{araujo+engstrom+jarlebring+2017}, where from the considered spectral window, $\om_M$ is the largest real part allowed for computations of eigenvalues.
\end{remark}

\begin{remark}\label{trunc_pml}
{\bf Truncation of the PML:} The PML is set up following the discussions in \cite{kim09,araujo+engstrom+2017}, which accounts for large enough $\ell$ and $\sigma_0$ such that the search region is feasible. 
Additionally, we use the space $S_0^{\fem}(\Omega):=\{u\in S^{\fem}(\Omega): u=0\,\,\,\hbox{for}\,\,\,x\in \partial\Om\}$ for computations with the PML formulation.
\end{remark}

Finally, we mention that all discretization methods presented in this work use the same FE space over $\Om_\dtn$.
\begin{remark}\label{fem_Om_a}
All formulations (LS, DtN, PML) use the FE triangulation $\mc T(\Om_\dtn)$, which is the restriction of $\mc T(\Om)$ to $\Om_\dtn$. 
\end{remark}
By using the FE mesh suggested in Remark \ref{fem_Om_a}, we ensure that the approximation properties in the physical domain are the same for all formulations.

\section{Numerical sorting of potentially spurious solutions}\label{sec:sorting}

In this section we derive a discrete indicator from equation \eqref{eq:lippSchw} that allow us to identify potentially spurious solutions once we have computed FE solutions $(\om^\fem_m,u^\fem_m)$ to \eqref{eq:matrix_nep} or to \eqref{eq:matrix_nep_piecewise}. 
The identification of potentially spurious solutions is important since adaptive finite element methods and quasimodal expansions in practice will fail if spurious solutions are included in the expansion or influence the mesh-refinement. The idea of sorting potentially spurious solutions is in the spirit of the standard residual error estimator and marking strategy used in adaptive finite element methods \cite{Giani2016}. Adaptive FE can be applied when the coarsest mesh is sufficiently fine. However, the pre-asymptotic regime is very large in resonance computations, in particular for cases with large air regions in the computational domain. A goal of the paper is therefore to supplement the information given by the PDE based residual estimator with information from an integral equation. The sorting scheme is based on a computationally cheap approximation of the condition $\|\chi_{\dtn}T(\om)\chi_{\dtn}u\|<\pseudo$.

Let $\{\varphi_j\}$ be a basis for $S^\fem(\Om_\dtn)$.
Then, the discrete Lippmann-Schwinger equation \eqref{eq:lipp_collocation} is written in the form
$$T(\om)u^\fem=u^\fem-K(\om)u^\fem,\,\,
	u^\fem:=\sum_{j=1}^{N} \xi_j\varphi_j,\,\,\, 
	\hbox{with}\,\,\,\|u^\fem\|_{L^2(\Om_\dtn)}=1. 
$$

\begin{definition}\label{def:sorting}
{\bf Pseudospectrum indicator:} The computed eigenvalue $\om^\fem$ belongs, for given $\pseudo>0$, to the $\pseudo$-\emph{psudospectrum} $\sigma_{\pseudo}(T^\fem)$ if the pair $(\om^\fem,u^\fem)$ satisfies $\|T^\fem(\om^\fem)u^\fem\|_{\Om_\dtn}<\pseudo$. Then, for a given domain $\Om\supseteq \Omega_r$, we define the pseudospectrum indicator as
\begin{equation}
	\pseudo^\fem(\Om):=\|T^\fem(\om^\fem)u^\fem\|_{\Om_\dtn}.
	\label{eq:pseudo_ind}
\end{equation}
\end{definition}
Particularly, we want to be able to measure whether the computed eigenpair $(\om^\fem,u^\fem)$ is converging to a physical pair. Naturally, non convergent pairs exhibit large $\pseudo^\fem$ values.
Additionally, we can identify and remove certain spurious eigenpairs in the PML based formulation by considering the following definition.
\begin{definition}\label{def:pml_eigs}
	{\bf PML added eigenpairs:} The use of the coordinate stretching technique in
	formulations \eqref{eq:pml1d_var}, \eqref{eq:pml2d_var} introduces new eigenpairs to problem \eqref{eq:master_eq}. These new eigenvalues accumulate close to the critical line of the modified PML problem \cite{araujo+engstrom+2017}, and the corresponding eigenfunctions $v_m$
	exhibit oscillations in $\Om_\PML$, but decay in the physical region $\Om_\dtn$. 
	Then, by using the normalization $\|v_m\|_{L^2(\Om)}=1$, the
	PML critical eigenvalues exhibit
	$$0 < \fr{\|v_m\|^2_{L^2(\Om_\dtn)}}{|\Om_\dtn|}<\fr{\|v_m\|^2_{L^2(\Om_\PML)}}{|\Om_\PML|},$$
	which can be successfully used as a filtering criterion for removing PML added eigenpairs.
\end{definition}
A similar idea was in \cite{ELBUSAIDY2020100161} used to study photoacoustic resonators.

\subsection{Numerical pseudospectra computation}\label{sec:pseudo_comp}
Computations of the \emph{pseudospectra} provide insight into the behavior of the resolvent of the discretized operator function $F^\fem$, allowing us to evaluate its spectral stability. In our computations, we use that $\sigma_\pseudo(F^\fem)$ is the set of all $z\in\mx C$ such that 
\begin{equation}
s_{\scriptsize \hbox{min}}\, F^\fem(z)<\epsilon,
\label{eq:pseudo_mat}
\end{equation} 
where $s_{\scriptsize \hbox{min}}\, F^\fem(z)$ denotes the smallest singular value of $F^\fem(z)$ \cite[Def. 2.10]{trefethen07}. For the singular value computations we used SLEPc \cite{slepc05+roman}.

\subsection{On spurious solutions in $\R$}\label{sec:slab1d}

In \cite{araujo+engstrom+2017} we introduced a sorting strategy based on Definition \ref {def:sorting} but performing computations only inside $\Om_r$. As suggested by \eqref{eq:pseudo_ind}, computations can be also performed in $\Om_0$,
and it is natural to ask whether or not the {sorting scheme performs worse if air is included in the evaluation.} 
Then, we numerically test the solutions to the Lippmann-Schwinger (LS) formulation by enlarging the computational domain to include air. If the LS operator function would exhibit undesired spurious eigenvalues, this would render the method unreliable for detection of spurious pairs.

The following problem has been considered by several authors including \cite{osting13,kim09}. Define for {$n_1\neq 1$} the piecewise constant function $n$ as
\begin{equation}\label{eq:slab_refractive}
n \left( x\right) =\left\{ 
\begin{array}{lcr}
n_1 & \hbox{if} & |x|\leq 1 \\
1& \hbox{if} & |x|> 1
\end{array} \right.
\end{equation}

\begin{figure}[h!]
	\hspace*{5mm}
	\begin{tikzpicture}[thick,scale=0.95, every node/.style={scale=0.95}]
	\draw (20.70, 4.20) node{\includegraphics[trim={700 0 0 0},scale=0.802]{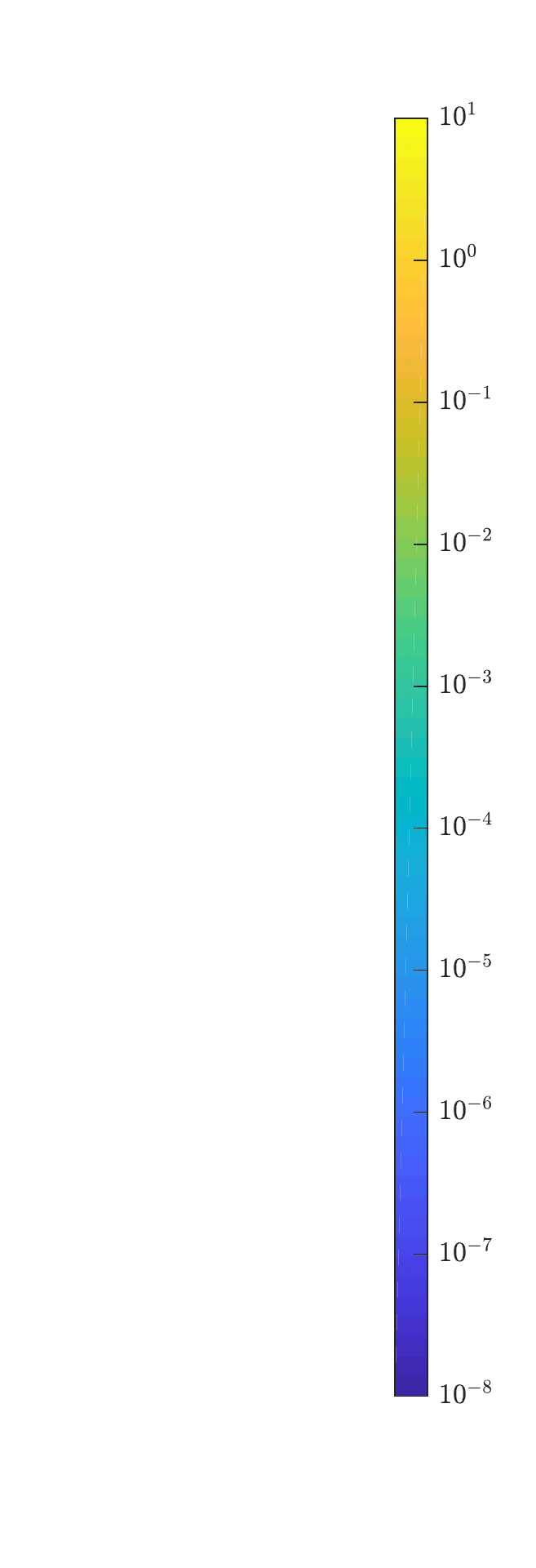}};
	\draw( 0.00, 9.0) node{\includegraphics[scale=0.43]{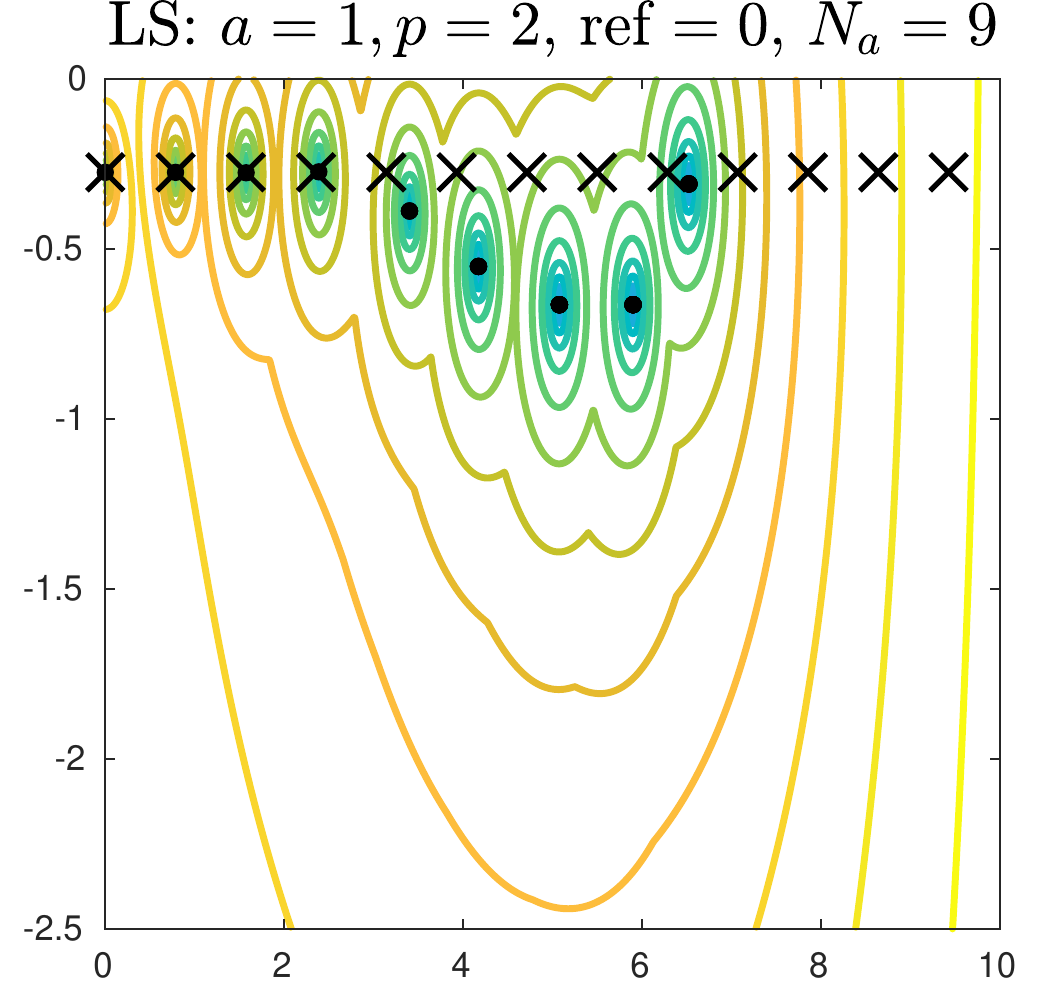}};
	\draw( 4.70, 9.0) node{\includegraphics[scale=0.43]{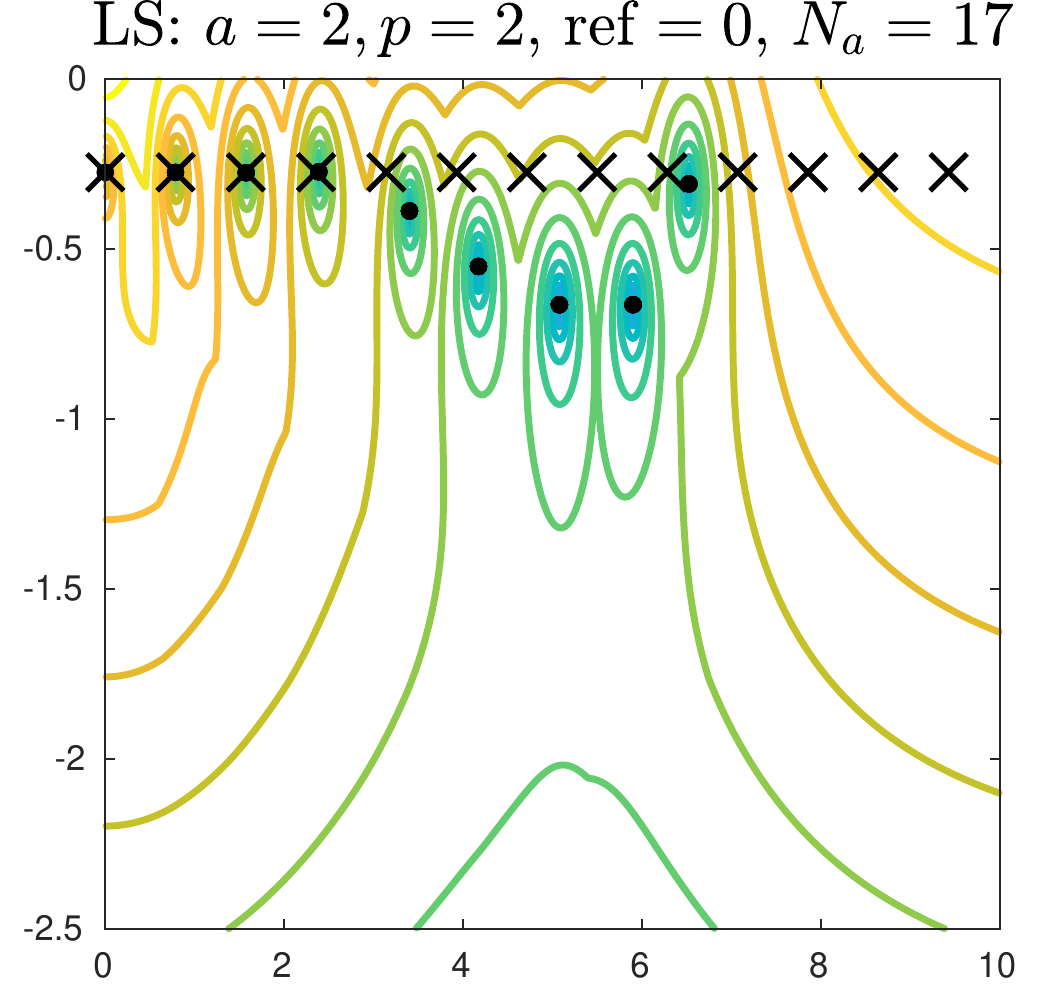}};
	\draw( 9.40, 9.0) node{\includegraphics[scale=0.43]{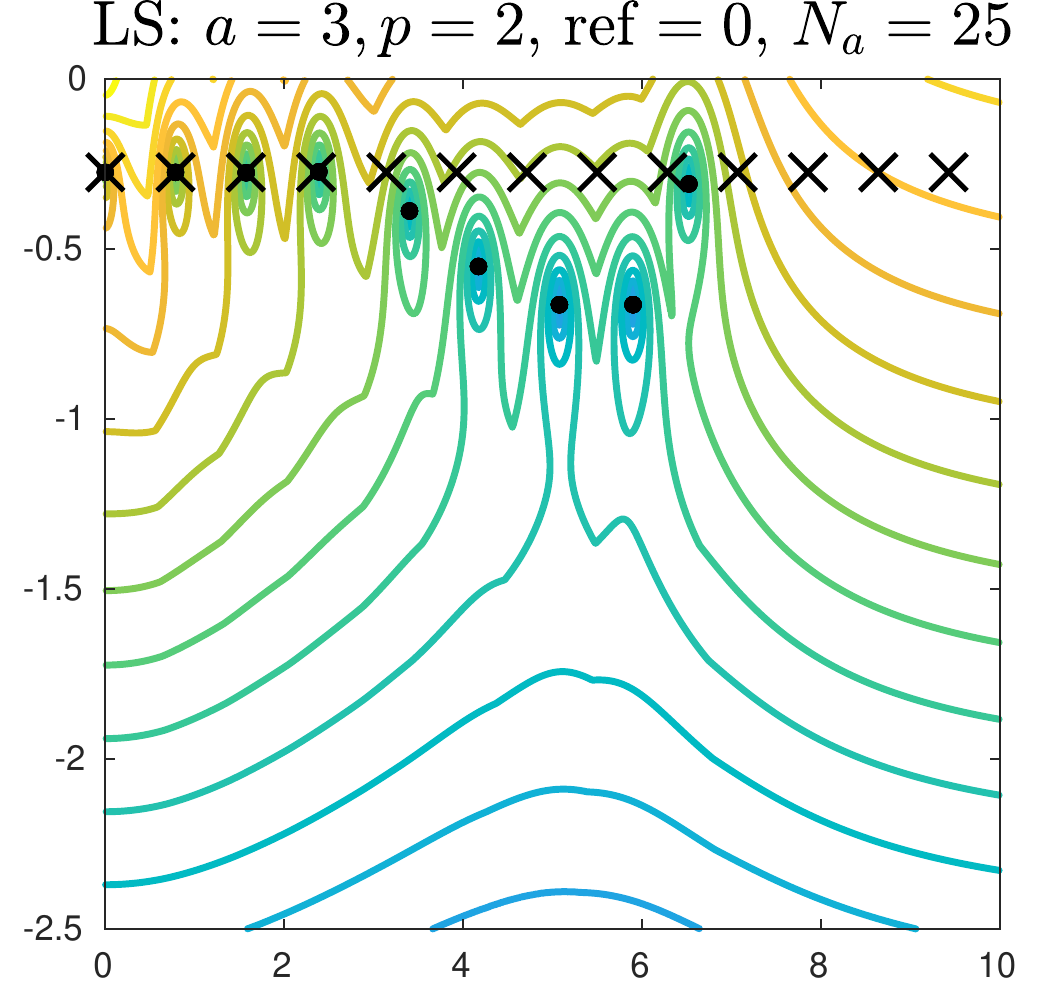}};
	
	\draw( 0.00, 4.5) node{\includegraphics[scale=0.43]{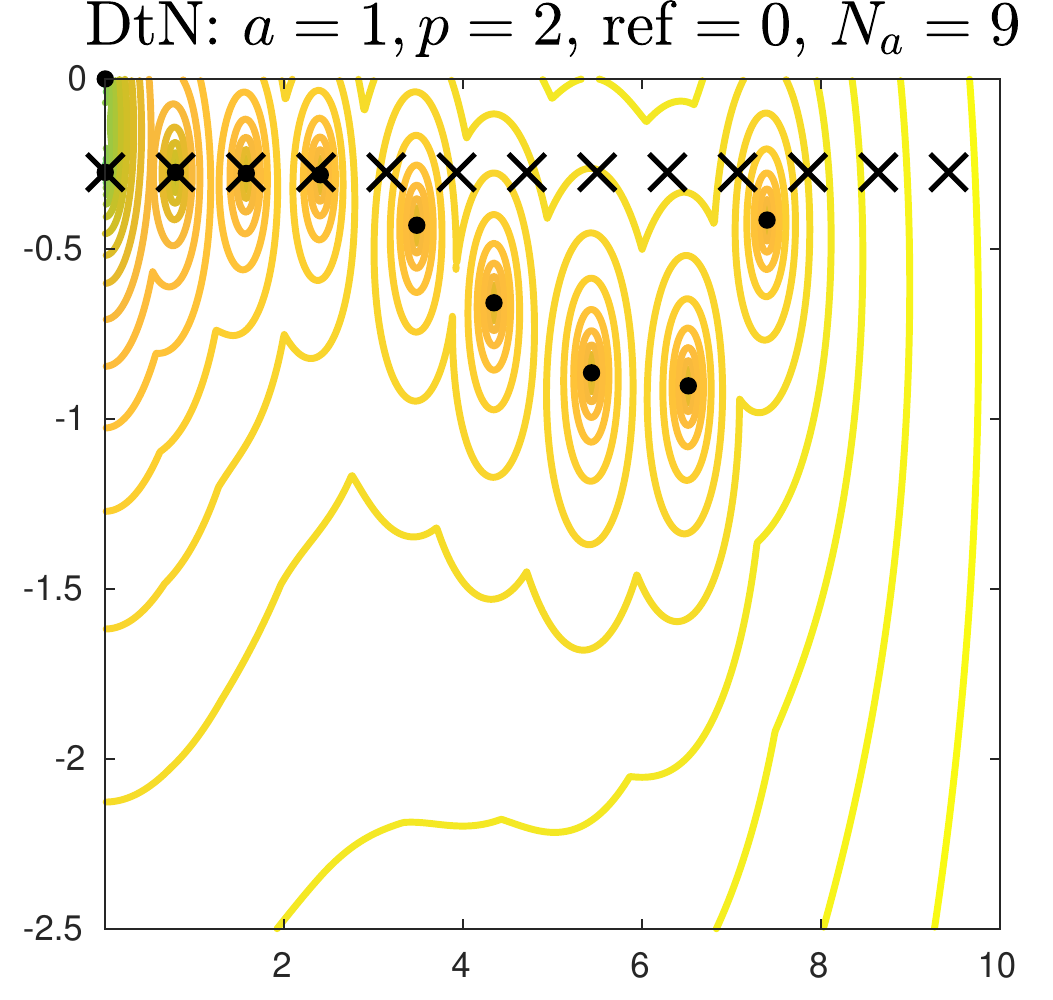}};
	\draw( 4.70, 4.5) node{\includegraphics[scale=0.43]{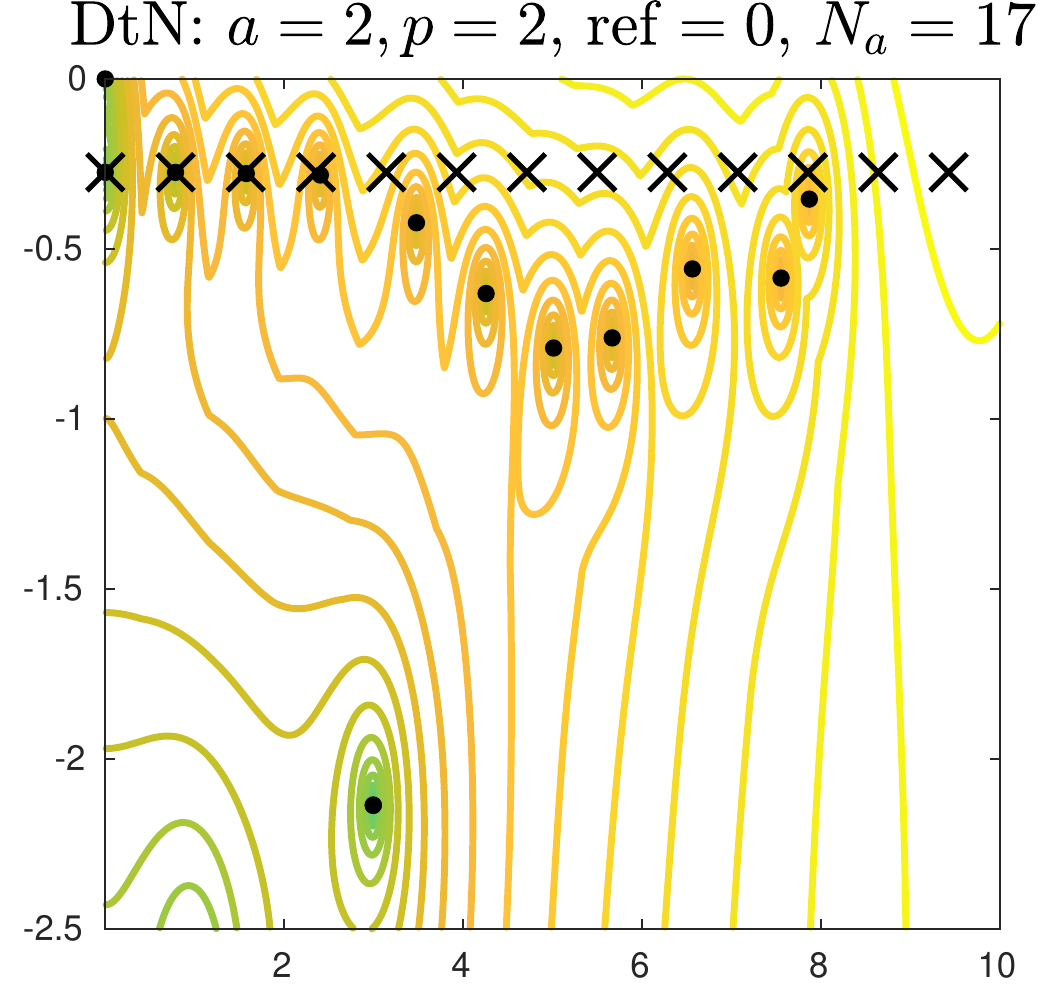}};
	\draw( 9.40, 4.5) node{\includegraphics[scale=0.43]{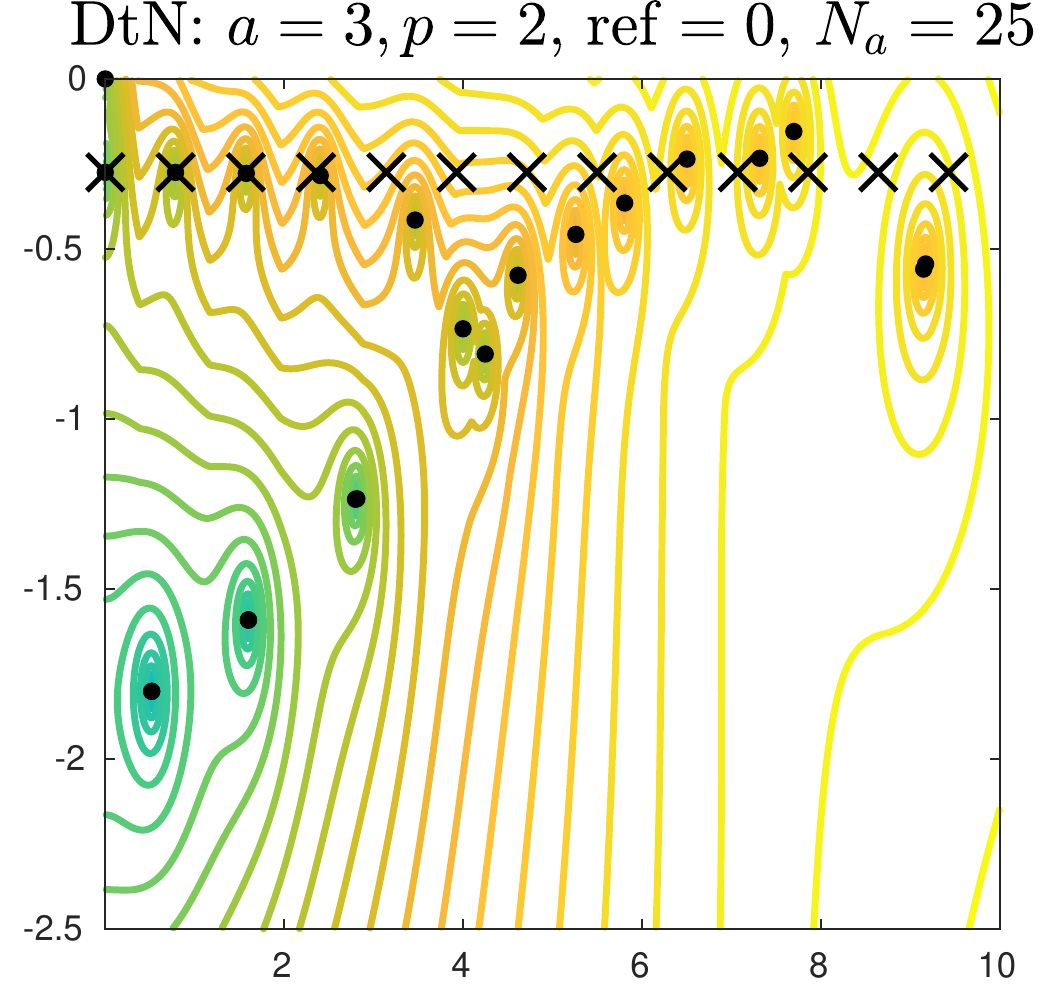}};
	
	\draw( 0.00, 0.0) node{\includegraphics[scale=0.43]{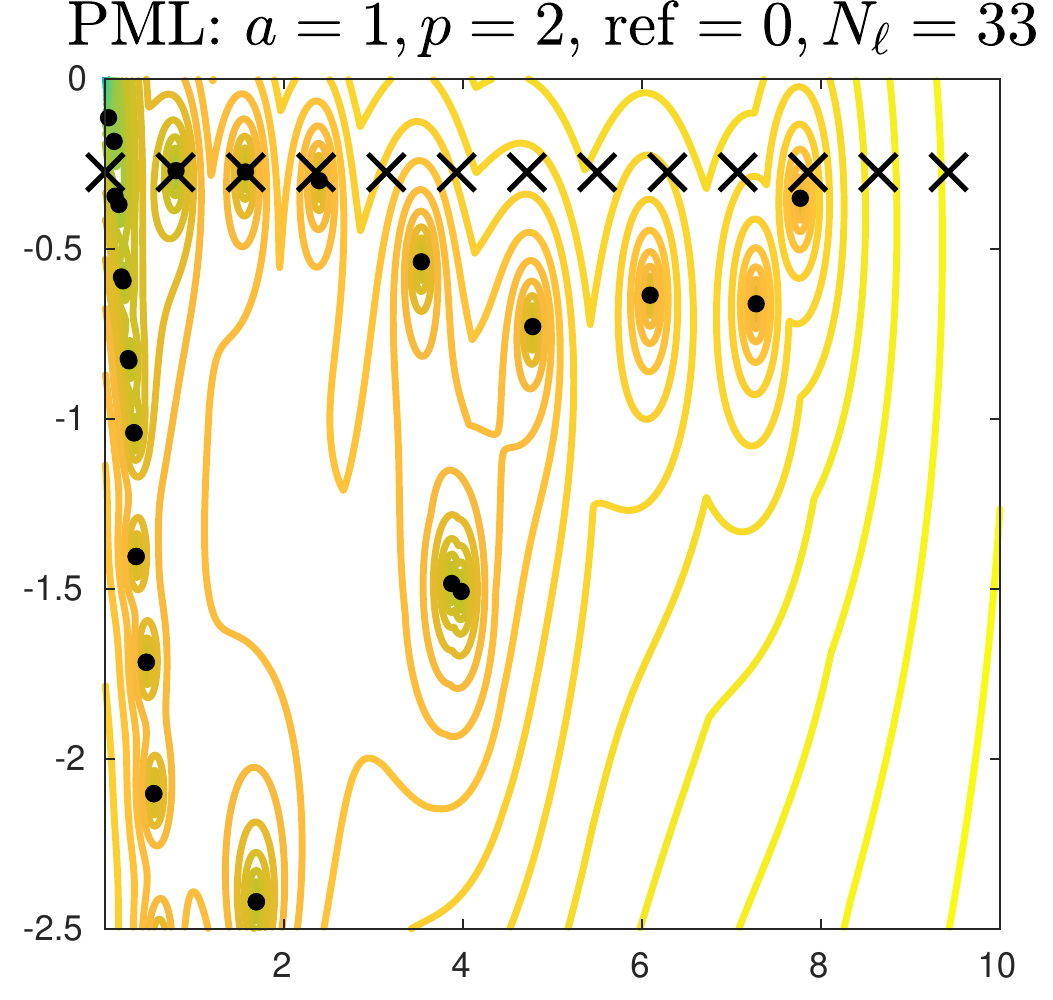}};
	\draw( 4.70, 0.0) node{\includegraphics[scale=0.43]{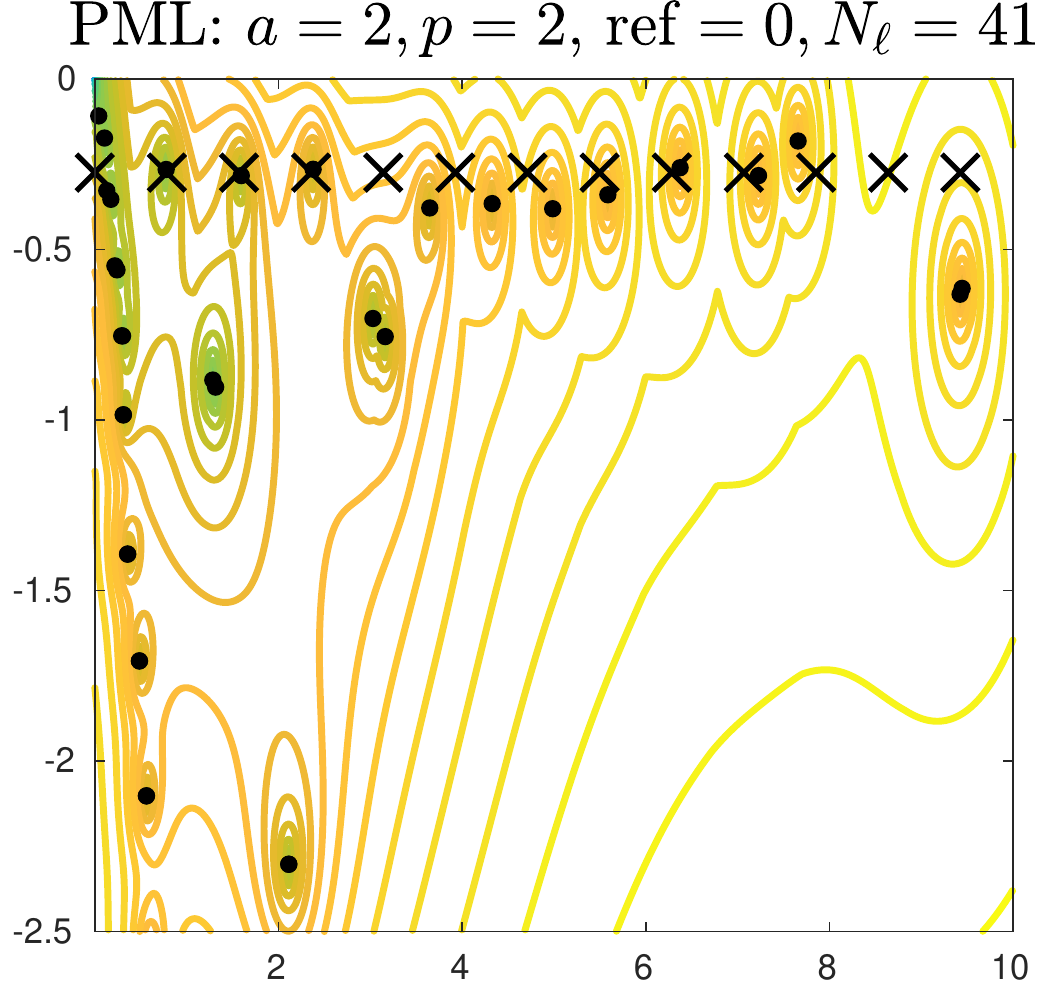}};
	\draw( 9.40, 0.0) node{\includegraphics[scale=0.43]{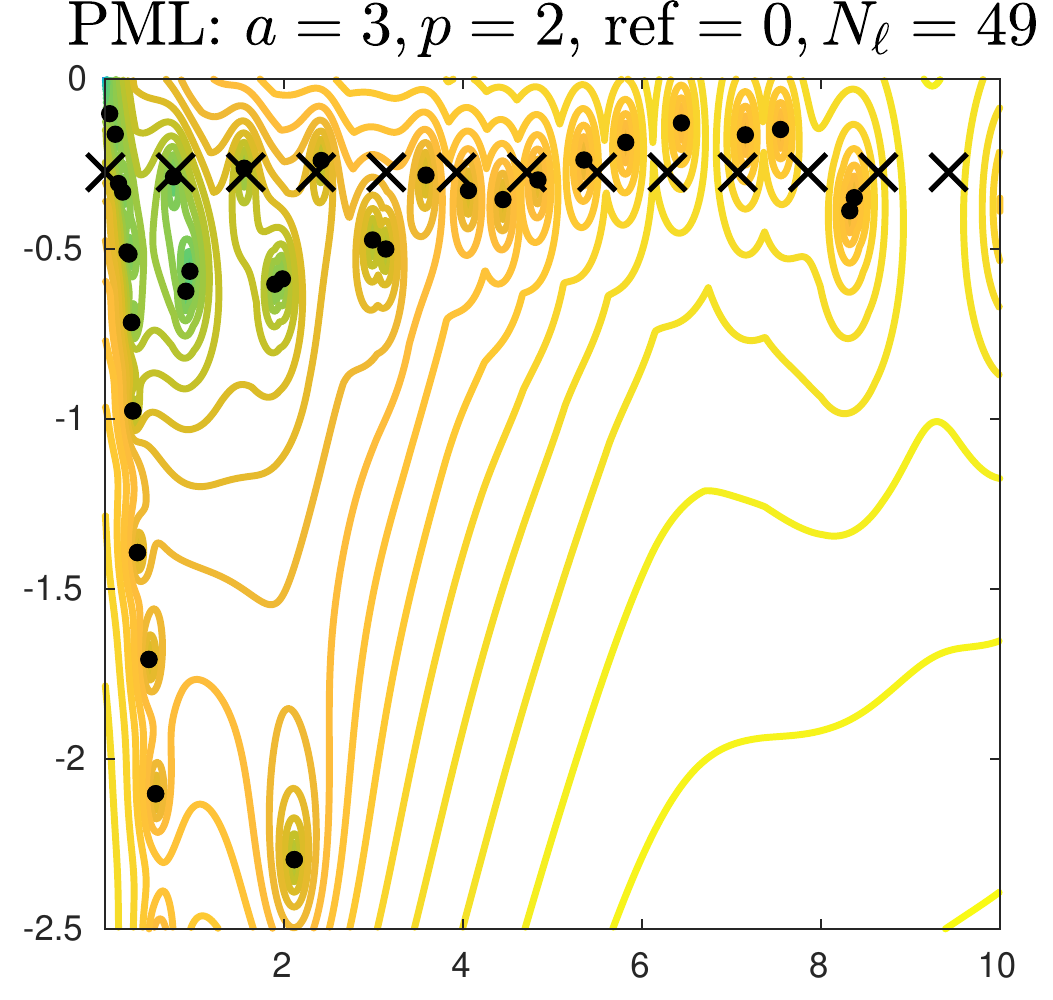}};
	
	\end{tikzpicture}
	\vspace*{-15mm}
	\caption{\emph{Pseudospectrum for the TM Slab problem with coarse mesh: We illustrate for discretizations with fixed $p,h,\ell$ the effect of including air regions for the DtN, PML and Lippmann-Schwinger formulations. For reference, we mark with crosses $(\times)$ exact eigenvalues.}}
	\label{fig:psudo_slab_r0}
\end{figure}


\begin{figure}[h!]
	\hspace*{5mm}
	\begin{tikzpicture}[thick,scale=0.95, every node/.style={scale=0.95}]
	\draw (20.70, 4.20) node{\includegraphics[trim={700 0 0 0},scale=0.802]{arXiv_img/colorbar}};
	
	\draw( 0.00, 9.0) node{\includegraphics[scale=0.43]{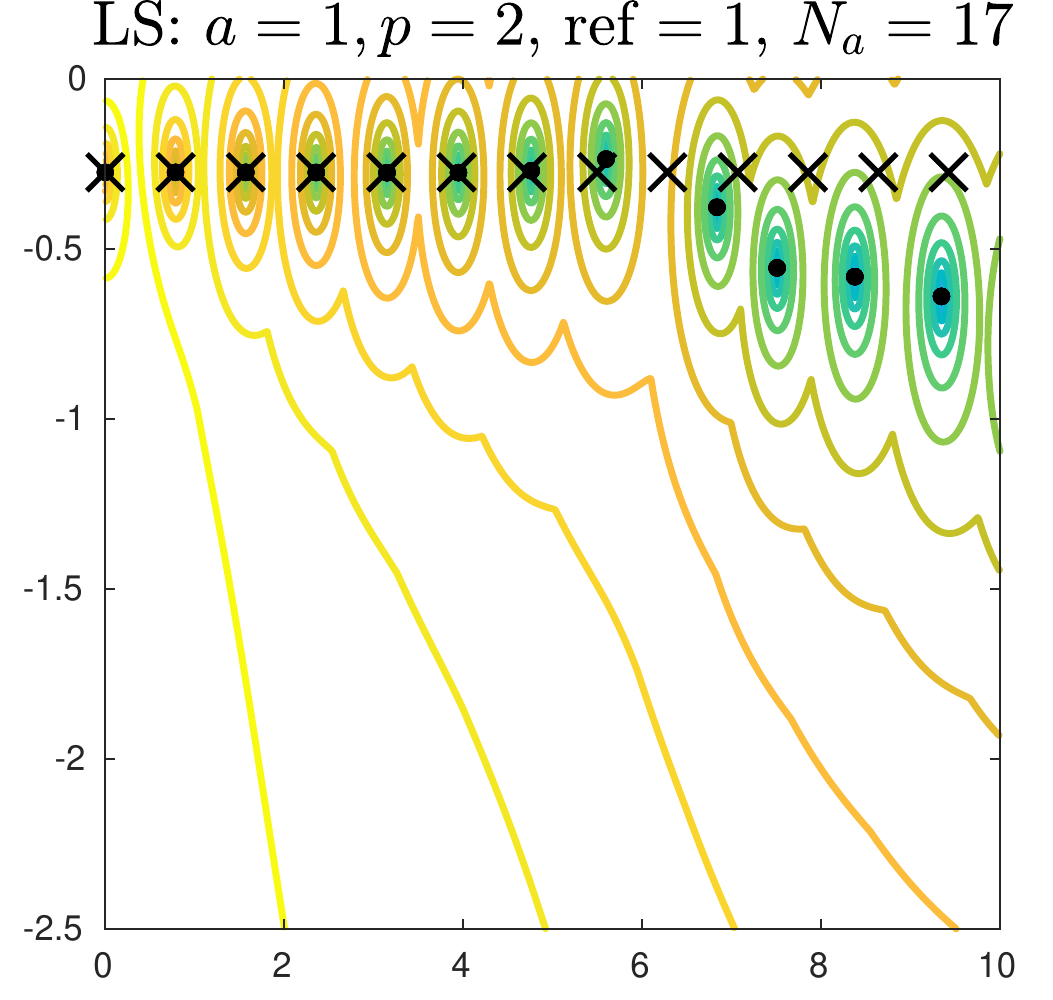}};
	\draw( 4.70, 9.0) node{\includegraphics[scale=0.43]{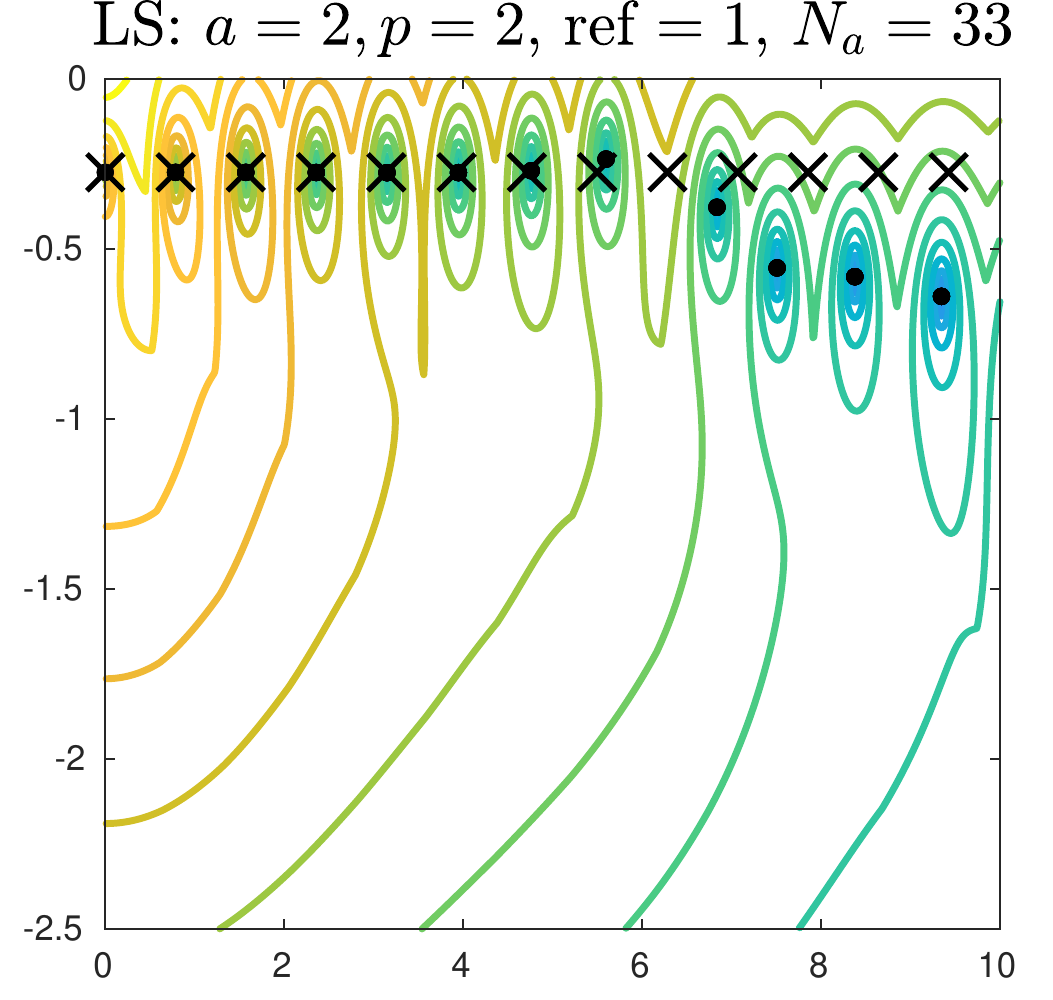}};
	\draw( 9.40, 9.0) node{\includegraphics[scale=0.43]{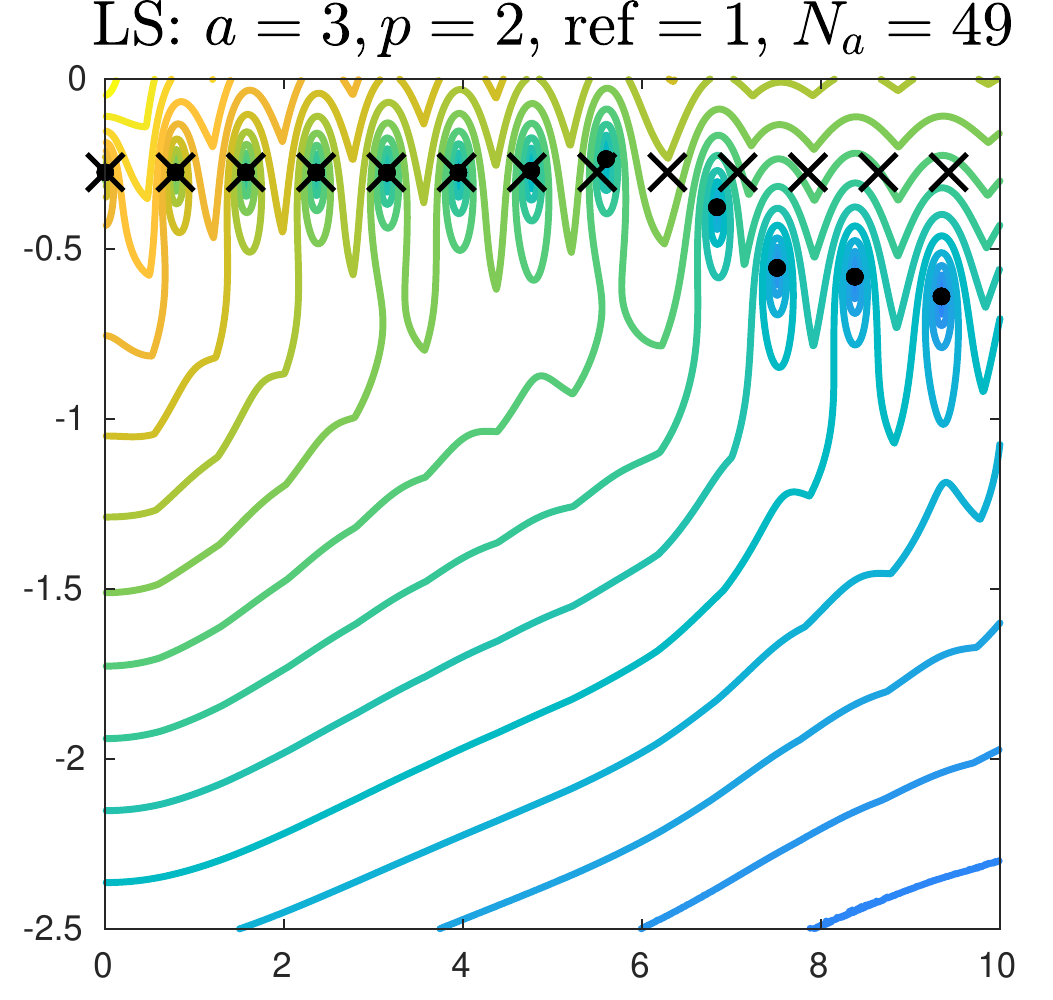}};
	
	\draw( 0.00, 4.5) node{\includegraphics[scale=0.43]{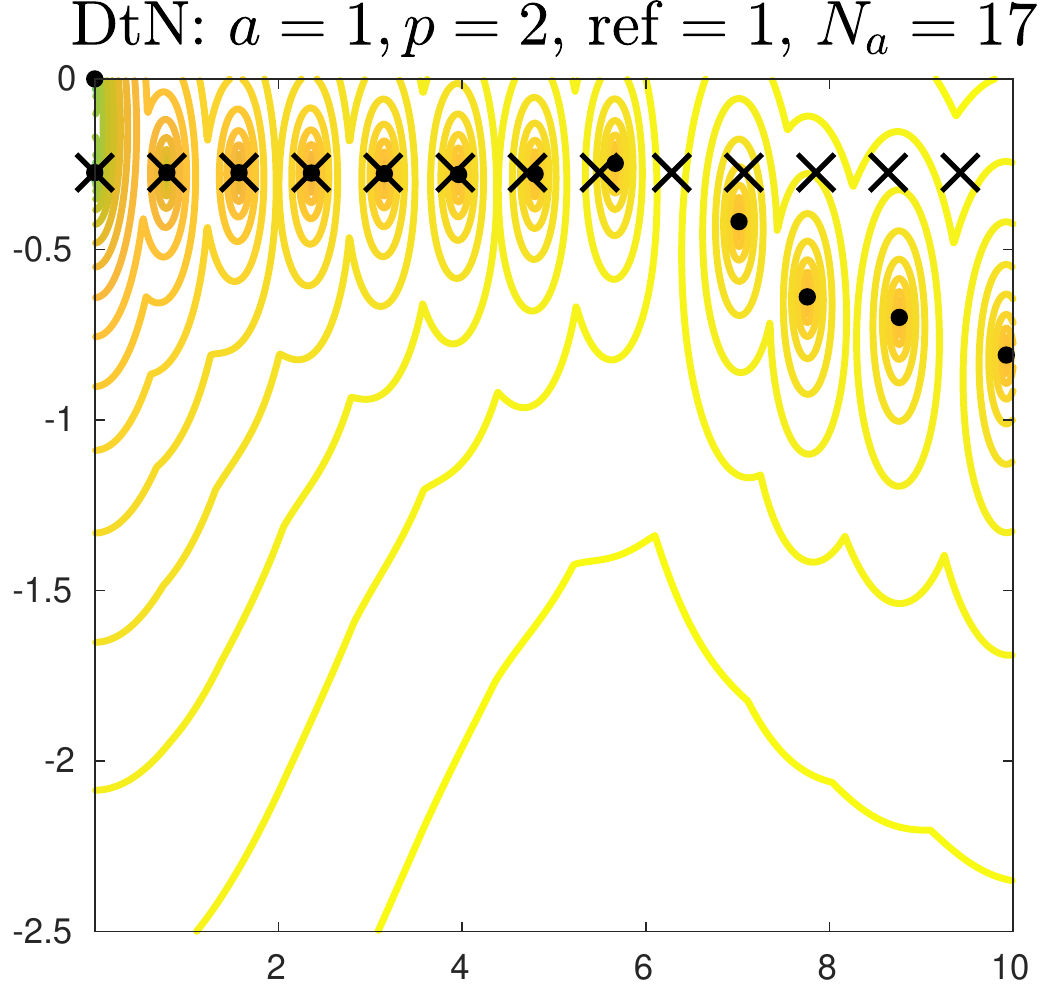}};
	\draw( 4.70, 4.5) node{\includegraphics[scale=0.43]{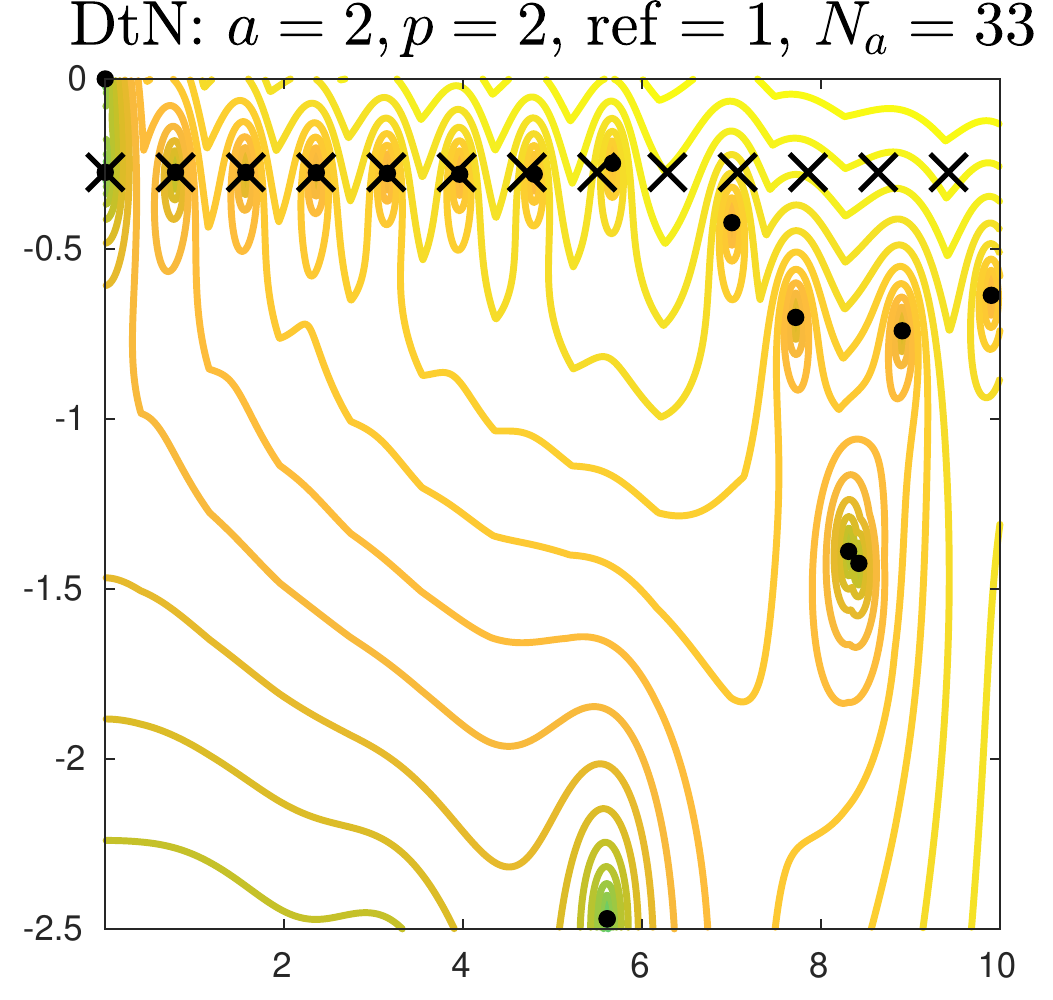}};
	\draw( 9.40, 4.5) node{\includegraphics[scale=0.43]{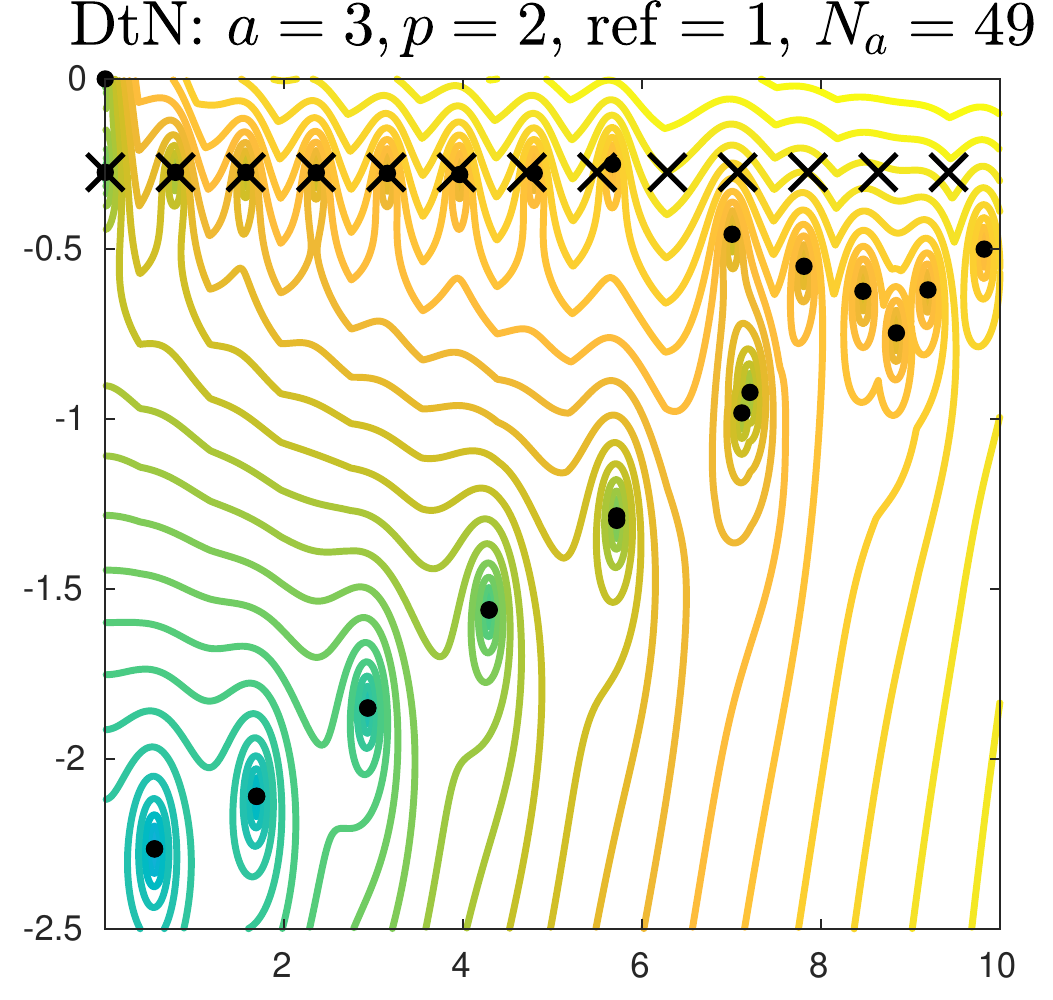}};
	
	\draw( 0.00, 0.0) node{\includegraphics[scale=0.43]{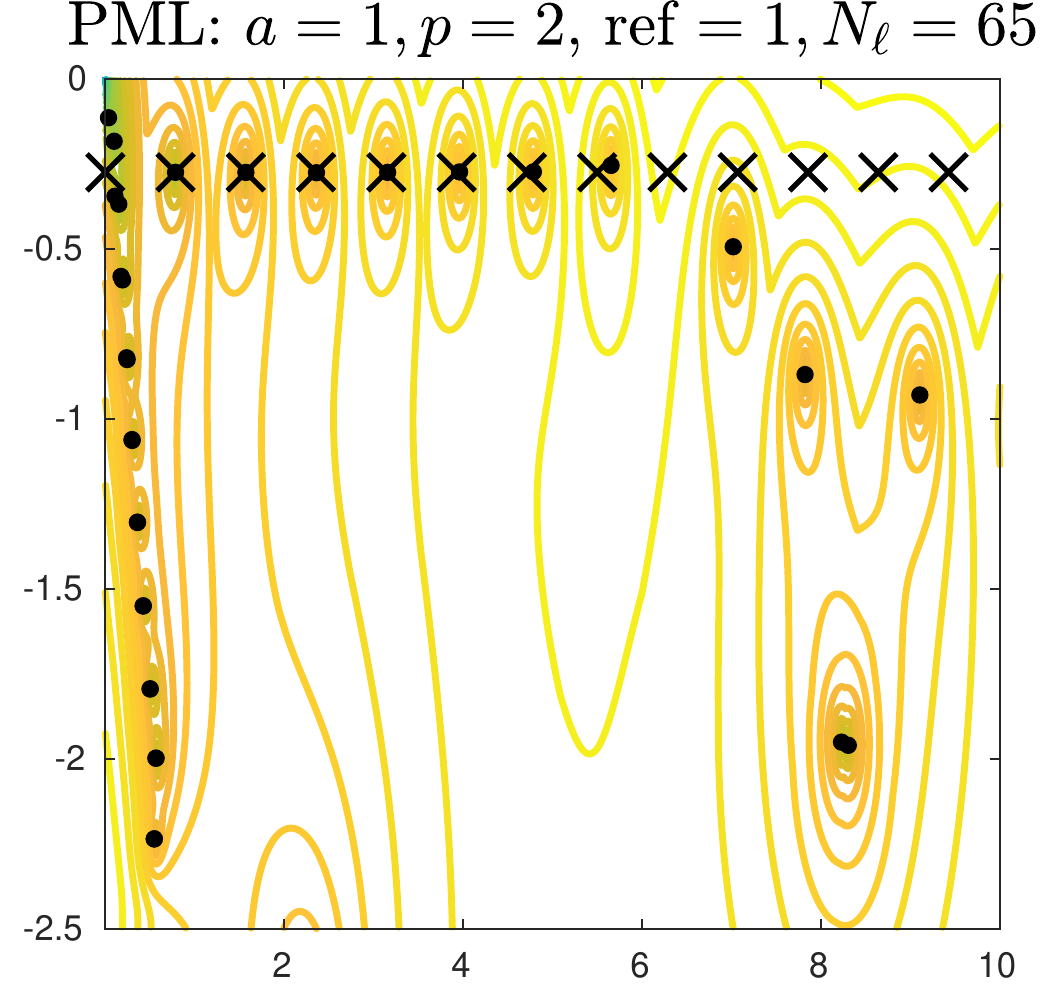}};
	\draw( 4.70, 0.0) node{\includegraphics[scale=0.43]{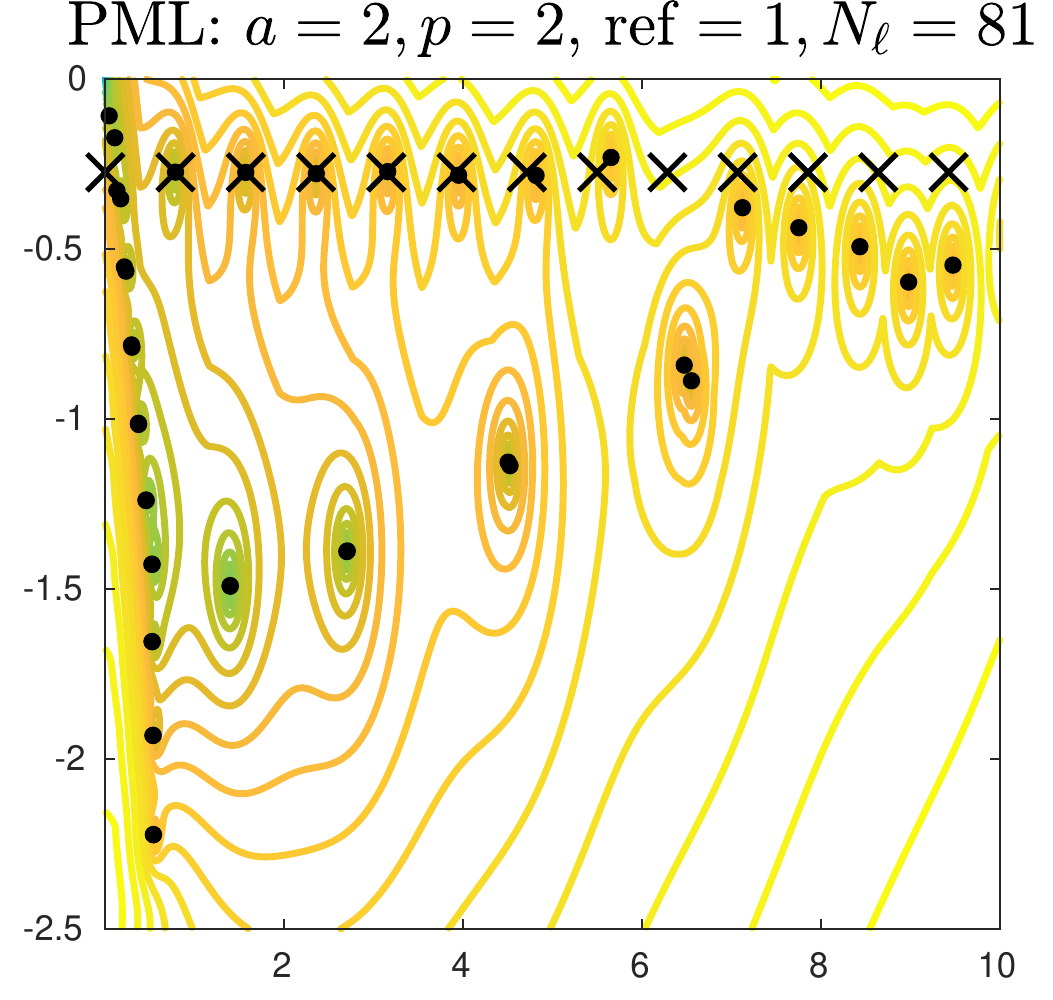}};
	\draw( 9.40, 0.0) node{\includegraphics[scale=0.43]{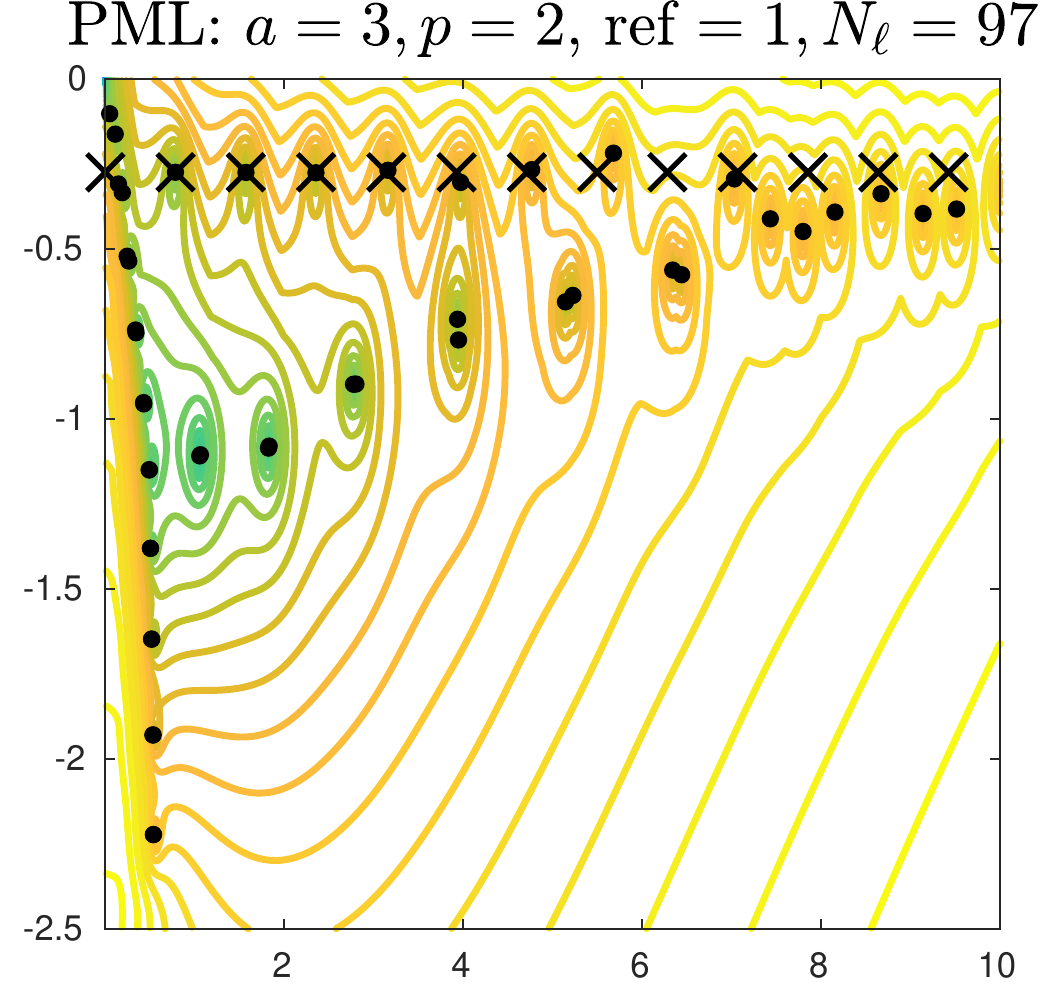}};
	
	\end{tikzpicture}
	\vspace*{-15mm}
	\caption{\emph{Pseudospectrum for the TM Slab problem with one mesh refinement: We illustrate for discretizations with fixed $p,h,\ell$ the effect of including air regions for the DtN, PML and Lippmann-Schwinger formulations. For reference, we mark with crosses $(\times)$ exact eigenvalues.}}
	\label{fig:psudo_slab_r1}
\end{figure}

The corresponding exact resonances of \eqref{eq:Helmholtz}-\eqref{eq:formalDtN} for TM 
polarization are given by
\begin{equation}
e^{4 i n_1 \om}=\mu^2,\,\, \om_m=\fr{\pi m-i\hbox{Log}(\mu)}{2n_1}, \,\, \mu=\fr{n_1+1}{n_1-1},
\label{eq:eig_slabTM}
\end{equation}

We start by discussing the stability of the spectrum of the DtN, PML and LS formulations. Namely, the discretization \eqref{eq:lipp_collocation} inherits the property described in Remark \ref{flexibility_ls}, from where it is expected that the spectrum of the LS formulation would not be sensitive to perturbations of $\Omega_a=(-a,a)$ in the sense discussed in Section \ref{sec:stability}.
For verification, we propose a similar experiment to the one presented in \cite[Figs.~4.1,5.2,5.5]{araujo+engstrom+2017}, but now we allow larger domains $\Om_\dtn\supseteq \Om_r$ in the computation of the Slab problem \eqref{eq:slab_refractive}. 
 
The FE approximations to eigenvalues and the pseudospectrum of the problem are
for $a=1,2,3$ obtained using the tools introduced in Section~\ref{sec:pseudo_comp}. 
We keep track of the number of eigenvalues in a fixed region of the complex plane, and the location of eigenvalues that remain for all perturbations. 

As it can be seen in Figure \ref{fig:psudo_slab_r0} for a coarse mesh, and in Figure \ref{fig:psudo_slab_r1} with one mesh refinement, the number of eigenvalues in the Lippmann Schwinger formulation remained constant for $a=1,2,3$. This means that no spurious eigenvalues are added due to the inclusion of air. Furthermore, from the plots we see that the computed eigenvalues remain unperturbed when increasing $a$, which shows that the eigenvalues are not sensitive to perturbations of the domain.

In the same Figure we follow the last discussion but for computations using the DtN and PML formulations.
It is observed that by increasing $a$, there is an increase of the number of computed eigenvalues in a fixed region of the complex plane.
Additionally, it can be seen that the location of eigenvalues is slightly modified when perturbing $a$.
These observations lead us to conclude that 
eigenvalues from the DtN and PML formulations are very sensitive to perturbations of the domain. 
Note that we for many 1D problems can use any of the methods to obtain very accurate approximations of the resonances without using large computer resources. However, the sensitivity to perturbations is in practice very important for problems in higher dimensions.

Before the next discussion, we briefly introduce quadrature rules for $d=1$, but postpone specific details until Section~\ref{sec:comp_details}. In the one dimensional case, the integration of $f\in C^{2m}(K),\,m\in \mx N$ over an element $K$ is approximated by formulas of the form 
$\int_{K} f(x)\,dx = \sum_{i=1}^{m} {\rm w}_i f({\rm x}_i) + E$, where ${\rm w}_i$ are quadrature weights, ${\rm x}_i$ quadrature nodes, 
and $E$ is the remainder \cite[Ch.~8]{hildebrand87}.
By employing $m$-point quadrature rules with large enough $m$
we can ensure that the remainder $|E|$ is smaller than the FE discretization error.
Finally, we gather the quadrature points defined over $\Om_0$ and $\Om_r$, in the sets $\mc X_0,\mc X_r$ respectively. 
Let 
$\mc X_0:=\cup_{i}\{\mr x_j\in K_i|K_i\subset\Om_0\}_{j=1}^{m}$,
$\mc X_r:=\cup_{i}\{\mr x_j\in K_i|K_i\subset\Om_r\}_{j=1}^{m}$, and $\mc X_a:=\mc X_0\cup \mc X_r$.

In the discretization of the Lippmann-Schwinger formulation \eqref{eq:lipp_collocation}, 
we encounter the situation $x_i, y\in K_l\subset \Om_r$,
where the evaluation of the kernel $\Phi(x_i, y)$ becomes problematic
at the evaluation point $y=x_i$.
In one-dimensional problems ($d=1$), the kernel $\Phi({x}, y)$
is continuous, but has a jump in the derivative at points $x=y$. In the troublesome element $K_l:=(x_l,x_{l+1})$, we can always split the integration interval $K_l \rightarrow (x_l,x_j)\cup (x_j,x_{l+1})$ and perform two separate quadrature integrations. Then, by using Gauss-type of quadratures, it is possible to avoid the evaluation of $\Phi(x_j,x_j)$ \cite{araujo+engstrom+2017}. 

\begin{figure}
	\centering
	\begin{tikzpicture}
	\draw(00.00,0.0) node {
		\includegraphics[scale=0.50,trim=0mm 0mm 0mm 0mm,clip]{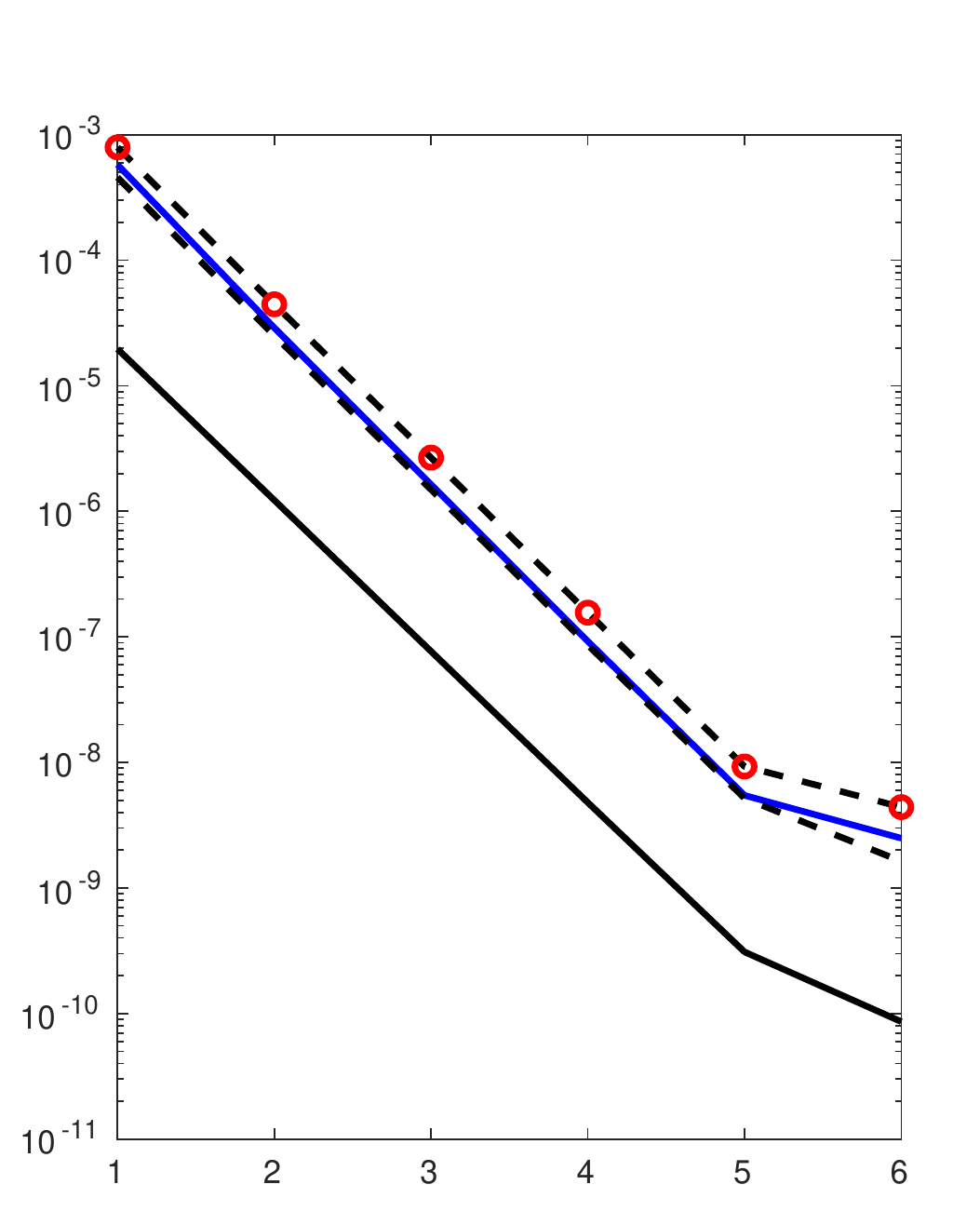} };
	\draw(05.50,0.0) node {
		\includegraphics[scale=0.50,trim=0mm 0mm 0mm 0mm,clip]{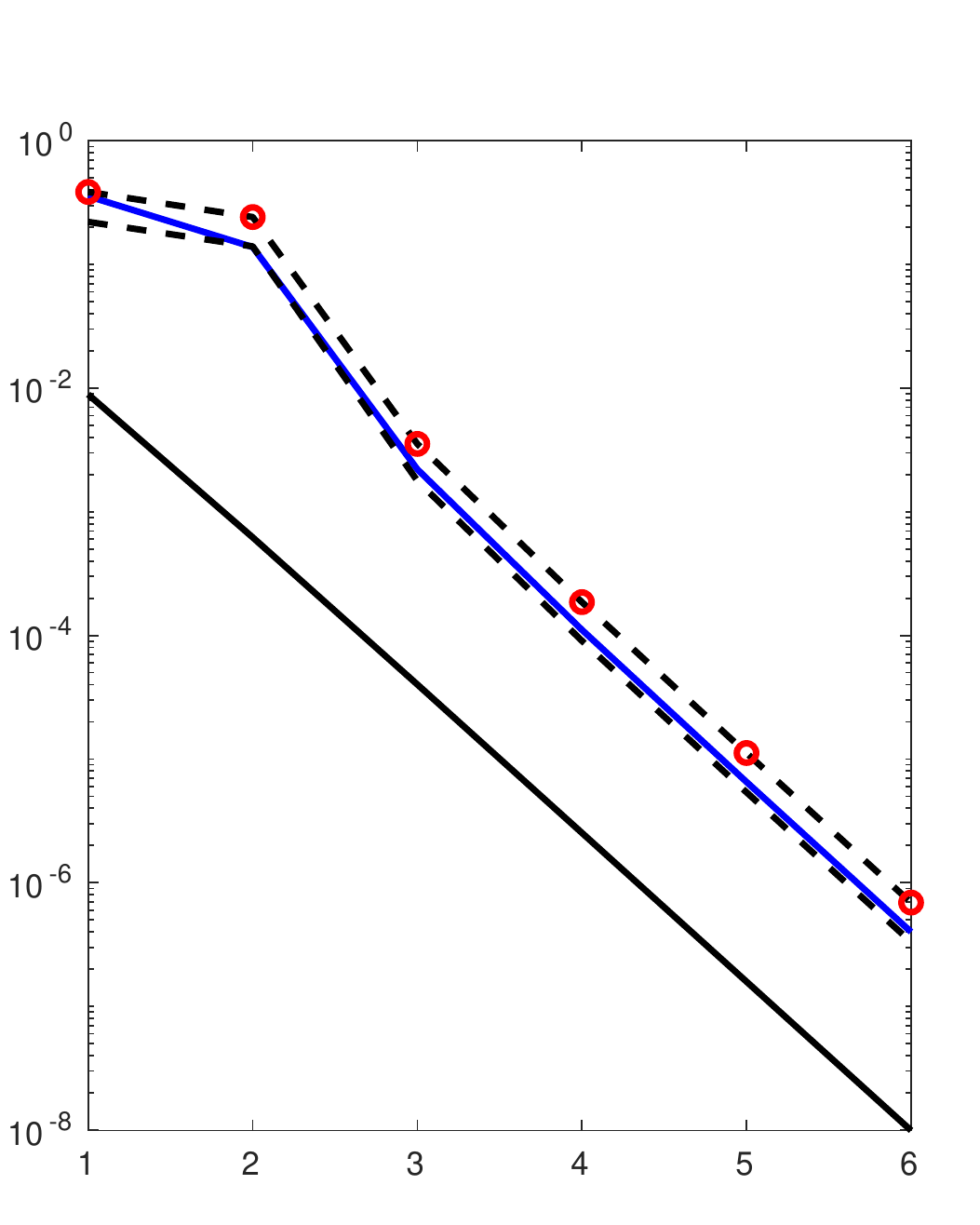} };
	\draw(11.00,0.0) node {
		\includegraphics[scale=0.50,trim=0mm 0mm 0mm 0mm,clip]{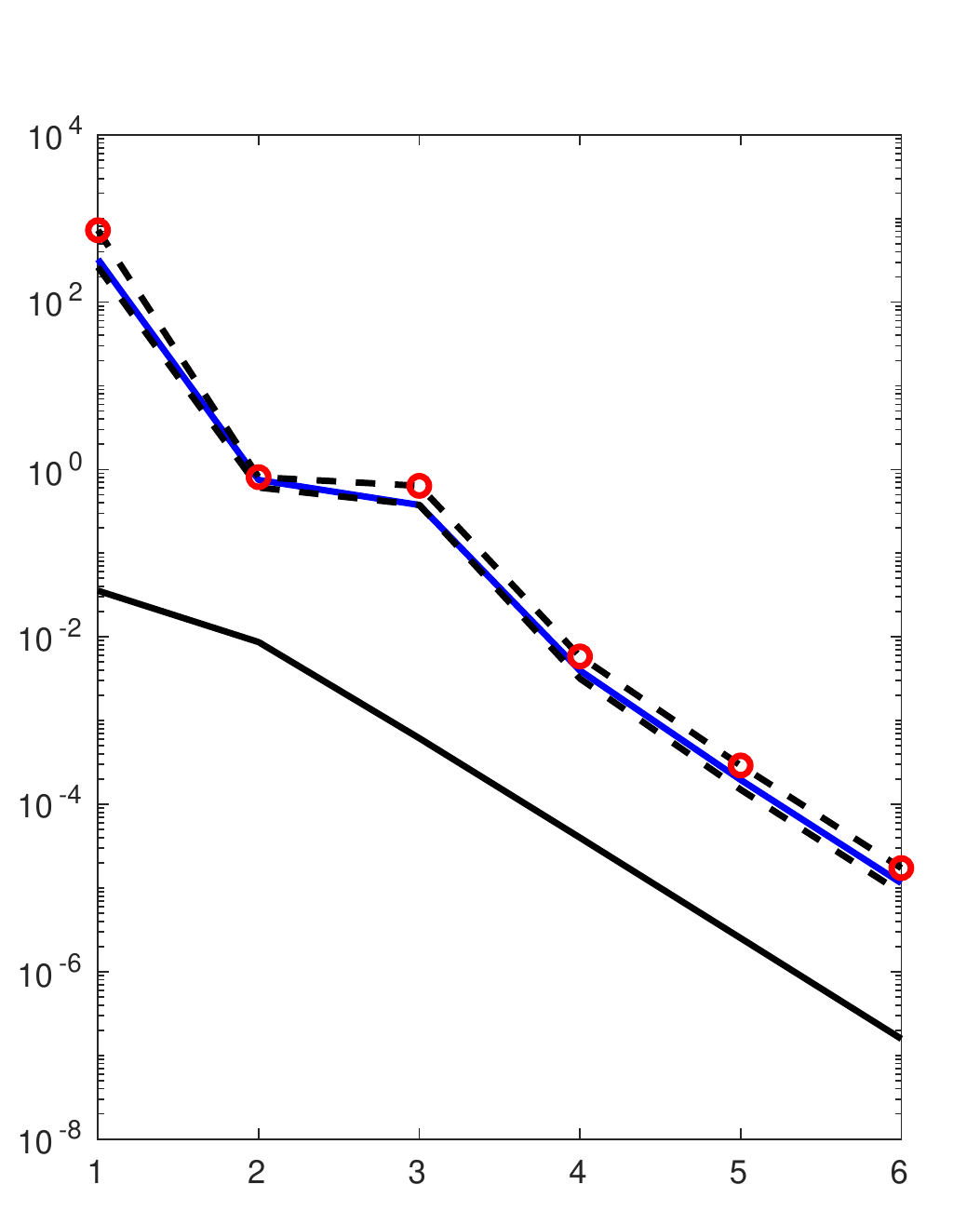} };
	\end{tikzpicture}
	\caption{
		Convergence of various quantities of interest for the development of the sorting strategy presented in Section~\ref{sec:sorting_estimates}.
		The plots are computed for the slab problem described in Section \ref{sec:slab1d} 
		for $a=3$, $p=2$ and featuring five levels of mesh refinement ref $=1,2,\ldots,6$.
		Horizontally, we show results for the eigenpairs corresponding to $m=1,5,10$ from equation \eqref{eq:eig_slabTM}.
		In black solid line we show $|(\om_m-\om^\fem)/\om_m|$ where $\om^\fem$ corresponds to the closest
		computed eigenvalue to $\om_m$, and $u^\fem$ its corresponding computed eigenfunction.
		In blue solid line we show the pseudospectrum indicator $\delta^\fem(\Om_a)$,
		and in dashed lines we show $\min_{{\rm x}\in\mc X_a} |T^\fem(\om^\fem) u^\fem({\rm x})|$ and 
		$\max_{{\rm x}\in\mc X_a} |T^\fem(\om^\fem) u^\fem({\rm x})|$. 
		In red circles we locate the values $\max_{{\rm x}\in\mc X_0} |T^\fem(\om^\fem) u^\fem({\rm x})|$.
	}
	\label{fig:pointwise_filter_p2a3slab}
\end{figure}

In Figure~\ref{fig:pointwise_filter_p2a3slab} we show convergence of various quantities 
of interest for the development of the sorting strategy to be presented in Section~\ref{sec:sorting_estimates}.
The plots are computed for the slab problem \eqref{eq:slab_refractive}
with $p=2$ and featuring five levels of mesh refinement ref $=1,2,\ldots,6$.
Horizontally, we show results for the eigenpairs corresponding to $m=1,5,10$ from equation \eqref{eq:eig_slabTM}.
In black solid line we show $|(\om_m-\om^\fem)/\om_m|$ where $\om^\fem$ corresponds to the closest
computed eigenvalue to $\om_m$, and $u^\fem$ its corresponding computed eigenfunction.
In blue solid line we show the pseudospectrum indicator $\delta^\fem(\Om_a)$,
and in dashed lines we show $\min_{{\rm x}\in\mc X_a} |T^\fem(\om^\fem) u^\fem({\rm x})|$ and 
$\max_{{\rm x}\in\mc X_a} |T^\fem(\om^\fem) u^\fem({\rm x})|$. 
Finally, in red circles we add $\max_{{\rm x}\in\mc X_0} |T^\fem(\om^\fem) u^\fem({\rm x})|$.

From the figure it is evident that as the mesh is refined, there is convergence of the eigenvalue error
$|(\om_m-\om^\fem)/\om_m|$, and residuals $\min_{{\rm x}\in\mc X_a} |T^\fem(\om^\fem) u^\fem({\rm x})|$ and 
$\max_{{\rm x}\in\mc X_a} |T^\fem(\om^\fem) u^\fem({\rm x})|$.
Additionally, the rate of convergence of $\min_{{\rm x}\in\mc X_a} |T^\fem(\om^\fem) u^\fem({\rm x})|$ and 
$\max_{{\rm x}\in\mc X_a} |T^\fem(\om^\fem) u^\fem({\rm x})|$ is the same as the rate of convergence of 
$|(\om_m-\om^\fem)/\om_m|$ and the residuals are around two order of magnitude higher than the eigenvalue error.
Moreover, the values $\min_{{\rm x}\in\mc X_a} |T^\fem(\om^\fem) u^\fem({\rm x})|$ and 
$\max_{{\rm x}\in\mc X_a} |T^\fem(\om^\fem) u^\fem({\rm x})|$ are tight bounds for $\delta^\fem(\Om_a)$.
Finally, the value $\max_{{\rm x}\in\mc X_0} |T^\fem(\om^\fem) u^\fem({\rm x})|$ coincides with
$\max_{{\rm x}\in\mc X_a} |T^\fem(\om^\fem) u^\fem({\rm x})|$ for the one-dimensional case.

The fact that the residuals $\min_{{\rm x}\in\mc X_a} |T^\fem(\om^\fem) u^\fem({\rm x})|$ and 
$\max_{{\rm x}\in\mc X_a} |T^\fem(\om^\fem) u^\fem({\rm x})|$ exhibit the same rate of convergence of the eigenvalue error, 
can be used as an indication of the eigenvalue convergence for problems where an exact solution is not available.

\section{On spurious solutions in $\R^d$}\label{sec:slabdd}

From Figure~\ref{fig:pointwise_filter_p2a3slab} it is, for the studied case, evident  that the quantity $\max_{{\rm x}\in\mc X_a} |T^\fem(\om^\fem) u^\fem({\rm x})|$ can be used to mark potentially spurious scattering resonance pairs. The proposed sorting strategy for higher dimensions is based on this observation. In the following subsections, we discuss computational details and computational cost before the new sorting indicator is introduced in formula \eqref{eq:max_indicator}.

\subsection{Computational details for the evaluation of integrals}\label{sec:comp_details}

{In this section we outline the computational details for the FE discretization of the DtN and PML based formulations in Section \ref{sec:fe_formulations}. Furthermore, we present results on integration of weakly singular kernels that are used in the new numerical sorting scheme in Subsection \ref{subsec:Comp_sorting}. For convenience of the reader we provide a summary of the used standard techniques.}

\subsubsection{Curved elements}\label{sec:curved}

Consider a physical element $K$ and let $\mc K:=(-1,1)^d$ denote the master element.
Numerical quadrature is used to integrate a {function} over $K$ and when high order polynomial spaces are used, it is convenient to compute information from the shape functions $\varphi_j,\,\nabla\varphi_j$ in the master element $\mc K$ and then store it. In this way we gain in performance as computations involving high order polynomials are expensive. 
Consequently, functions defined over a, possibly curved, physical element $K$ are mapped to act over $\mc K$, where numerical quadratures become simpler.
Then, the mapping $X_K:\R^d\rightarrow \R^d$ transforms coordinates as $K = X_K(\mc K)$. The action of the mapping is enforced by the Jacobian's determinant $J=\det(D X_K)$ \cite[Sec.~3.3]{Strang2008}, \cite[Sec.~3.4]{pavel-solin04}, where integrals transform as
\begin{equation}
\int_K f(x)\,dx=\int_{\mc K} f\circ X_K (y)\,J(y)\,dy.
\label{eq:jacobian}
\end{equation}
In the case where $K$ is a line, quadrilateral or a brick element, the explicit expression for $X_K$ is a known linear transformation. When $K$ has curved edges, then $X_K$ can be described by the so-called theory of Transfinite Interpolation \cite{Gordon1973}, and the implementation 
and computational details can be {found} in \cite{Gordon73}, \cite[Sec.~3.2]{pavel-solin04}. 
A general rule of thumb is that the bending of the edges must be small compared to the diameter of the element, and that the angles at the element corners should be close to $\pi/2$. 
For further details and explicit error estimates on curved elements the reader is referred to \cite[Sec.~3.3]{Strang2008}, \cite[Sec.~6.7]{Oden76}. For the description on how $\varphi_j,\,\nabla\varphi_j$ transform from $K$ to $\mc K$, and other related details, the reader is referred to \cite[Sec.~3.3]{pavel-solin04}.

\subsubsection{Numerical quadrature}\label{sec:integration} 

In this subsection, we briefly revise numerical integration by \emph{Gauss-Legendre} quadratures \cite[Ch.~4]{solin04}. 
In the one dimensional case, integration over the master element $\mc K$ is approximated by formulas of the form 
$\int_{\mc K} f(x)\,dx = \sum_{i=1}^{m} w_i f(x_i) + E$, with $w_i$ the quadrature weights and $x_i$ the quadrature nodes.
The coefficients $w_i$ are all positive \cite[Sec.~8.4]{hildebrand87}. 
The so-called quadrature error or remainder is denoted $E$, and  
under the assumption that $f\in C^{2m}(\mc K),\,m\in \mx N$, then 
the $m$-point Gauss quadrature's remainder satisfies
\begin{equation}
|E| \leq c|f^{(2m)}(\xi)|,\,\,\,\text{for}\,\,\,\xi\in \mc K\,\,\,\text{and}\,\,\,c>0.
\label{eq:quad_err}
\end{equation}
Then, the Weierstrass approximation theorem \cite[Sec.~1.2]{hildebrand87} guarantees the existence of a polynomial $P(x)$ such that $\sup_{x\in\mc K} |f(x)-P(x)|\leq \delta$, for an specified $\delta>0$. 
In this way $w_i,x_i$ can be set to minimize $|E|$, and $P(x)$ is integrated exactly. 
Particularly, if $f$ is a polynomial of order $p<2m$, the remainder vanishes and the quadrature gives the exact value of the integral.

An effective way of reducing $|E|$ is by increasing the polynomial degree $p$ until the residual is below $\delta$. In the quadrature formula, increasing $p$ is equivalent to increasing the number of evaluation points $m$.

In $\R^d$, $d>1$, it is possible to derive similar quadrature formulas for the integration of functions that are in $C^{2m}(\mc K)$. The resulting formulas are the so-called \emph{composite Gauss quadratures} \cite[Sec.~4.2.1]{solin04}, which can be constructed in our case by using tensor products of one dimensional quadrature rules.
Particularly, when a physical element $K$ is allowed to be curved,
we approximate integration with the formula 
\begin{equation}
	\begin{array}{rl}
		\dis\int_{K} f(y)\,dy = & \!\!\!\! \dis\int_{\mc K} f\circ X_K (x) J(x) \,dx =
		\dis\sum_{j=1}^{m^d} w_j f\circ X_K (x_j)J(x_j) + E(\mc K) \\[2mm]
		= & \!\!\!\! \dis\sum_{j=1}^{m^d} {\rm w}_j f({\mr x}_j) + E(\mc K),
	\end{array}
	\label{eq:tensor_quadrature}
\end{equation}
where ${\rm w}_j = w_j J(x_j)$ and ${\mr x}_j=X_K (x_j)$
corresponding to a composite Gauss quadrature 
with $m^d$ weights and nodes $w_j, x_j$.
Finally, we discuss quadrature rules when integration is performed over a domain 
$\Om\supseteq \cup_i K_i$ defined as the union of several elements $K_i$. 
The integrand is now required to be piecewise smooth $f\in C^{2m} (K_i)$, for $i=1,2,\ldots,N_{\text{elements}}$.
{Then, we obtain}
\begin{equation}
\int_{\Om} f(x)\,dx=\sum_i \int_{K_i} f(x)\,dx=\sum_i \sum_j \mr w_j (K_i) f(\mr x_j (K_i)) + E(\Om),
\label{eq:composite_quad}
\end{equation}	
where for each element $K_i$, we have the quadrature pairs $\mr x_j (K_i),\mr w_j (K_i)$, similarly as in expression \eqref{eq:tensor_quadrature}.
The polynomial spaces that we use for $d=2,3$ are based on the tensor product of one dimensional finite element spaces \cite[Sec.~2.2]{solin04}, \cite{dealII90}.
As we work with piecewise smooth coefficients, we make sure that the jumps of the integrand coincide with the possibly curved element edges $\partial K_i$, such that $f\in C^{2m}(K_i)$. This allow us to use quadrature rules in each individual element and guarantee convergence of the error of the numerical integration.

Further details on Gaussian quadratures can be revised in e.g. \cite[Ch.~8]{hildebrand87}, and implementation details are provided in \cite[Ch.~4]{solin04}.
\subsubsection{Integrating weakly singular kernels}\label{sec:weak_singular}

In the discretization of the Lippmann-Schwinger formulation \eqref{eq:lipp_collocation}, 
we encounter the problematic situation where the kernel $\Phi(x_i, y)$ 
is required to be evaluated at $y=x_i$.
In the one-dimensional case ($d=1$), this obstacle is easily overcome by
splitting the integration interval and performing two separate integrations
by using Gauss-type of quadratures rules.

In higher dimensions ($d=2,3$) the kernel is weakly singular \cite[Sec.~2.3]{Colton+Kress1983}, what makes the integration in \eqref{eq:lipp_collocation} more demanding. 
This difficulty can be overcome by specializing the quadratures \cite{Kaneko1994,Duan2009,Anand2016}.
Usually, extra effort is spent in refining adaptively elements $K_l$ containing the singularity. 
Then, a Nystr\"om type of high order quadratures, combined with interpolation in polar coordinates 
along with other techniques are used in order to keep $|E(\Om_r)|$ small to desired order. 
As expected, the challenge becomes more pronounced in higher dimension \cite{Duan2009,Anand2016}. 
These techniques where successfully tested \cite{Duan2009,Anand2016} for the solution of scattering problems 
for a given incoming wave employing the Lippmann-Schwinger formulation. However, our case is very different as we look for scattering resonances where the corresponding eigensolver is computationally more demanding than a linear solve.


\subsection{Solution of the nonlinear eigenvalue problems}\label{sec:matrx_solve}

The approximation of resonances based on the DtN formulation for $d=2$ leads to the matrix problem in \eqref{eq:matrix_nep}. The solution of this NEP is based on the solution strategy presented by Ara\'ujo et al.~\cite{araujo+engstrom+jarlebring+2017}, where we use a specialization of the Infinite Arnoldi method \cite{Jarlebring:2012:INFARNOLDI,Jarlebring:2015:WTIARTR} called the tensor infinite Arnoldi method (TIAR). 
In particular we introduce a pole cancellation technique in order to increase the radius of convergence for computation of eigenvalues that lie close to the poles of the matrix-valued function.

For the approximation of resonances when the permittivity function is described by the rational model \eqref{eq:drude_lorentz}, we prefer to solve the corresponding matrix NEP by the techniques presented by 
Ara\'ujo et al.~\cite{araujo+campos+engstrom+roman+2020}, which is a specialization of the solver by G{\"u}ttel et al.~\cite{guttel17} implemented in the SLEPc library \cite{slepc05+roman}.

\subsection{Properties of volume integral equations for resonance computation}\label{sec:ls_properties}

We discretize \emph{volume} integral equations by using the 
scheme presented in \eqref{eq:lipp_collocation}, which does not require any boundary conditions as mentioned in Remark \ref{flexibility_ls}.
In Section \ref{sec:slab1d} numerical computations illustrate that the spectrum of the resulting discrete operator
exhibits desired stability properties for perturbations of $a$.

\begin{remark}\label{ls_coll_dense}
	The resulting system matrices has dimensions comparable to FE matrices: $N_{LS}\times N_{LS}$, with $N_{LS}\leq c h^{-d}$. {However, the} 	matrices are dense ($\mc O(N_{LS}^2)$ storage),
	non-symmetric, and 	the elements of $T(\om)$ are transcendental functions of $\om$.
\end{remark}

{The consequences of Remark \ref{ls_coll_dense} in a NEP solution strategy is that
the matrix $T(\om_{j})$ from \eqref{eq:lipp_collocation}
must be re-assembled for each new iteration $\om_{j+1}=\om_j+\delta\om$, which is computationally expensive, especially for problems of dimension $d>1$.}

\subsection{Memory requirements}\label{sec:assem_mem}

Let $N_K$ be the number of cells in a triangulation in space dimension $d$, $p$ the polynomial degree of the basis functions in use, and $w=16$ bytes is the memory required to store a complex number in double precision. Given the number of non zeros elements $N_z$ in a matrix, the memory required to store it is $W=N_z\times w$.

In the collocation method given in Section~\ref{sec:lipp_collocation}, matrices are dense and we get $N^D_z\approx [N_K\times p^{d}]^2$. In turn, the FE matrices from Section~\ref{sec:assem} are sparse, and each cell in the triangulation contributes with a block of size $[(p+1)\times (p+1)]^d$ support points.
Additionally, we have scattered connections of order $(p+1)^d$ with neighboring cells that we omit for simplicity. A simple estimation gives $N_z^S\approx N_K\times (p+1)^{2d}$.
Furthermore, by assuming $m$ the division in a one dimensional partition, it is reasonable to have 
$N_K=m^d$, and $m=10,10^2,10^3$ for a small, moderate and large problem respectively.

\begin{table}
	\centering
	\begin{tabular}{ | c | c | c | c | l | l | l | l |}
		\hline
		$d$ & $p$  & $m$    & $N_K$  & $N^S_z$       & $N^D_z$        & $W^S(bytes)$ & $W^D(bytes)$ \\ \hline
		\hline
		$1$ & $2$ & $10$ & $10^1$ & $9.0\times 10^1$   & $4.0\times 10^{2}$ & $1.4\times 10^{3}$  & $6.4\times 10^{3}$ \\ \hline
		$2$ & $2$ & $10$ & $10^2$ & $8.1\times 10^{3}$ & $1.6\times 10^{5}$ & $1.3\times 10^{5}$  & $2.6\times 10^{6}$\\ \hline
		$3$ & $2$ & $10$ & $10^3$ & $7.3\times 10^{5}$ & $6.4\times 10^{7}$& $1.2\times 10^{7}$ & $1.0\times 10^{9}$\\ \hline
		\hline
		$1$ & $2$ & $10^2$ & $10^2$ & $9.0\times 10^2$   & $4.0\times 10^{4}$ & $1.4\times 10^{4}$  & $6.4\times 10^{5}$ \\ \hline
		$2$ & $2$ & $10^2$ & $10^4$ & $8.1\times 10^{5}$ & $1.6\times 10^{9}$ & $1.3\times 10^{7}$  & $2.5\times 10^{10}$\\ \hline
		$3$ & $2$ & $10^2$ & $10^6$ & $7.3\times 10^{8}$ & $6.4\times 10^{13}$& $1.2\times 10^{10}$ & $1.0\times 10^{15}$\\ \hline
		\hline
		$1$ & $2$ & $10^3$ & $10^3$ & $9.0\times 10^3$   & $4.0\times 10^{6}$ & $1.4\times 10^{5}$  & $6.4\times 10^{7}$ \\ \hline
		$2$ & $2$ & $10^3$ & $10^6$ & $8.1\times 10^{7}$ & $1.6\times 10^{13}$& $1.3\times 10^{9}$  & $2.6\times 10^{14}$ \\ \hline
		$3$ & $2$ & $10^3$ & $10^9$ & $7.3\times 10^{11}$& $6.4\times 10^{19}$& $1.2\times 10^{13}$ & $1.0\times 10^{21}$ \\ 
		\hline
	\end{tabular}
	\caption{\emph{Memory consumption estimation for matrices in \eqref{eq:matrix_nep}, and \eqref{eq:lipp_collocation} for $d=1,2,3$. }}
	\label{tab:matrix_memory}
\end{table}

{We list our simple estimations in table \ref{tab:matrix_memory} for $p=2$, for both FE discretization methods in Section~\ref{sec:assem} and collocation methods for volume integral equations in Section~\ref{sec:lipp_collocation}.
As expected, dealing with FE matrices is a standard way of discretizing wave problems and results in manageable sparse matrices for current computer memory constraints. It is evident that the load becomes larger with higher space dimension and number of cells, from
where matrices for $d=3$ can be stored only for small and moderate problems. 
In the case of matrices from the collocation strategy in \eqref{eq:lipp_collocation}, we conclude that 
storage becomes computationally unfeasible in higher space dimensions, for moderate and large problems.
Additionally, working with higher polynomial degree rules out even small problems.

\subsection {Computational platform and details}\label {sec:comp_details}
All numerical experiments have been carried out using the finite element library \verb+deal.II+ \cite{dealII90} with Gauss-Lobatto shape functions \cite[Sec.~1.2.3]{solin04}. For fast assembly and computations with complex numbers the package PETSc \cite{petsc-efficient} is used.

The computational platform was provided by the High-Performance Computing Center North (HPC2N) at Ume\aa\, University, and all experiments were run on the distributed memory system Abisko. The jobs were run in serial on an exclusive node: during the process, no other jobs were running on the same node. Node specifications: four AMD Opteron 6238 processors with a total of 48 cores per node.

\subsection {Computational details of the sorting scheme}\label{subsec:Comp_sorting}
In order to evaluate the sorting scheme, we are interested in computing the integrals from \eqref{eq:lipp_collocation} as accurately as possible. The available FE machinery for computing integrals over $\Om_0$ facilitates the numerical integration, which is done similarly as described in Section \ref{sec:integration}.

Due to the growth of most resonant modes, the point-wise residual $|(T^\fem(\om^\fem)u^\fem)(x_j)|$ is expected to be larger for $x_j\in\Om_0$. 
Additionally, as discussed in Section \ref{sec:weak_singular}, 
due to the unboundedness of $\Phi(x_j,x_j)$, 
the computation of \eqref{eq:lipp_collocation}, and \eqref{eq:pseudo_ind} for $x_j\in\Om_r$ requires considerable more effort compared to its evaluation for $x_j\in\Om_0$. The apparent reason for this is that for $x_j\in\Om_r$ and $d>1$, we have to numerically compute an integral with a weak singularity as discussed in Section \ref{sec:weak_singular}.

Below, we write explicitly the steps involved in computing $\pseudo^\fem(\Om_\dtn)$ 
from Definition \ref{def:sorting}. 
First, we split the integration into separate parts over $\Om_0,\Om_r$ 
and use the composite quadrature rules \eqref{eq:composite_quad} for evaluating the integrals. We need the following definitions.
\begin{definition}\label{def:quad_sets}
Let $\mc K_0:=\{i:K_i\subset \Om_0\}$ and $\mc K_r:=\{i:K_i\subset \Om_r\}$ 
be index sets defined over $\Om_0$ and $\Om_r$, respectively. We define the sets 
$\mc X_0:=\cup_{i\in\mc K_0}\{\mr x_j\in K_i\}_{j=1}^{m^d}$,
$\mc X_r:=\cup_{i\in\mc K_r}\{\mr x_j\in K_i\}_{j=1}^{m^d}$, and denote by
$\mc I_0$, $\mc I_r$ the resulting extracted index sets from the new ordering.
Additionally, let $\mc X_a:=\mc X_0\cup \mc X_r$ and $\mc I_a$ its corresponding ordering.
In this way we obtain the new quadrature rules
\begin{equation}
	\int_{\Om_q} f(x)dx=\sum_{j\in \mc I_q} \mr w_j f(\mr x_j)+E(\Om_q), \,\,\, \text{with}\,\,\,q=0,r,a.
\end{equation}
\end{definition}
Then, we have %
\begin{equation}
	\begin{array}{rl}
	\pseudo^\fem(\Om_\dtn)^2 \!\! & = \|T^\fem(\om^\fem)u^\fem\|^2_{\Om_\dtn} \\[2mm]
	& = \int_{\Om_0} |(T^\fem(\om^\fem)u^\fem)(x)|^2\,dx 
			+ \int_{\Om_r} |(T^\fem(\om^\fem)u^\fem)(y)|^2\,dy \\[2mm]
	& = \sum_{k\in \mc I_0} \mr w_k \alpha^2_{k} + \sum_{l\in \mc I_r} \mr w_l \beta^2_{l} + E(\Om_0)+E(\Om_r),
\end{array}
	\label{eq:filter_quadrature}
\end{equation}
where $\alpha_{k}:=|(T^\fem(\om^\fem)u^\fem)(\mr x_k)|,\,\,
\beta_{l}:=|(T^\fem(\om^\fem)u^\fem)(\mr y_l)|$, for $\mr x_k\in\Om_r,\,\mr y_l\in\Om_0$.
From \eqref{eq:lipp_collocation}, the evaluation of $T(\om)u$ involves
an integration over $\Om_r$, which we refer to as \emph{inner loop}. Then,
for each $k$ in $\alpha_{k}$, and each $l$ in $\beta_{l}$ we compute an inner loop additional to the explicit integration shown in \eqref{eq:filter_quadrature}.

\begin{definition}\label{def:quad_indicator}
	For a given quadrature, we define the discrete pseudospectrum indicator as 
	\begin{equation}\label{eq:pseudo_ind_h}
		\pseudo_h^\fem(\Om_\dtn)^2:= \sum_{k\in \mc I_0} \mr w_k|(T^\fem(\om^\fem)u^\fem)(\mr x_k)|^2  + 
								   \sum_{l\in \mc I_r} \mr w_l |(T^\fem(\om^\fem)u^\fem)(\mr y_l)|^2.
	\end{equation}
\end{definition}

\subsubsection {Computational costs}\label{sec:filter_costs}

In this subsection, we estimate the computational cost for performing the operations involved in \eqref{eq:pseudo_ind_h}.
Computationally, the errors in \eqref{eq:filter_quadrature} require special treatment as discussed in \ref{sec:weak_singular}. However, for simplicity of the estimations, we disregard additional costs from integration of weakly singular kernels in higher dimensions.

We estimate the costs in terms of the evaluation of $u^\fem(x_j)$ and $\Phi(x_i,x_j)$ in complex double precision, 
which combined account for the heaviest work load
in each individual term of \eqref{eq:pseudo_ind_h}. 
The evaluation of these two operations account for a rough computational time $t_q\approx 10^{-6}s$ bench-marked on the processor \verb+Intel Core i7-3770, CPU: 3.40GHz+.

For the estimation, assume that we have $N_K$ cells $K_i\subset \Om_\dtn$ and choose a finite element space of degree $p$. Then, we have $N_K\times (p+1)^d$ terms in the outer loop and the inner loop requires $N_r\times (p+1)^d$ terms, where $N_r$ denotes the number of cells in $\Om_r$.
The estimate for the computational cost for $\|T^\fem(\om^\fem)u^\fem\|^2_{\Om_\dtn}$ is about $N_K\times (p+1)^d\times N_r\times (p+1)^d=N_K\times N_r\times (p+1)^{2d}$ evaluations of the kernel.
Assume $m$ as the size of a one dimensional partition,
from where it is reasonable to have 
$N_r=m^d$, $N_K=(cm)^d$, for $c>1$. 
Then, the cost is given by $c^d\times( m\times(p+1))^{2d}$.

Aiming at getting a better intuition of the requirements of the computation,
we can check estimations for the cost, inner time $t_i$, total time $t$ running on a single processor. 
We set $c=2$ and for a small, moderate and large problem
we use $m=10,10^2,10^3$ respectively. 
In Table \ref{tab:check_cost} we show the estimations for the computational costs and times required by \eqref{eq:filter_quadrature}.

The presented sorting scheme is fully parallelizable and the total time of execution can be reduced by a factor of ten by using additional cores. However, the conclusion of Table \ref{tab:check_cost} is that the computational cost is extremely high for realistic computations.
Basically, evaluating $\pseudo^\fem_h(\Om_\dtn)$ as in \eqref{eq:pseudo_ind_h} results in sorting schemes that are far more expensive than the solution of the NEP \eqref{eq:matrix_nep}.

\begin{table}
  \centering
  \begin{tabular}{ | c | c | c | c | l | l | l |}
    \hline
    $d$ & $p$  & $m$    & $N_r$  & $\text{cost}$       & $t_i(s)$             & $t(s)$ \\ \hline
    \hline
    $1$ & $2$  & $10^1$ & $10^1$ & $1.8\times 10^3$    & $9.0\times 10^{-5}$  & $1.8\times 10^{-3}$ \\ \hline
    $2$ & $2$  & $10^1$ & $10^2$ & $3.2\times 10^{6}$  & $8.1\times 10^{-3}$  & $3.2\times 10^{0}$  \\ \hline
    $3$ & $2$  & $10^1$ & $10^3$ & $5.8\times 10^{9}$  & $7.3\times 10^{-1}$  & $5.8\times 10^{3}$  \\ \hline
    \hline
    $1$ & $2$  & $10^2$ & $10^2$ & $1.8\times 10^5$    & $9.0\times 10^{-4}$  & $1.8\times 10^{-1}$ \\ \hline
    $2$ & $2$  & $10^2$ & $10^4$ & $3.2\times 10^{10}$ & $8.1\times 10^{-1}$  & $3.2\times 10^{4}$  \\ \hline
    $3$ & $2$  & $10^2$ & $10^6$ & $5.8\times 10^{15}$ & $7.3\times 10^{2}$   & $5.8\times 10^{9}$  \\ \hline
    \hline
    $1$ & $2$  & $10^3$ & $10^3$ & $1.8\times 10^7$    & $9.0\times 10^{-2}$  & $1.8\times 10^{2}$  \\ \hline
    $2$ & $2$  & $10^3$ & $10^6$ & $3.2\times 10^{14}$ & $8.1\times 10^{1}$   & $3.2\times 10^{8}$  \\ \hline
    $3$ & $2$  & $10^3$ & $10^9$ & $5.8\times 10^{21}$ & $7.3\times 10^{5}$   & $5.8\times 10^{15}$ \\ 
    \hline
  \end{tabular}
  \caption{\emph{Cost and time estimation for computing \eqref{eq:pseudo_ind_h} for $d=1,2,3$. }}
  \label{tab:check_cost}
\end{table}

\subsubsection{Sorting strategy}\label{sec:sorting_scheme}
A pseudospectrum strategy based on Definition \ref{def:quad_indicator}, consist of the
sorting of computed pairs $(\om_m^\fem,\xi_m)$, solution to $\eqref{eq:matrix_nep}$, according to their respective indicator $\pseudo^\fem_h(\Om_\dtn)$. 
As discussed in Section \ref{sec:weak_singular}, the evaluation of the $\beta_{l}$ 
in \eqref{eq:filter_quadrature} requires special treatment such as non-standard quadrature rules similar to the ones introduced in \cite{Kaneko1994,Duan2009,Anand2016}. Additionally, as estimated in Section~\ref{sec:filter_costs} and Table \ref{tab:check_cost}, the evaluation of 
$\pseudo^\fem_m(\Om_\dtn):=\|u_m-Ku_m\|_{L^2(\Om_\dtn)}$ 
in higher dimensions
is prohibitively expensive since $Ku_m$ is a volume integral operator.
With these issues in mind, our aim is to propose an approximated version of $\pseudo^\fem_h(\Om_\dtn)$ from Definition \ref{def:quad_indicator} such that we improve in performance compared to the results in Table \ref{tab:check_cost}, and we avoid the use of specialized quadrature schemes for the evaluation of singular kernels. 
Our goal is to reduce the complexity of the sorting scheme, such that the cost of the new strategy scales linearly with the cost of the computation of the inner loop.

In the remaining of the section, we present an alternative sorting alternative based on Definition \ref{def:quad_indicator} with computational cost that scales with the cost of evaluating the inner loop.

\subsubsection{Sorting estimations}\label{sec:sorting_estimates}

The orderings given in Definition \ref{def:quad_sets} are used to group quadrature 
pairs over $\Om_0$, $\Om_r$ and $\Om_a$ separately.
The resulting pairs are written as $\{\mr x_k,\mr w_k\}$ for index $k\in \mc I_q$
with $q=0$ corresponding to quadrature pairs from $\Om_0$,
     $q=r$ from $\Om_r$, and similarly for $q=a$ from $\Om_a$.
Then, from Definition \ref{def:quad_indicator} we have
\begin{equation}
	\pseudo_h^\fem(\Om_\dtn)^2
  \leq \max_{j\in \mc I_\dtn}|(T^\fem(\om^\fem)u^\fem)({\rm x}_j)|^2 \cdot |\Om^h_\dtn|.
	\label{eq:filter_steps}
\end{equation}
In the estimate \eqref{eq:filter_steps}, we used the properties that the quadrature weights ${\rm w}_j$ are positive, 
and that $|\Om^h_q|=\sum_{j\in\mc I_q} \mr w_j$.

We base our sorting strategy in the following definition.
A function $u\in C(\Om_\dtn)$ satisfies
\begin{equation}\label{eq:estimate}
	 \min_{\mr x\in\mc X_a} |u(\mr x)-K^\fem u(\mr x)|\cdot |\Om^h_a|^{1/2} \leq \pseudo_h^\fem(\Om_\dtn) 
	 \leq \max_{\mr x\in\mc X_a} |u(\mr x)-K^\fem u(\mr x)|\cdot |\Om^h_a|^{1/2}.
\end{equation}
Since both the FE approximations and the Nystr\"om approximations converge as we increase the dimension of the
discretization, then we expect that the estimate \eqref{eq:estimate} wraps around 
$\pseudo_h^\fem(\Om_\dtn)$ tighter for finer discretizations.

\begin{definition}\label{def:sort_str} 
	{\bf Sorting indicator:} For a given eigenpair $(\om_m^\fem,v_m^\fem)$ we define
\begin{equation}
	\tilde \pseudo_m:=\max_{x\in \Om_a} |v^\fem_m(x)-K^\fem v^\fem_m(x)|.
	\label{eq:max_indicator}
\end{equation}
\end{definition}

Then, \eqref{eq:estimate} suggests an alternative strategy
as an approximation of the pseudospectrum indicator in definition \ref{def:sorting}. 
Basically, the goal is to test points hoping that when the number of evaluations is large we have a good approximation to 
the upper bound \eqref{eq:max_indicator}.
In order to make the most of each evaluation, we consider the following heuristic ordering of the evaluation points:
\begin{itemize}
	\item We assume that $|T^\fem(\om^\fem)u^\fem (x)|$ is large at points $x$ where $|u^\fem (x)|$ is large.
	\item Due to the rapid growth of eigenfunctions, we assume that 
	non-convergent/spurious pairs exhibit $\max_{y\in\Om_r} |u(y)-K^\fem u(y)| < \max_{x\in\Om_0} |u(x)-K^\fem u(x)|$.
	\item We want to avoid large errors in the quadrature rules used. Then, it is convenient to evaluate residuals away from
the singularities of the kernel $\Phi(x,y)$. These are located at points $x=y$.
\end{itemize}
The last reasons suggest that 
for identification of spurious pairs, it is convenient to compute first the residuals at points outside the resonator.
This is, for the computation of \eqref{eq:max_indicator}, we should start by evaluating points in $\Om_0$. 

\begin{definition}\label{def:sort_str} 
{\bf Sampling set:} For given constants $\disG_m,\,c>0$ we define the sampling set 
	\begin{equation}
		X_m:=\{x\in\Om_{0}\,: \inf_{y\in\Om_r} |x-y|>\disG_m\,\,\hbox{and}\,\,|v_m(x)|>\,c\}.
	\end{equation}
\end{definition} 

\begin{remark}\label{def:approx_int_check}
The proposed strategy consists in evaluating
the residual $|(T^\fem(\om^\fem)u^\fem)(x)|$ in points $x\in X_m$ as suggested by
\eqref{eq:max_indicator}, where the evaluation points are not clustered together. 
Additionally, we want to exclude points ${x}_l$, such that $|u^\fem({x}_l)|\approx 0$, and
we want to avoid integrating over cells with a singular kernel. For this, we select points ${x}_j\not\in \Om_r$ such that $\inf_{y\in\Om_r} |x_j-y| \geq \disG_m$, in order to avoid the singularity peak. Consecutively, we select $\disG_m$ such that $|\Phi (\om_m\disG_m)|\leq C_1$, which guarantees that the integrand is bounded and quadrature rules remain accurate.
The benchmarks in $\R^d$, $d=1,2,3$ presented in this paper indicate that for practical computations
using $N_s\approx 10$ evaluations is a good choice.
Ultimately, we use the normalization $v_m:=u_m/\|u_m\|_{\Om}$, with 
$\Om:=\Om_\dtn\cup \Om_\PML$ for the PML formulations. Then, we filter out added PML eigenvalues 
described in Definition \ref{def:pml_eigs} by 
requiring the condition $\|v_m(x)\|^2_{\Om_\dtn}/|\Om_\dtn|>\|v_m\|^2_{\Om_\PML}/|\Om_\PML|$.
\end{remark}

The result of applying Remark~\ref{def:approx_int_check} is an effective and inexpensive way of testing the computed pair $(\om^\fem,\xi)$, where the cost scales linearly with the inner loop. 
If the resulting sorting indicator is larger than a user pre-defined threshold, then we disregard the eigenpair and 
continue testing the next eigenpair in the collection.

\section{Applications to metal-dielectric nanostructures}\label{sec:benchmarks}

In this section we study four interesting metal-dielectric configurations, from where numerical approximations to resonances and resonant modes are computed. Consecutively, eigenpairs are tested and solutions are sorted according to their corresponding pseudospectrum indicator \eqref{eq:max_indicator}. The sorting strategy is tested on problems where exact pairs are known.
Additionally, we {consider} as reference problem a test case used in \cite{nannen2018}.

The first three configurations serve as benchmarking strategies for non-dispersive and piecewise constant material properties. {Then, we apply the sorting algorithm on a} configuration introduced in \cite{araujo+campos+engstrom+roman+2020} where a metal coating is motivated from realistic applications in nano-photonics. Here, three different relative permittivity models are used:
$\eps_v:=1$ (\emph{Vacuum}), $\eps_s:=2$ (\emph{Silica}), and $\eps_{metal}$ (\emph{Gold}), modeled by a sum of Drude-Lorentz terms \eqref{eq:drude_lorentz}. For $\eps_{metal}$ we use the data given in table \ref{gold_data} gathered in \cite{Rakic98}. This model of Gold has been extensively tested and has validity for $\om\in[0.5,6.5]\,eV$, where $eV$ denotes \emph{electron volt}. 

We introduce a demanding configuration where the refractive index is a continuous function of space motivated from the so-called \emph{graded materials}.
Finally, we consider an acoustic benchmark problem in $\mx R^3$.

\subsection{Modeling details}\label{sec:scaling}

In finite precision arithmetic we prefer to work with dimensionless quantities, where we transform from dimensionless variables to physical variables (denoted with \textasciitilde). We use common physical constants in SI units: $\hbar$ is the scaled Planck's constant, $c$ is the speed of light in vacuum, and 
$e$ is the electron charge. 
In the numerical computations, we use the scaling factors $W=eV/\hbar$ in \emph{Hertz} and $L=2\pi c/W$ in \emph{meters}. Then, we define the dimensionless quantities
\begin{equation}
	x =\fr{\tilde x}{L},\quad {\omega}=\fr{\tilde \om}{W}\quad \hbox{satisfying}\quad LW=2\pi c.
	\label{scaling}
\end{equation}
The resulting length factor is $L=1239.842\,nm$, from where our spectral window becomes numerically equivalent to $eV$ scaling.

\begin{table}[]
\centering
\begin{tabular}{lll}
$\eps_\infty=1$  & $\om_p=$ 9.03           &            -          \\
$f_0=$ 0.76             & $\om_0=$ 0              & $\gamma_0=$ 0.053     \\
$f_1=$ 0.024            & $\om_1=$ 0.415          & $\gamma_1=$ 0.241     \\
$f_2=$ 0.01             & $\om_2=$ 0.83           & $\gamma_2=$ 0.345     \\
$f_3=$ 0.071            & $\om_3=$ 2.969          & $\gamma_3=$ 0.87      \\
$f_4=$ 0.601            & $\om_4=$ 4.304          & $\gamma_4=$ 2.494     \\
$f_5=$ 4.384            & $\om_5=$13.32           & $\gamma_5=$ 2.214   
\end{tabular}
\caption{Drude Lorentz data for Gold, taken from \cite{Rakic98}, with time convention $e^{-i\om t}$.}
\label{gold_data}
\end{table}


\subsection{Benchmarks in 2D}\label{sec:models_2D} 

The next two problems have radial symmetry centered at the origin, and the solutions expressed 
in polar coordinates $(r,\theta)$, will be written in terms of Bessel and Hankel functions of integer order $m$.
In this simple case outgoing solutions of \eqref{eq:master_eq} satisfy
\begin{equation}
	u=H_m^{(1)}(a \om )\left(
	\begin{array}{c}
		\cos m\theta \\
		\sin m\theta
	\end{array} \right),
	\,\,\,\hbox{for}\,\,\, x\in\partial B(0,a),\,\,\,\hbox{and}\,\,\, m\in \mx Z,
	\label{eq:outgoing}
\end{equation}
where \,supp$\,(n-1)\subset B(0,a)$. 
In subsections \ref{sec:SD} and \ref{sec:SCD}, we present solutions satisfying \eqref{eq:master_eq} and \eqref{eq:outgoing} for specific permittivity profiles.

\subsubsection{Standard benchmark: Single disk problem (SD)}\label{sec:SD}

\begin{figure}
	\centering
	\begin{tikzpicture}
		\draw(00.00,18.0) node {
			\includegraphics[scale=0.80,trim=5mm 0mm 5mm 0mm,clip]{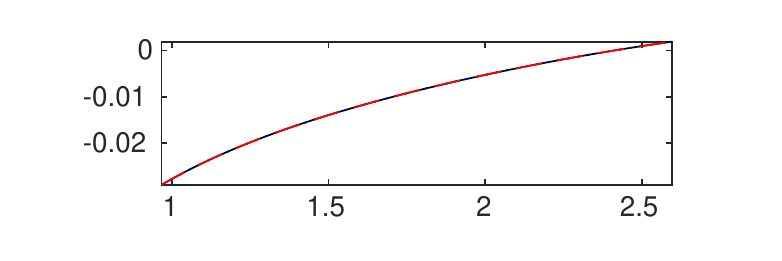} };
		\draw(05.50,18.0) node {
			\includegraphics[scale=0.80,trim=5mm 0mm 5mm 0mm,clip]{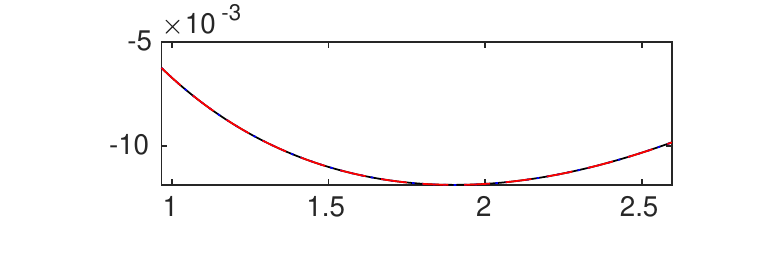} };
		\draw(11.00,18.0) node {
			\includegraphics[scale=0.80,trim=5mm 0mm 5mm 0mm,clip]{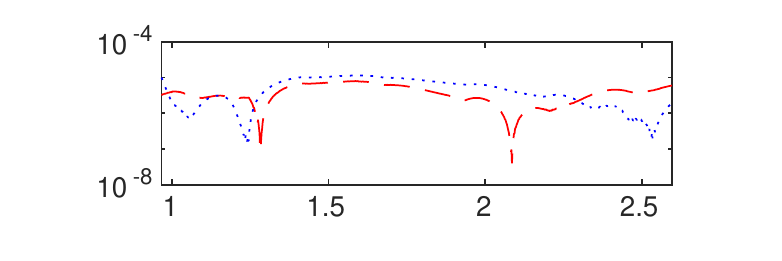} };
		\draw(00.00,16.0) node {
			\includegraphics[scale=0.80,trim=5mm 0mm 5mm 0mm,clip]{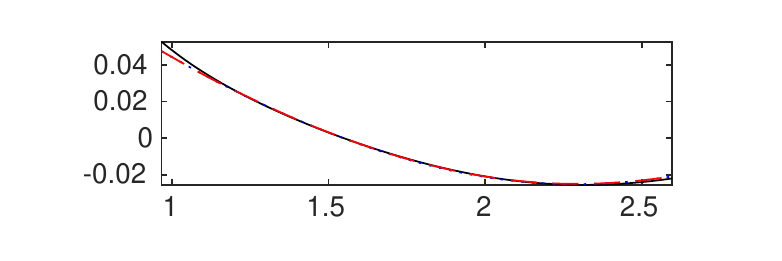} };
		\draw(05.50,16.0) node {
			\includegraphics[scale=0.80,trim=5mm 0mm 5mm 0mm,clip]{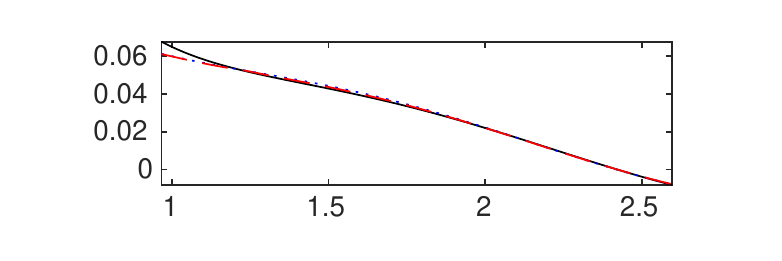} };
		\draw(11.00,16.0) node {
			\includegraphics[scale=0.80,trim=5mm 0mm 5mm 0mm,clip]{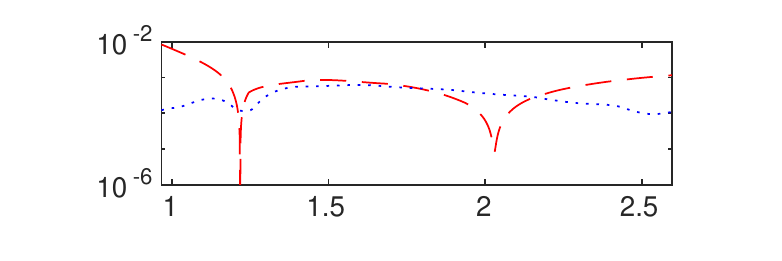} };
		\draw(00.00,14.0) node {
			\includegraphics[scale=0.80,trim=5mm 0mm 5mm 0mm,clip]{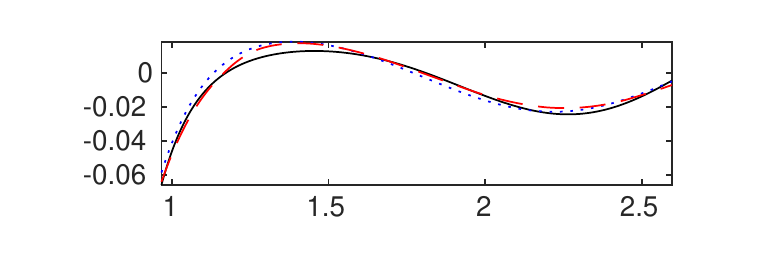} };
		
		\draw(05.50,14.0) node {
			\includegraphics[scale=0.80,trim=5mm 0mm 5mm 0mm,clip]{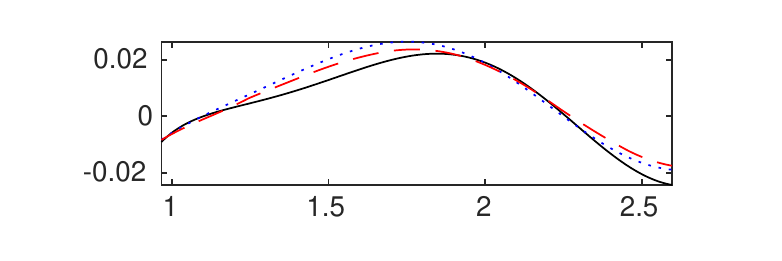} };
		\draw(11.00,14.0) node {
			\includegraphics[scale=0.80,trim=5mm 0mm 5mm 0mm,clip]{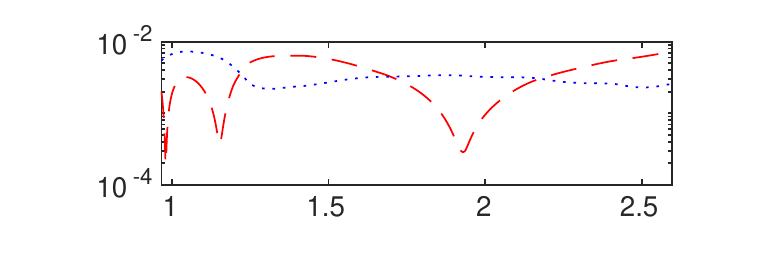} };
		\draw(00.00,12.0) node {
			\includegraphics[scale=0.80,trim=5mm 0mm 5mm 0mm,clip]{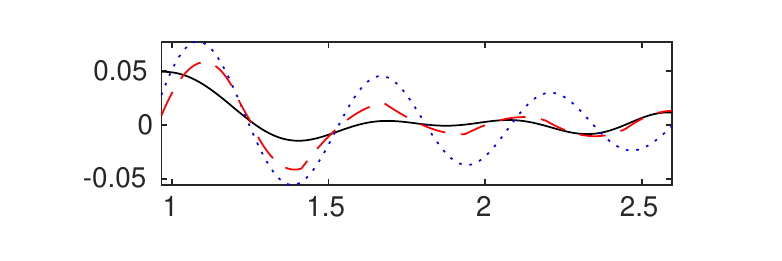} };
		\draw(05.50,12.0) node {
			\includegraphics[scale=0.80,trim=5mm 0mm 5mm 0mm,clip]{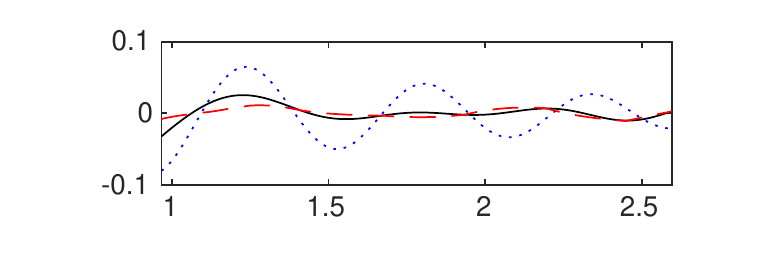} };
		\draw(11.00,12.0) node {
			\includegraphics[scale=0.80,trim=5mm 0mm 5mm 0mm,clip]{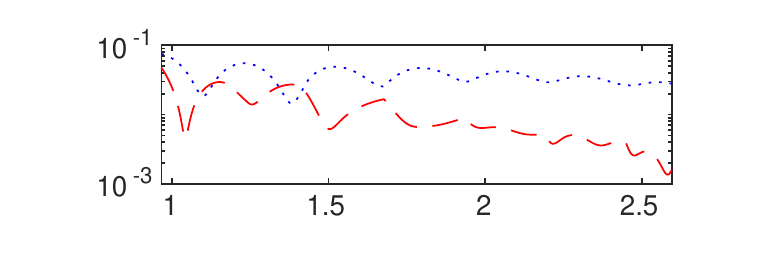} };		
		\draw(00.00,10.0) node {
			\includegraphics[scale=0.80,trim=5mm 0mm 5mm 0mm,clip]{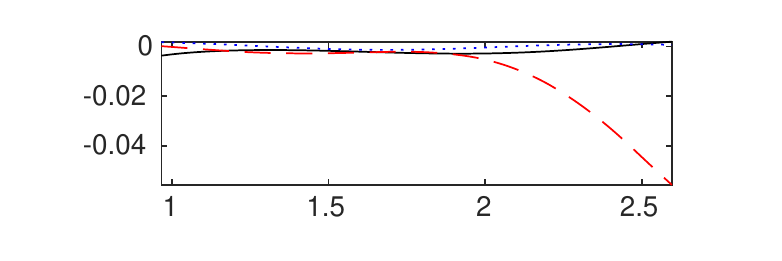} };
		\draw(05.50,10.0) node {
			\includegraphics[scale=0.80,trim=5mm 0mm 5mm 0mm,clip]{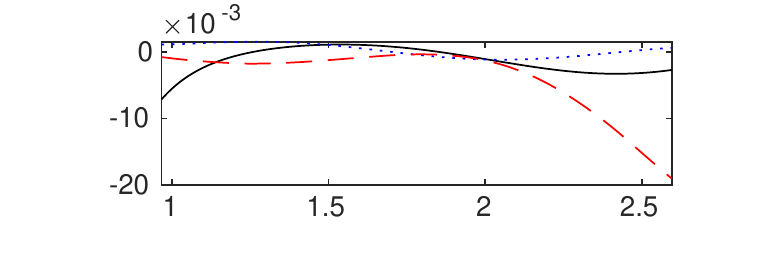} };
		\draw(11.00,10.0) node {
			\includegraphics[scale=0.80,trim=5mm 0mm 5mm 0mm,clip]{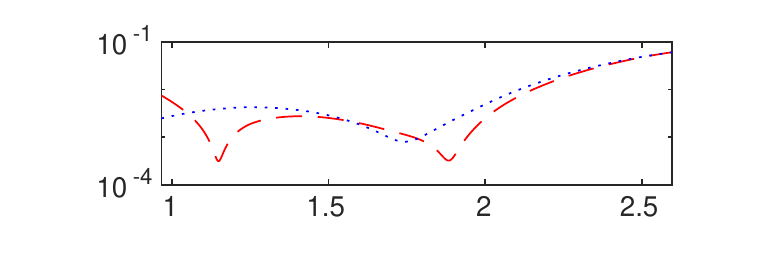} };
	\end{tikzpicture}
	\caption{\emph{Functions $u(x), u^\fem(x),K^\fem u^\fem(x)$ and corresponding residuals restricted to the straight line \eqref{eq:line}, from SD problem described in Section \ref{sec:SD} for TM polarization with a discretization with $p=2$ and one mesh refinement.
	Vertically we show different representative eigenfunctions. In the left panel of the figures we present real parts of $u$ (black), $u^\fem$ (red), $Ku^\fem$ (blue), and imaginary parts in the middle panel. 
In the right most panel we present the residuals $|u(x)-u^\fem(x)|$ in red dashed 
and $|T^\fem u^\fem(x)|=|u^\fem(x)-K^\fem u^\fem(x)|$ in blue dotted lines.
	} }
	\label{fig:pointwise_filter_p2r1sd}
\end{figure}


\begin{figure}
	\begin{tikzpicture}
		\draw(0.00,12.2) node {\includegraphics[scale=0.50]{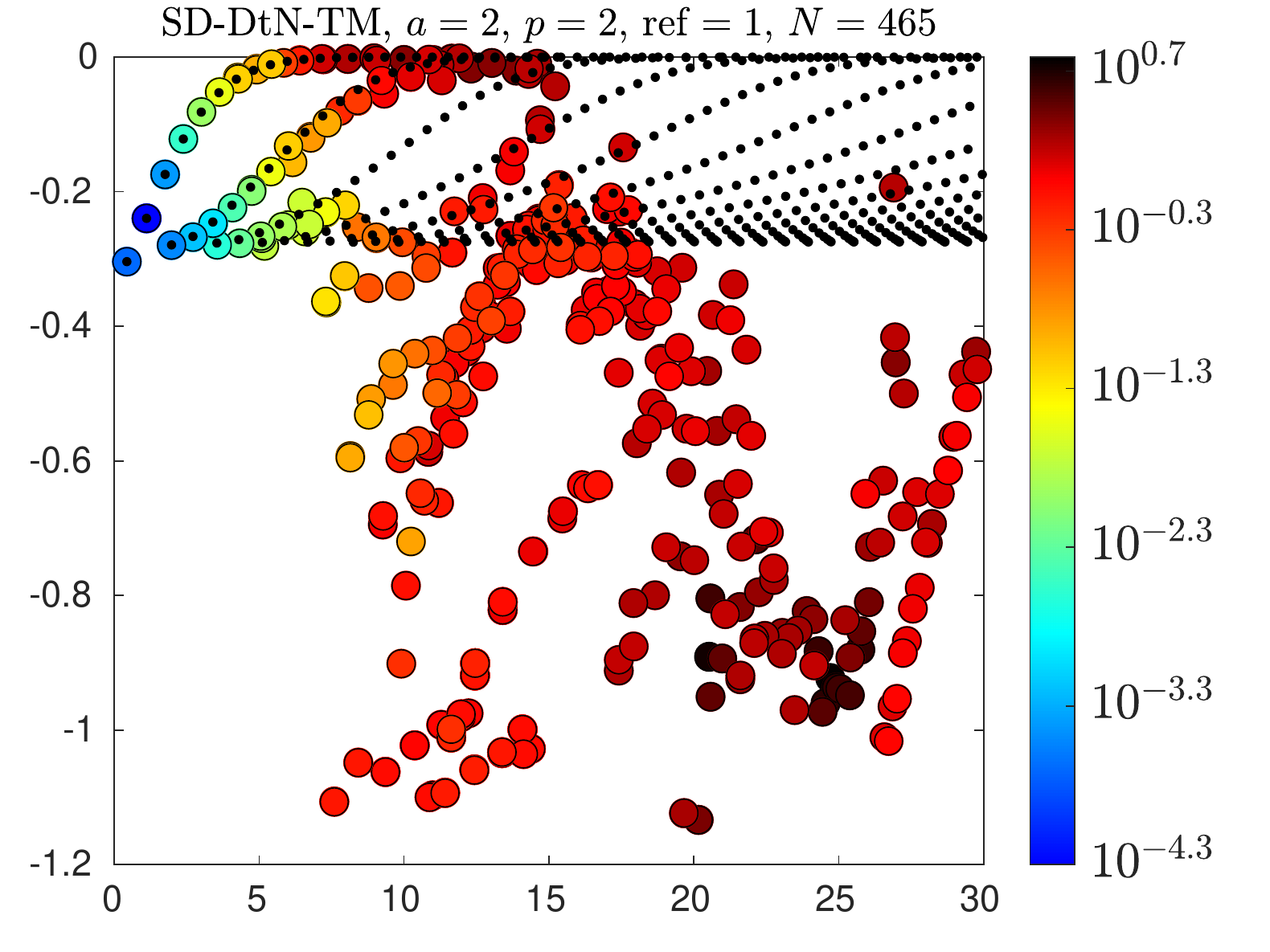} };
		\draw(8.00,12.2) node {\includegraphics[scale=0.50]{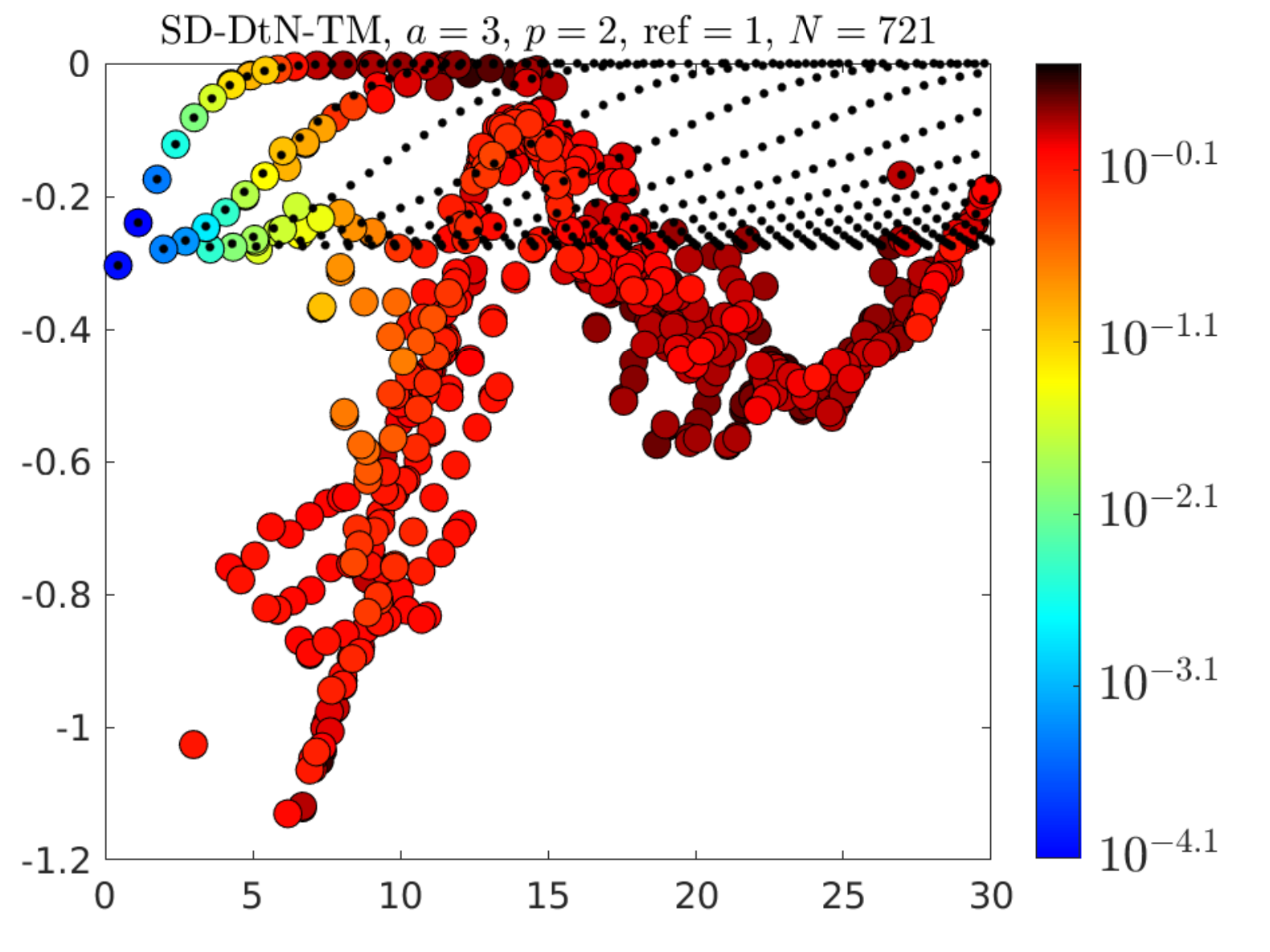} };
		\draw(0.00, 6.1) node {\includegraphics[scale=0.50]{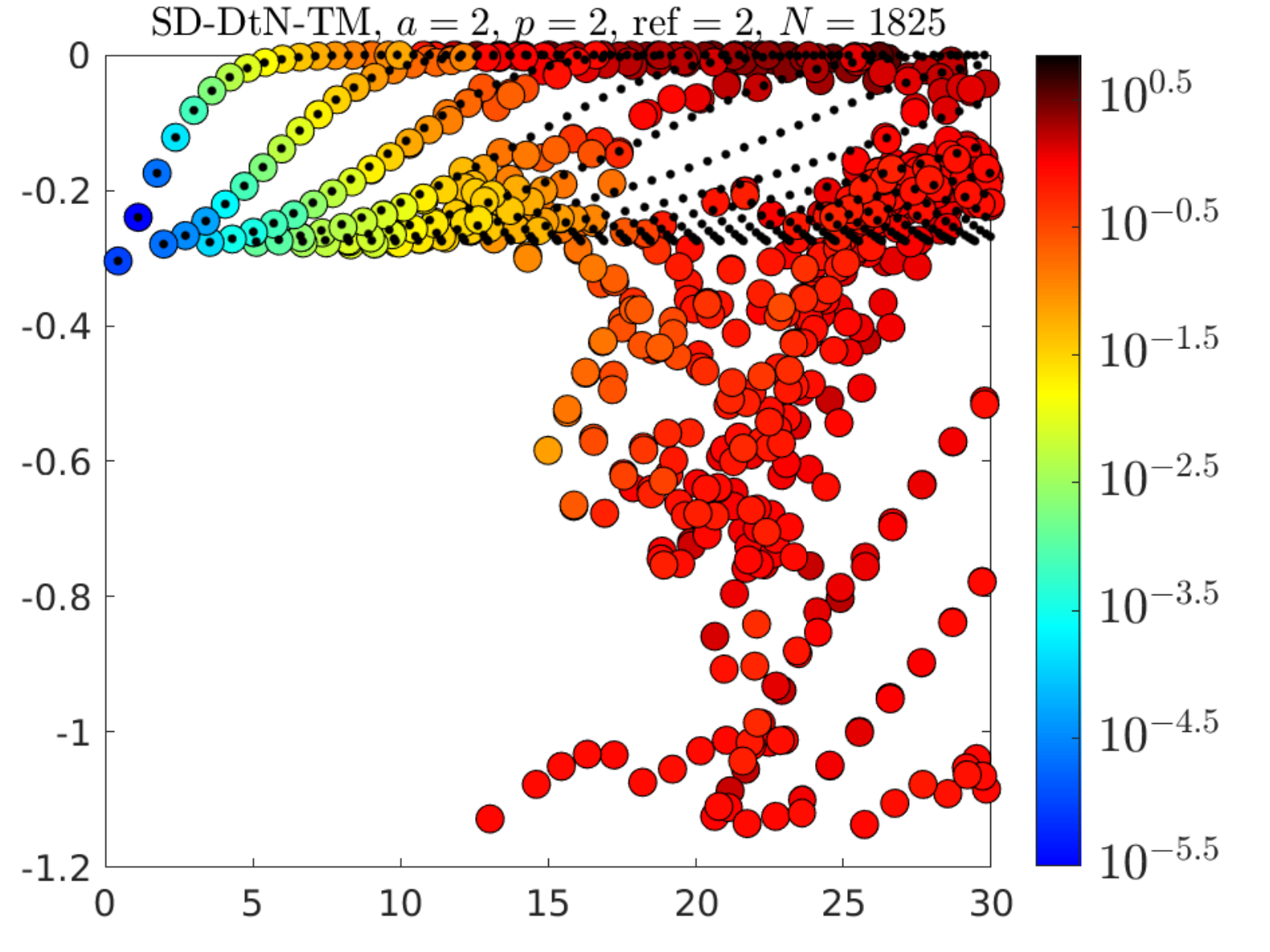} };
		\draw(8.00, 6.1) node {\includegraphics[scale=0.50]{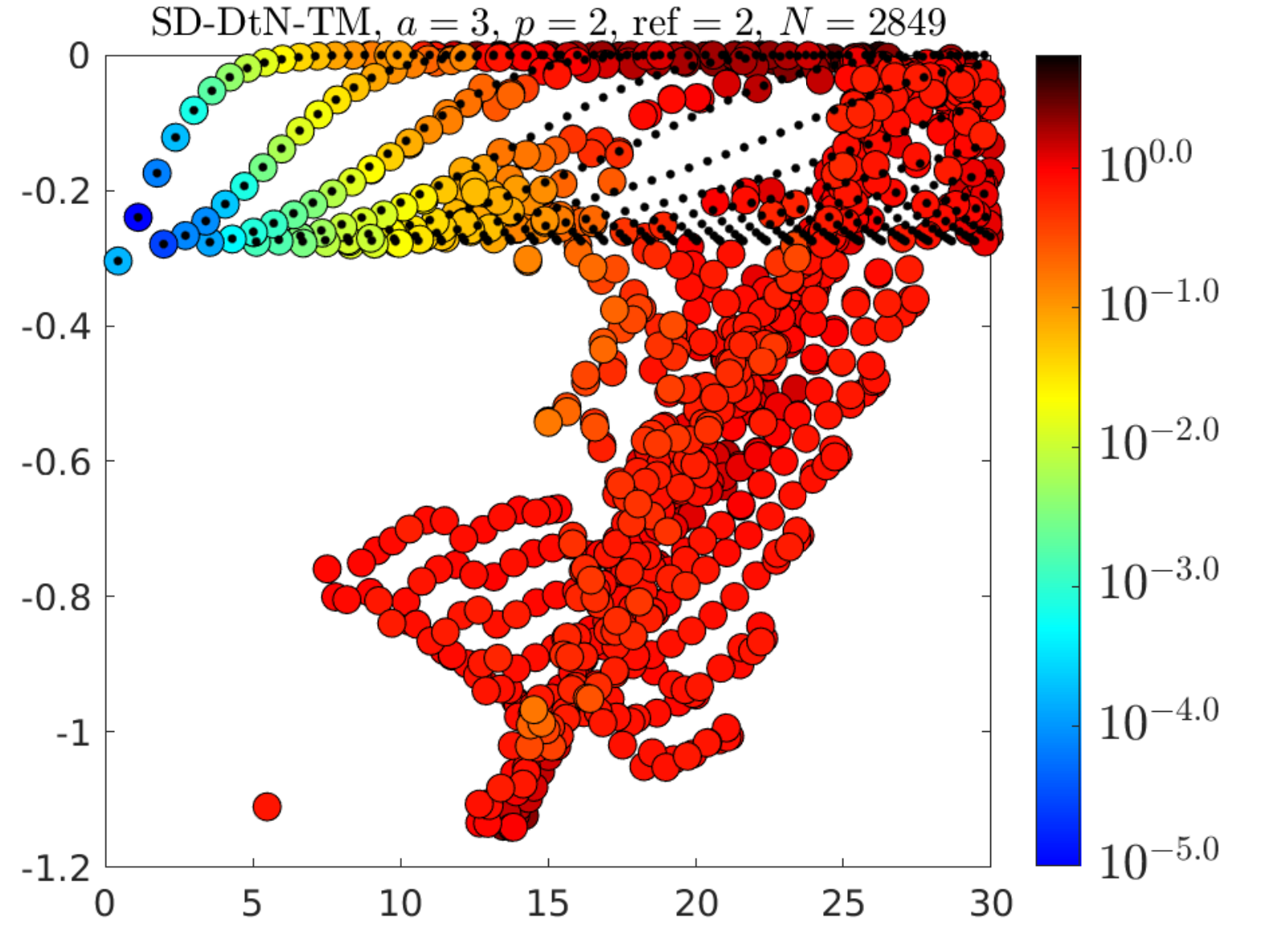} };
		\draw(0.00, 0.0) node {\includegraphics[scale=0.50]{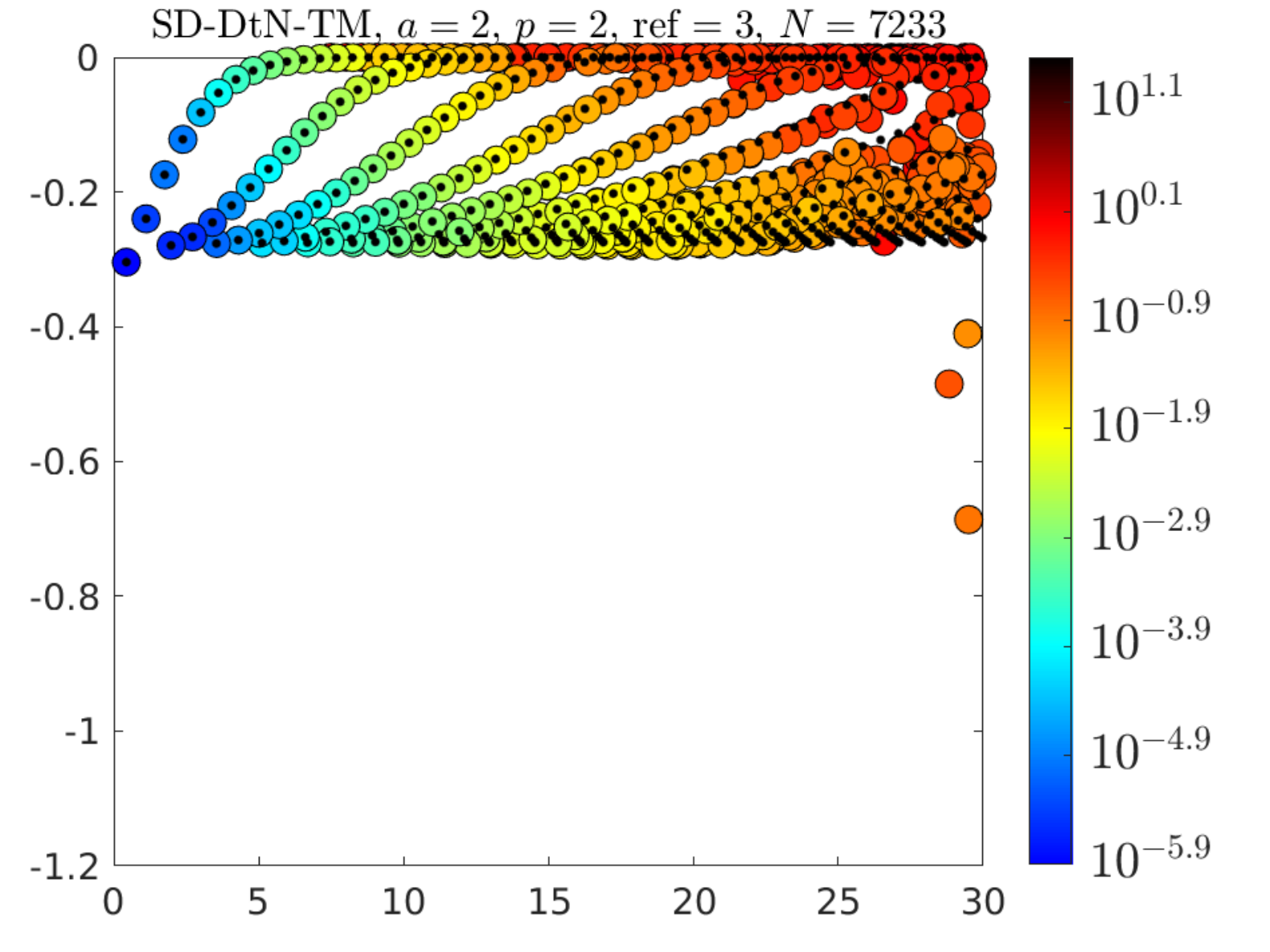} };
		\draw(8.00, 0.0) node {\includegraphics[scale=0.50]{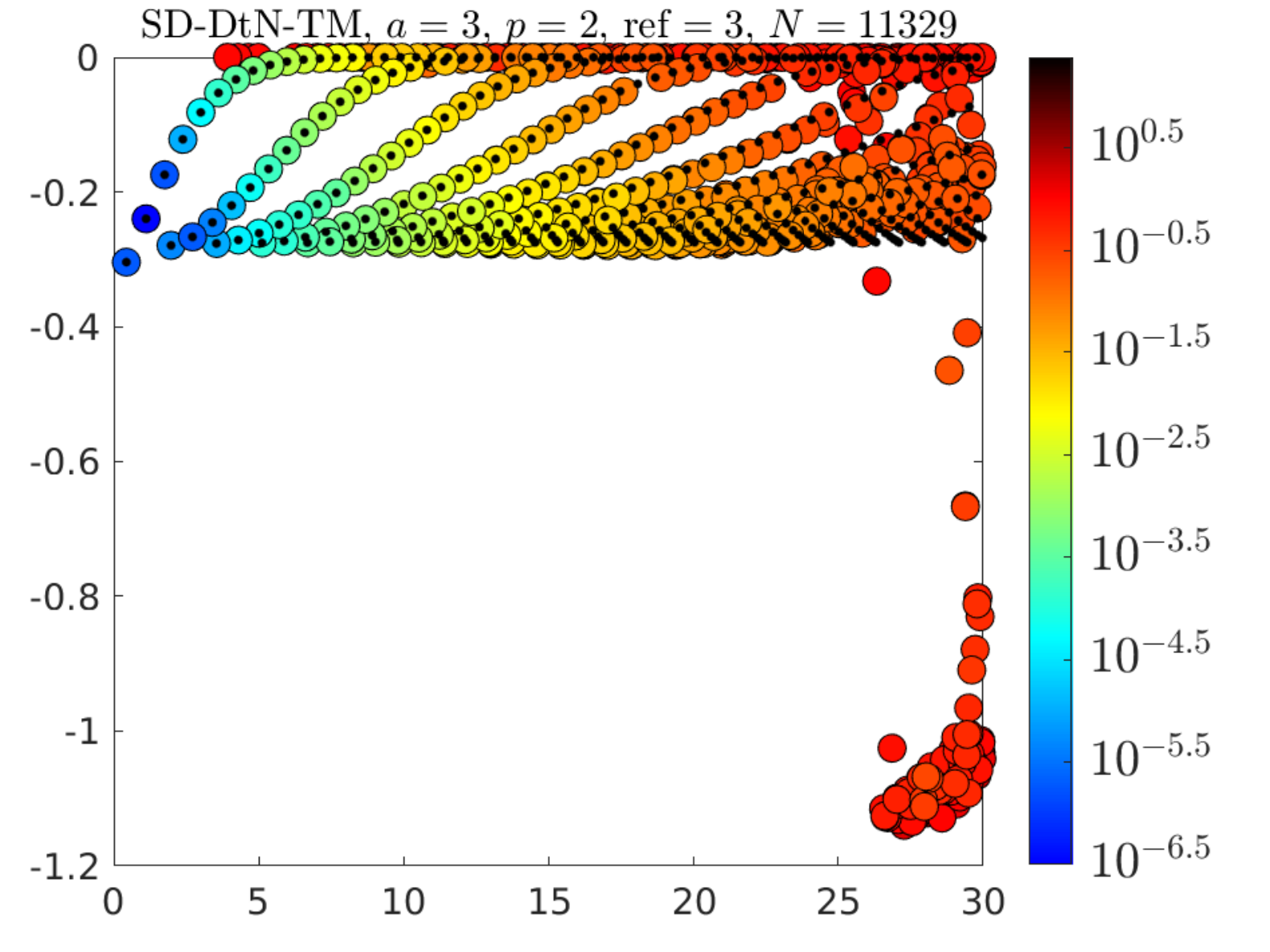} };
	\end{tikzpicture}
	\caption{\emph{Spectral window showing exact (dots) and FE eigenvalues (circles) the problem described in Section \ref{sec:SD} for TM polarization featuring three levels of mesh refinement. In colors we give $\tilde \pseudo_m$ corresponding to $\om_m^\fem$ from computations over ten points.} }
	\label{fig:filter_p2r2sd}
\end{figure}

Denote by $u=u_1,\,n=n_1$ the restrictions of $u,n$ to $\Om_1:=B(0,R)$, and set $n=n_2=1$ elsewhere.
The corresponding exact eigenfunctions to \eqref{eq:master_eq} and \eqref{eq:outgoing} read:
\begin{equation}
	u_1=
	N_m J_m(n_1 \om r)\left(
	\begin{array}{c}
		\cos m\theta \\
		\sin m\theta
	\end{array} \right),\,\, 
	u_2=
	H_m^{(1)}(\om r)\left(
	\begin{array}{c}
		\cos m\theta \\
		\sin m\theta
	\end{array} \right)
	,\,\, N_m:=\fr{H_m^{(1)}(\om R)}{J_m(n_1 \om R)}.
	\label{eq:exact}
\end{equation}
The eigenvalues $\om$ corresponding to $m=0$ are \emph{simple} and those corresponding to $m>0$ 
are \emph{semi-simple} and have multiplicity $\alpha=2$.
The exact eigenvalue relationship for TM and TE can be written as
\begin{equation}
	J_m(n_1 \om R) H_m^{(1)\prime}(\om R)-g\,J'_m(n_1 \om R) H_m^{(1)}(\om R) =0,
	\label{eq:reson_SD}
\end{equation}
where $g=n_1,\,g=1/n_1$ corresponds to the TM polarization and TE polarization, respectively.
In the case of a semi-simple eigenvalue $\om_j$, the exact solution consist of two
eigenfunctions $u_{j,1}$ and $u_{j,2}$. Then, we define the linear combination $v_j=c_1u_{j,1}+c_2u_{j,2}$
for the constants $c_1,c_2\in\mx C$. In this way, after we have computed an eigenpair $(\om^\fem,u^\fem)$, we
compare $u^\fem$ with the closest exact solution $v_j$. That is, we find $c_1,c_2$
by solving the minimization problem
\begin{equation}\label{eq:min_degenerated}
	\min_{c_1,c_2} \|c_1u_{j,1}+c_2u_{j,2}-u^\fem\|^2.
\end{equation}
This can be done in the discrete setting with a standard \emph{least-squares} optimization routine.

For numerical computations we use $R=1$, and $a=2,3$. Additionally, we place the disk center a distance $s=0.2$ from the origin such that many terms are needed in the DtN for approximation of resonances. \\

\emph {Comparison with respect to exact solutions and LS-projections:}
We denote with $u^\fem$ the computed FE eigenfunction, with $u$ the best fitted exact solution and with $K^\fem u^\fem$ the Lippmann-Schwinger projected eigenfunction.
The best fitted exact eigenfunction $u$ is obtained by solving the minimization problem \eqref{eq:min_degenerated}. 

For convenient visualization and comparison of the eigenfunctions we define the straight line $\mc L\subset \Om_0$ as 
\begin{equation}\label{eq:line}	
	\mc L:=\{(x,y)\in \Om_0\,|\,x\in (s+\eps+\cos(\pi/6),a\cos(\pi/6)-\eps),y=\sin(\pi/6)\},
\end{equation}
where $s=0.2$.
The vertices defining $\mc L$ are located a small distance away from the resonator and the boundary $\Gamma_\dtn$ by setting $\eps=10^{-3}$. This line has been chosen in order to avoid computations over an edge of an element such that we avoid super-convergence of residuals. 

In Figure \ref{fig:pointwise_filter_p2r1sd}, we gather representative results for the single disk problem. 
The depicted solutions in the figure correspond to the TM polarization with $p=2$, ref$\,=1$, and $a=3$.
Vertically we show different representative eigenfunctions that are drawn for $x\in\mc L$. 
In the left panel of the figures we present real parts of $u$ (black), $u^\fem$ (red), $Ku^\fem$ (blue), and imaginary parts in the middle panel. 
In the right most panel we present the residuals $|u(x)-u^\fem(x)|$ in red dashed 
and $|T^\fem u^\fem(x)|=|u^\fem(x)-K^\fem u^\fem(x)|$ in blue dotted lines.

\emph {Results:} 
In Figure~\ref{fig:pointwise_filter_p2r1sd}
	we observe that for eigenpairs that are good approximations then $|T^\fem u^\fem(x)|$ follows very closely $|u(x)-u^\fem(x)|$, showing an important resemblance. This presents a measure convergence of FE computations.
	Additionally, some $u^\fem$ are seen to grow exponentially away from the resonator.
	However, the corresponding $K^\fem u^\fem$ grows at a slower rate and seems to agree more with $u$ than $u^\fem$ itself. 
	Finally, in most cases, $|T^\fem u^\fem(x)|$ is smoother and more flat than $|u(x)-u^\fem(x)|$. 

The application of the sorting scheme described in Remark \ref{def:approx_int_check} to this problem gives the results presented in Figures \ref{fig:filter_p2r2sd} and \ref{fig:filter_p2r2sdte}. Exact eigenvalues are marked with dots and computed eigenvalues are marked with colored circles, where the color is given by approximating the sorting indicator \eqref{eq:max_indicator} as described in Remark \ref{def:approx_int_check}. 
In Figure \ref{fig:filter_p2r2sd} we present plots for the TM polarization and in Figure \ref{fig:filter_p2r2sdte} results for the TE polarization. The left panels are computed by placing the DtN at $a=2$, while $a=3$ in the panels on the right, and both discretizations have the same FE up to $a=2$.

From the figures we observe that increasing $a$ results in an increase in the number of eigenvalues in the given spectral window, which is expected from the discussion in Section~\ref{sec:slab1d}. Moreover, the added eigenvalues from $a=3$ pollute larger regions in the spectral window compared to $a=2$. This conclusion can also be obtain by noticing that the minimum in $\tilde \pseudo$ increases for larger $a$, which is related to the fact that exponential growth of eigenfunctions becomes a challenge for FE discretizations.

Additionally, we see that computed eigenvalues with small sorting indicator resemble the pattern drawn by the exact eigenvalues. As the value of the indicator increases the corresponding eigenvalues loose this property and become erratic. In this way we can choose a threshold and for example keep eigenvalues $\om^\fem_m$ with $\tilde \pseudo_m \leq 10^{-2}$ to obtain a set of eigenvalues resembling the behavior of the exact eigenvalues. 
Moreover, in Figure \ref{fig:filter_p2r2sd} we illustrate results for computations on a sequence of finer meshes.
We observe that computation on finer meshes improve the accuracy of computed eigenvalues and exhibit less spurious eigenvalues than computation on coarser meshes. Additionally, we obtain lower values for $\tilde \pseudo_m$ for finer discretizations compared with coarser discretizations.

Finally, it should be noted that the indicator $\tilde \pseudo$  resembles the FE discretization error. Namely, $\tilde \pseudo_m$ increases with $\re \om_m$, which is the expected behavior
for the FE discretization error of non-dispersive Helmholtz problems \cite{ihl98,Ainsworth04,araujo+campos+engstrom+roman+2020}. Additionally, eigenvalues with small indicator values appear very close to the exact eigenvalues of the problem.

\subsubsection{Reference configuration: Single square problem (SS)}\label{sec:SS} 
For this problem, we set $\Om_1:=[-s,s]^2$, $n=n_1$ for $x\in\Om_1$ and $n=1$ elsewhere. 
The correct parameters for obtaining the results presented in \cite[Fig.~10]{nannen2018} are $s=1.5$ and $n_1=0.2$ for TM polarization.\\

\emph{Results:} 
For this problem, we compute all eigenvalues in the spectral window only once, and cheaply test each computed eigenpair. The results are plotted in Figure \ref{fig:filter_square}, by using a DtN in the left panel, and PML in the right panel, where both formulations use the same FE basis in $\Om_a$. 
Again we can see the effectiveness of using the indicator $\tilde \pseudo$. Correct approximations to resonances are easily identified from the overwhelming rest of computed eigenvalues. Additionally, we see that the FE-DtN produces better approximations to resonances than the FE-PML. 
For example, we mark with dots the two reference values given in \cite[Fig.~9]{nannen2018}, and 
we see that for the presented discretizations in Figure~\ref{fig:filter_square} 
the reference eigenvalues have a lower indicator $\tilde \pseudo_m$ for the
FE-DtN than with the FE-PML.

Comparison from the results in Figure \ref{fig:filter_square} against those in \cite[Figs.~10, 11, 12]{nannen2018} not only illustrates the reliability of the identification of true approximation to resonances by using the scheme in Definition \ref{def:sorting}, but it also shows its simplicity, flexibility and large coverage in the complex plane.

\begin{figure}
	\begin{tikzpicture}
		\draw(0.00, 0.0) node {\includegraphics[scale=0.50]{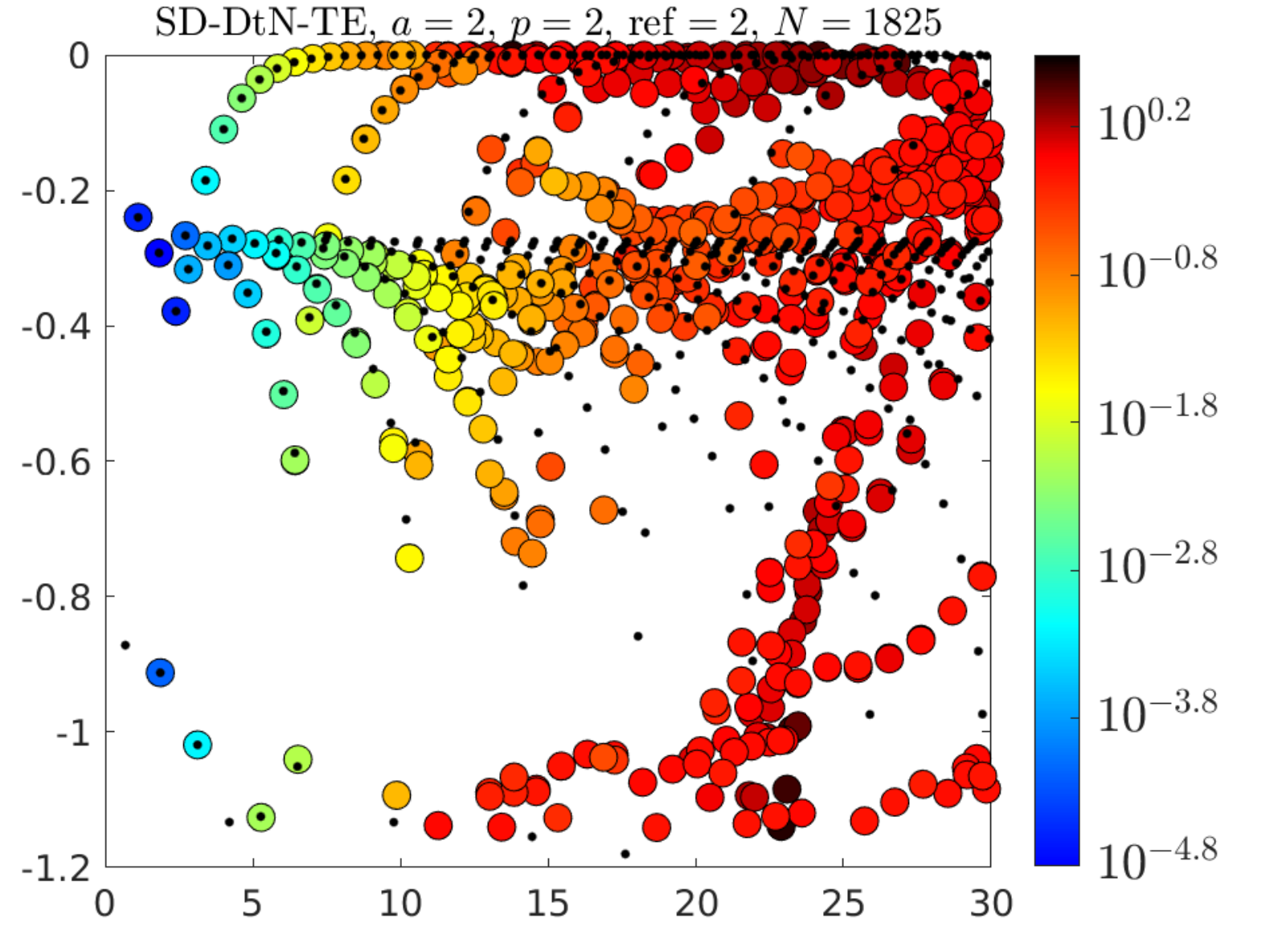} };
		\draw(8.00, 0.0) node {\includegraphics[scale=0.50]{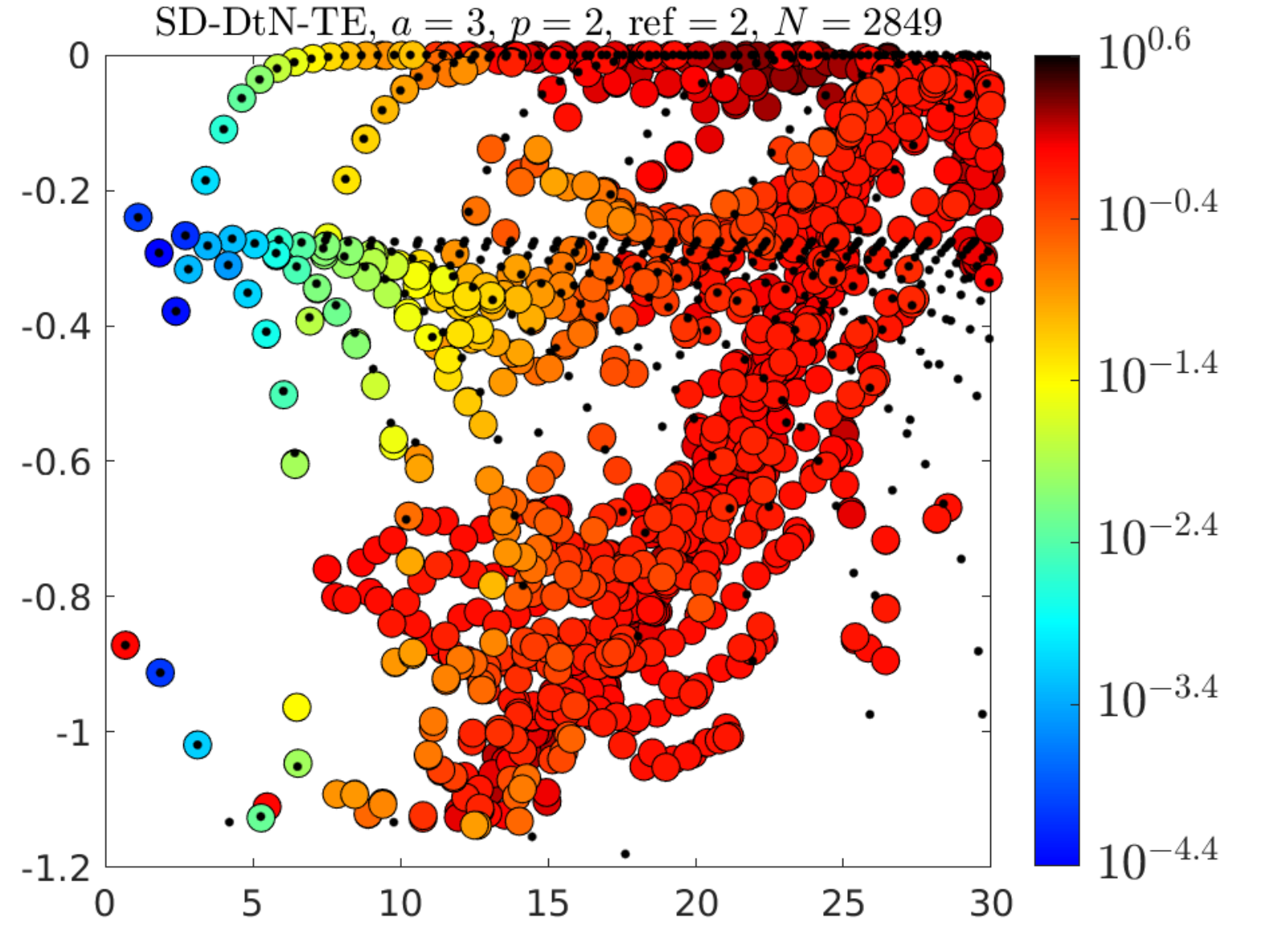} };
	\end{tikzpicture}
	\caption{\emph{Spectral window showing exact (dots) and FE eigenvalues (circles) the problem described in Section \ref{sec:SD} for TE polarization. In colors we give $\tilde \pseudo_m$ corresponding to $\om_m^\fem$ from computations over ten points.
} }
	\label{fig:filter_p2r2sdte}
\end{figure}

\begin{figure}
	\begin{tikzpicture}
		
	\draw(0.00,0.0) node{\includegraphics[scale=0.50]{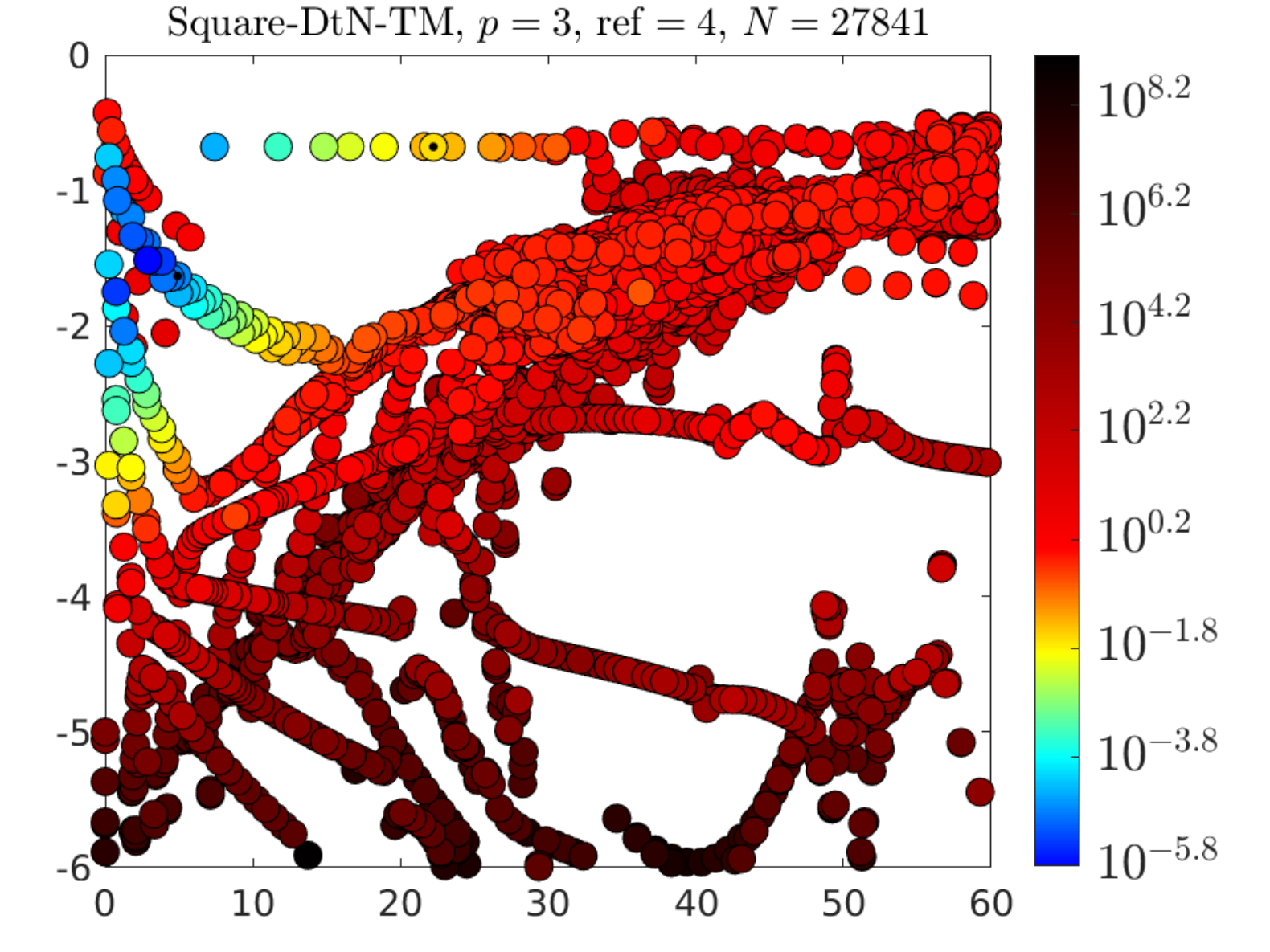}};
	\draw(8.00,0.0) node{\includegraphics[scale=0.50]{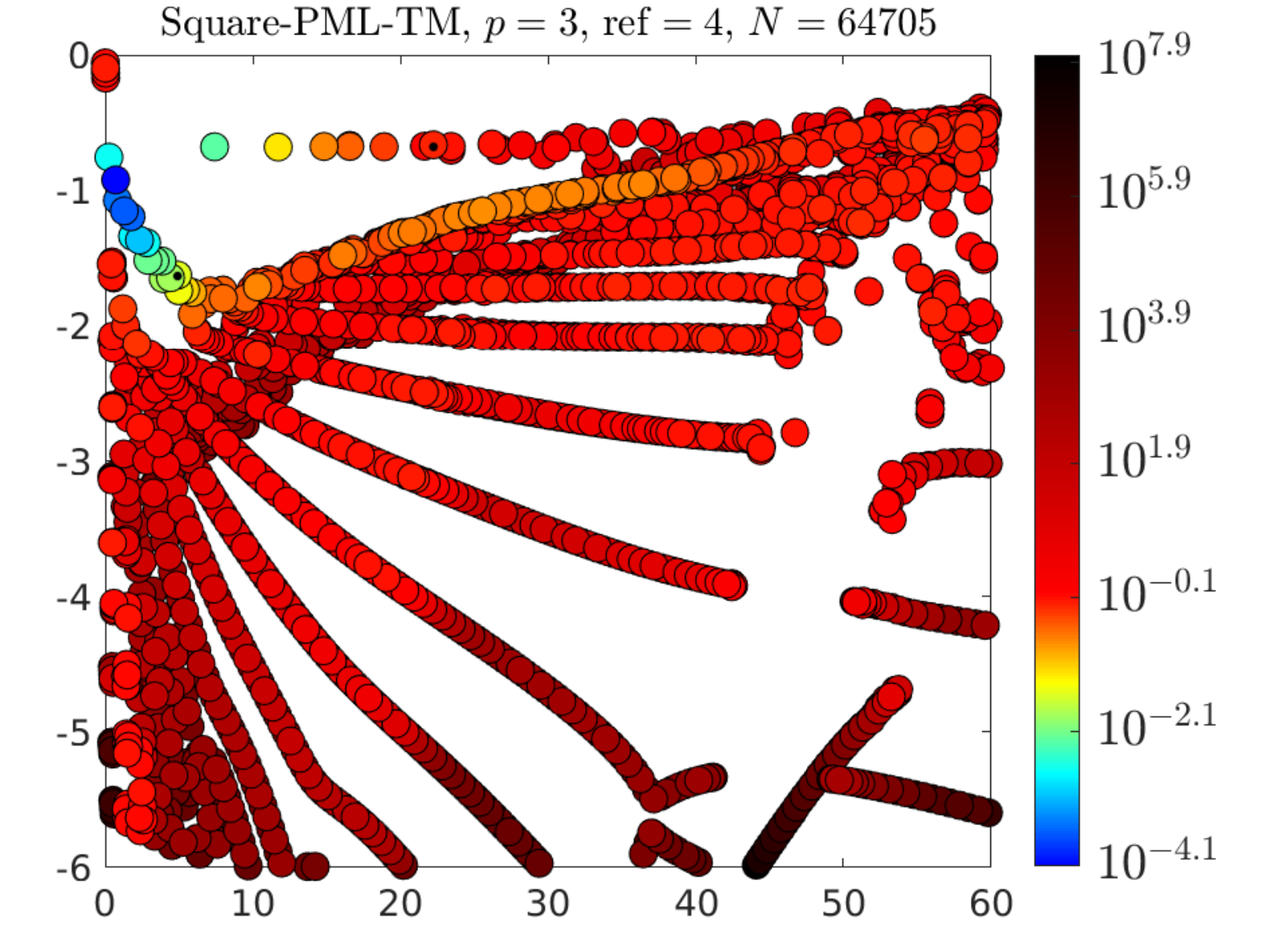}};
		
	\end{tikzpicture}
	\caption{\emph{Spectral window showing two reference eigenvalues (dots) from \cite{nannen2018} and FE eigenvalues (circles) of the Single Square problem described in Section \ref{sec:SS} for TM polarization. In colors we give $\tilde \pseudo_m$ corresponding to $\om_m^\fem$ from computations over ten points.} }
	\label{fig:filter_square}
\end{figure}

\subsubsection{Benchmark with dispersion: Single coated disk problem (SCD)}\label{sec:SCD} 

\begin{figure}
	\begin{tikzpicture}
		
		\draw( 0.00, 0.0) node {\includegraphics[scale=0.50]{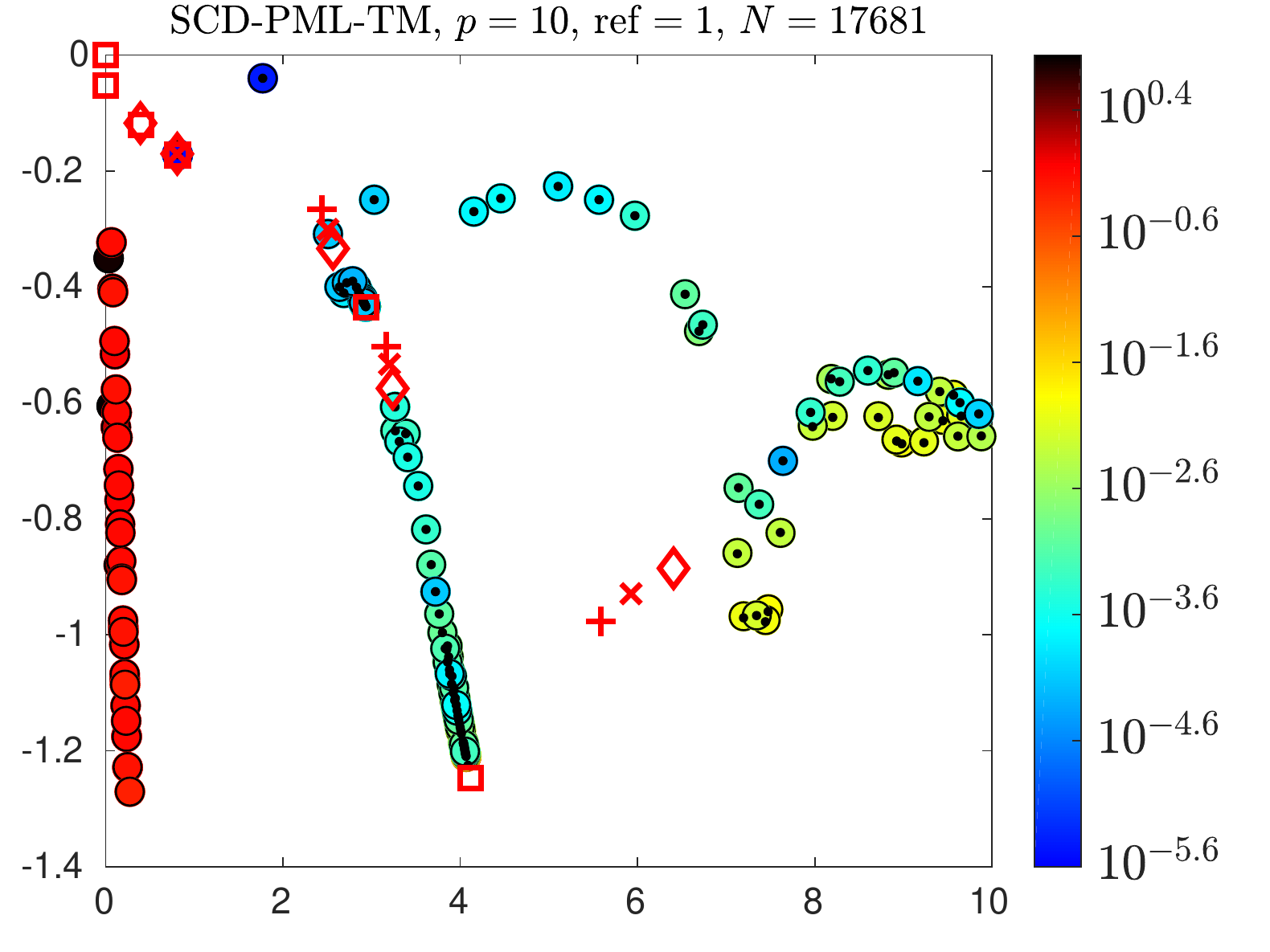} };
		\draw( 8.00, 0.0) node {\includegraphics[scale=0.50]{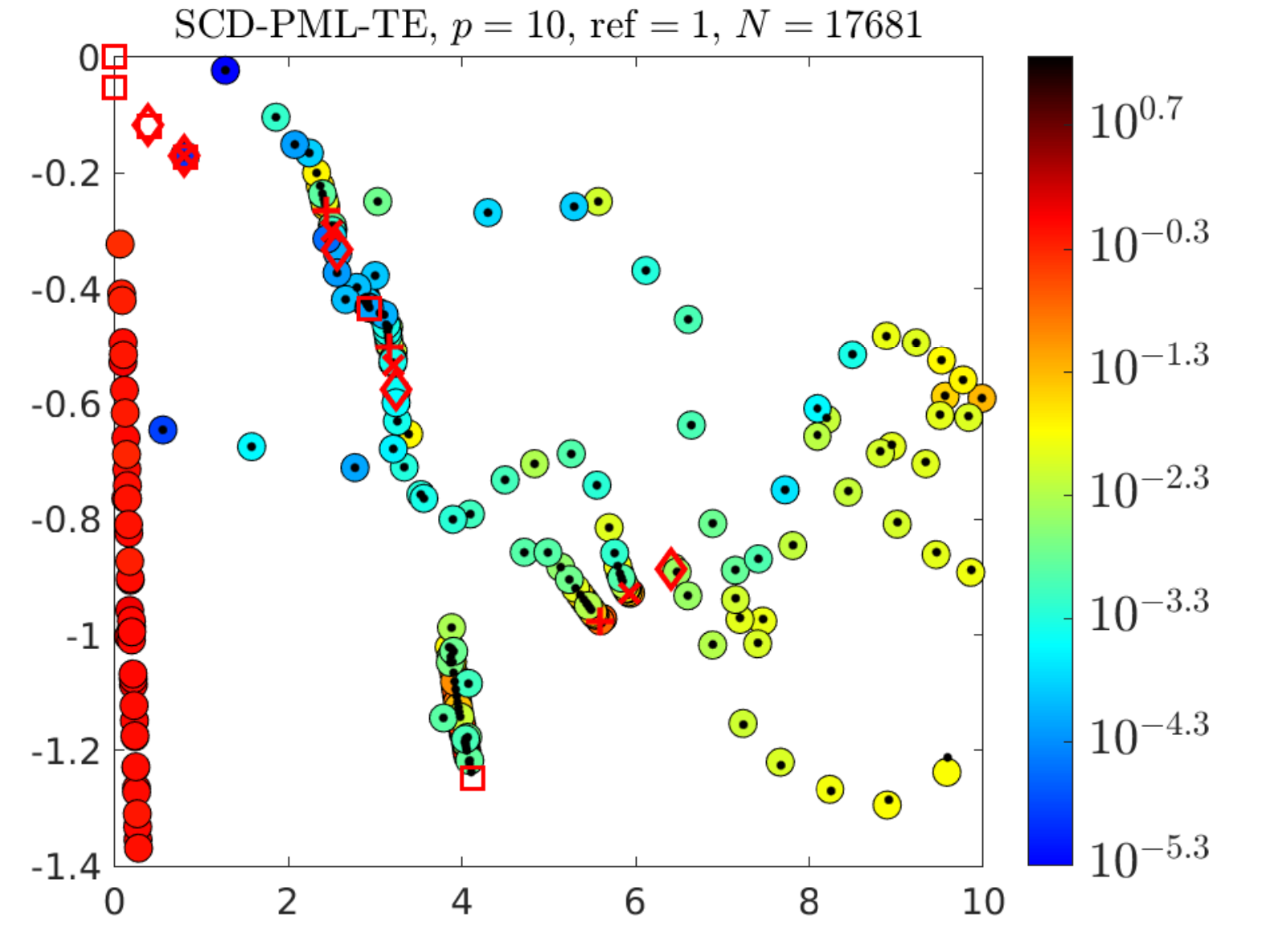} };
		
	\draw(-0.84,-3.6) node {\includegraphics[scale=0.50]{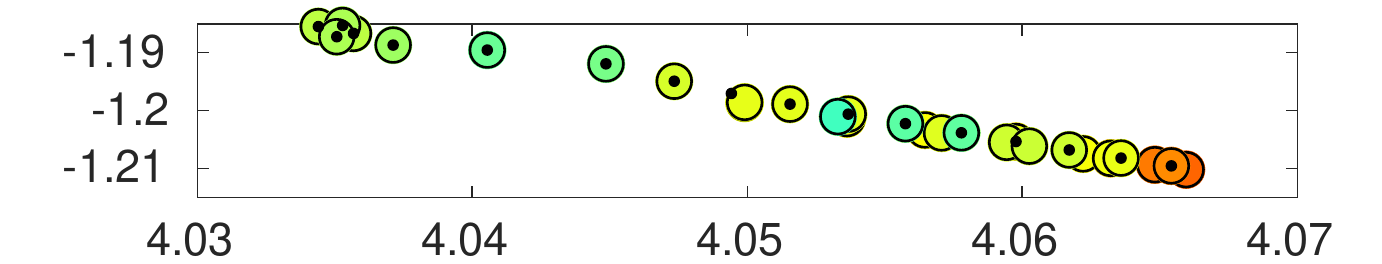} };
	\draw( 7.20,-3.6) node {\includegraphics[scale=0.50]{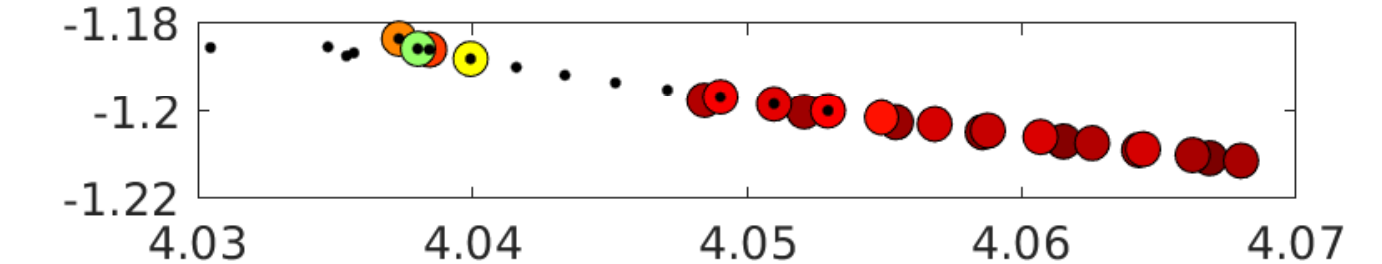} };
	\end{tikzpicture}
	\caption{\emph{Spectral window showing exact $\bullet$ and FE eigenvalues $\bigcirc$ of the Single Coated Disk problem described in Section \ref{sec:SCD} for TM and TE polarizations. The poles of $\eps(\om)$ are given by $\Box$ and it's roots $\eps(\om)=0$ with $\Diamond$. Additionally, the plasmonic branch points $\eps(\om)=-1$ is marked with ($\times$), and $\eps(\om)=-2$ with ($+$).
In colors we give $\tilde \pseudo_m$ corresponding to $\om_m^\fem$ from computations over ten points.} }
	\label{fig:filter_SCD}
\end{figure}

In this configuration, we consider a resonator consisting of a dielectric disk with a uniform coating layer. The geometry is described by two concentric circumferences of radii $0<R_1<R_2$, with vacuum as surrounding medium. The inner disk has constant relative permittivity index, and is coated by a layer of gold. We set $n_1=\sqrt{\eps_s},$ and $n_2:=\sqrt{\eps_{metal}}$ is the value such that $\im\{n_2\}$ (absorption coefficient) is positive.  

The exact solutions satisfy \eqref{eq:master_eq} and \eqref{eq:outgoing} with $R\geq R_2$. 
The resonance relationship reads
\begin{equation}
	\begin{array}{lcl}
	f^m_{1}(\om) &=& 	g_1 J^\prime_m(\om n_1 R_1) H^{(1)}_m(\om n_2  R_1) - 
										g_2 J_m(\om n_1 R_1) H^{(1)\prime}_m(\om n_2 R_1) , \\
	f^m_{2}(\om) &=& 	g_3 J_m(\om n_1 R_1) H^{(2)\prime}_m(\om n_2 R_1) -  
										g_4 J^\prime_m(\om n_1 R_1) H^{(2)}_m(\om n_2 R_1), \\
	f^m_{3}(\om) &=& 	g_5 H^{(1)}_m(\om n_2 R_2) H^{(1)\prime}_m(\om R_2) -   
										g_6 H^{(1)\prime}_m(\om n_2 R_2) H^{(1)}_m(\om R_2), \\
	f^m_{4}(\om) &=& 	g_7 H^{(1)}_m(\om R_2) H^{(2)\prime}_m(\om n_2 R_2) - 
										g_8 H^{(1)\prime}_m(\om R_2) H^{(2)}_m(\om n_2 R_2), \\[2mm]
	F_m(\om) &:=& (f^m_{1} f^m_{4} - f^m_{2} f^m_{3})(\om)=0,
	\end{array}
	\label{eq:reson_CSD}
\end{equation}
where for TM, $g:=(n_1,n_2,n_2,n_1,1,n_2,n_2,1)$, and for TE, $g:=(n_2,n_1,n_1,n_2,n_2,1,1,n_2)$. 
The parameters used for the computation are $R_1=0.8$, $R_2=1.0$ with scaling factor $L=1239.842\,nm$. 

A complex Newton root finder \cite{Yau98} is then used to compute very accurate approximations of the resonances. 
For each $m$ in equation \eqref{eq:reson_CSD}, we search numerically the resonances $\om_{m, 1},\om_{m, 2},\ldots$
with machine precision stopping criterion. 
In \cite[Table 2]{araujo+campos+engstrom+roman+2020}, we list a selection of resonances computed from \eqref{eq:reson_CSD}. 

\emph{Results:} The results for both polarizations are gathered in Figure \ref{eq:reson_CSD}. Here, we observe typical behavior of dispersive resonators $i$) there exist clustering of eigenvalues close to the poles and zeros of the Drude-Lorentz model \eqref{eq:drude_lorentz}, $ii$) in the TE polarization we have clustering of eigenvalues to the so-called \emph{plasmonic branch points} of the model, which are those values $\om$ such that $\eps_{\text metal}(\om)=-\eps_j$, with $j=0,1$. In other words, $-\eps_{\text metal}(\om)$ matches the value of a neighboring dielectric constant (juncture). In the figure, the bottom panels are close up windows showing such accumulations and the corresponding values for the resulting eigenpair indicator $\tilde \pseudo_m$ as described in Remark \ref{def:approx_int_check}.
As expected, the value of $\tilde \pseudo_m$ increases when approaching a critical value (pole, or zero), this behavior is expected since close to a critical point the value $|n(\om)\om|$ increases and the resulting eigenfunction oscillates more rapidly, and the FE error increases. 
This is further discussed in \cite{araujo+campos+engstrom+roman+2020}.


\subsubsection{Configuration with continuous $n(x)$: Bump problem}\label{sec:bump}
\begin{figure}
	\begin{tikzpicture}
	\draw( 0.00, 0.0) node {\includegraphics[scale=0.50]{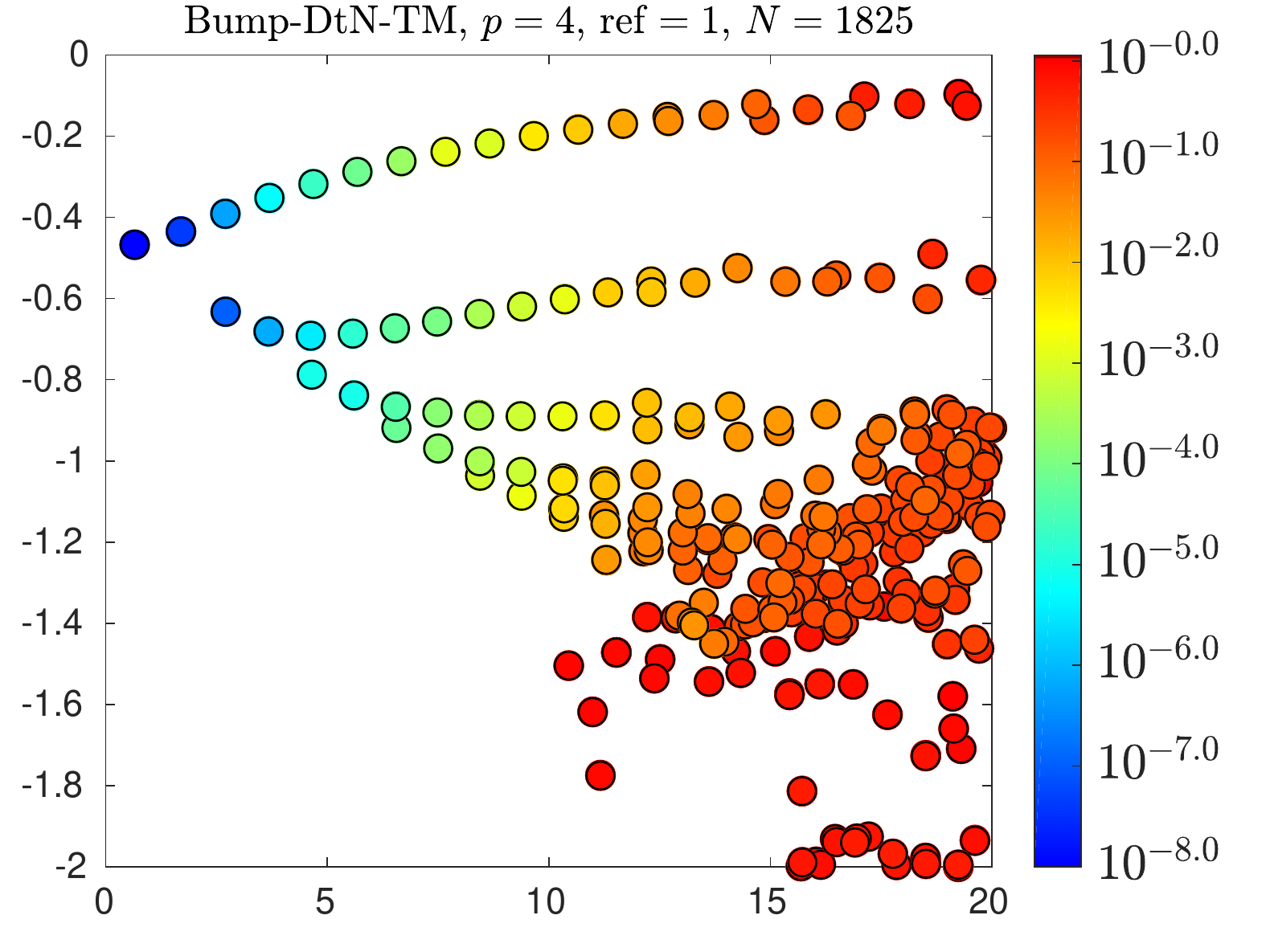} };
	\draw( 8.00, 0.0) node {\includegraphics[scale=0.50]{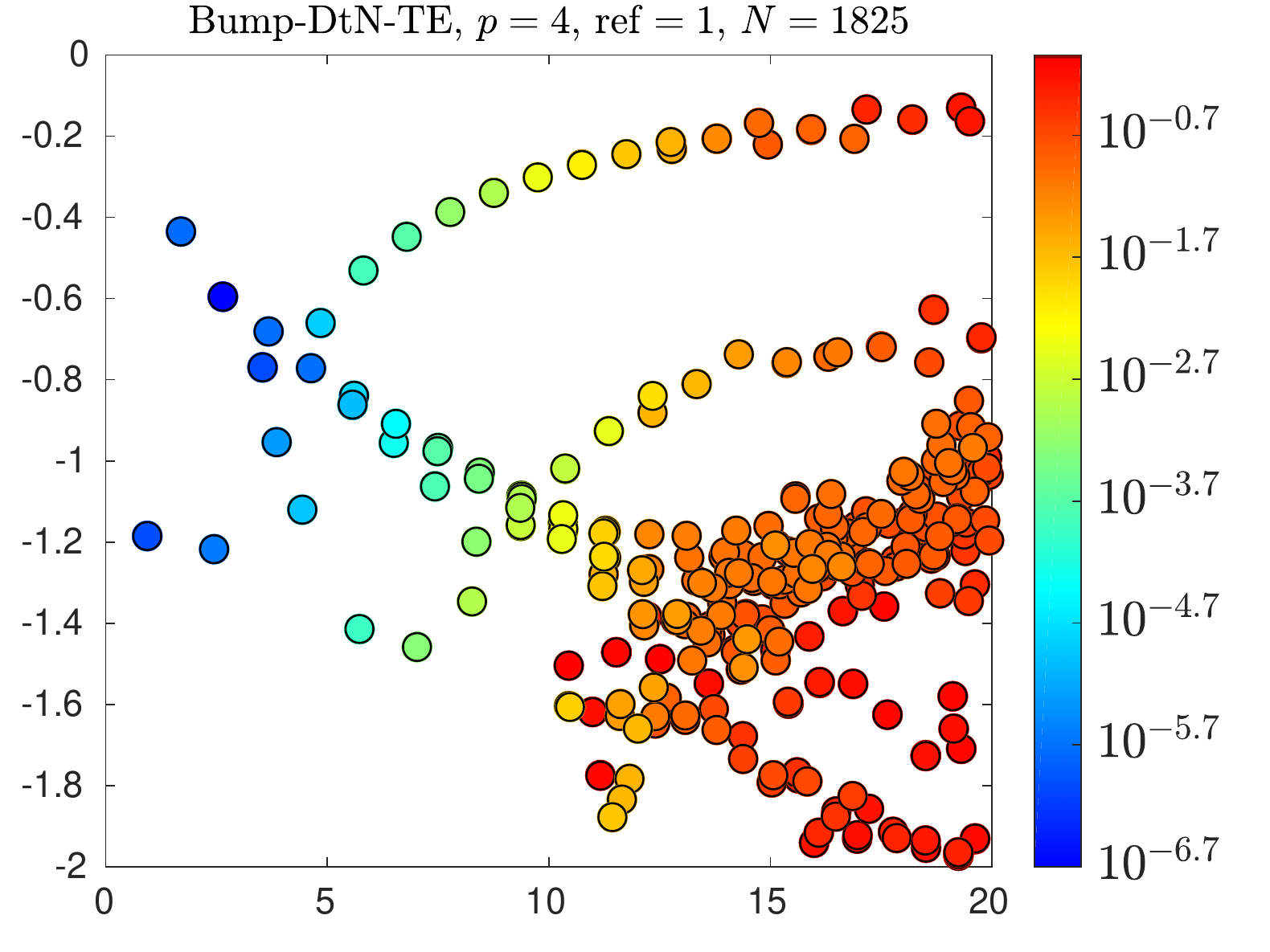} };		
	\end{tikzpicture}
	\caption{\emph{Spectral window showing computed FE eigenvalues (circles) of the Bump problem described in Section \ref{sec:bump} for TM and TE polarizations. In colors we give $\tilde \pseudo_m$ corresponding to $\om_m^\fem$ from computations over ten points.} }
	\label{fig:filter_bump}
\end{figure}

Consider the following 
refractive index:
\begin{equation}
	n(x)=\left\{
	\begin{array}{ll}
		1+P_3(|x|) & 0\leq |x| \leq R \\
		1 & |x|>R
	\end{array}
	\right.
	\label{eq:nonconst_n}
\end{equation}
subject to the compatibility conditions: $P_3(0)=1,\,P_3(R)=0,\,P'_3(0)=0,\,P'_3(R)=0$, and $R=1$.

\subsection{Acoustic benchmark in 3D: Single ball problem (SB)}\label{sec:SB} 
This case is analogous to the Single Disk problem \ref{sec:SD} with $d=3$.

Denote by $u=u_1,\,n=n_1$ the restrictions of $u,n$ to $\Om_1:=B(0,R)$, and set $n=n_2=1$ elsewhere. The corresponding exact outgoing resonant modes of \eqref{eq:master_eq} read:
\begin{equation}
	u_{m\ell}(r,\theta,\phi)=\left\{
	\begin{array}{rr}
		N_{m} j_m (n_1 \om r)\,Y_\ell^m (\theta,\phi), & r\leq R \\
		h_m^{(1)}(\om r) \,Y_\ell^m (\theta,\phi), & r > a
	\end{array} \right.,\,\, |\ell|<m
	,\,\, N_m:=\fr{h_m^{(1)}( \om R)}{j_m(n_1 \om R)}.
	\label{eq:exact}
\end{equation}
The eigenvalues $\om$ corresponding to $m=0$ are \emph{simple} and those corresponding to $m>0$ are \emph{semi-simple} and
have multiplicity $\alpha=2 m+1$.
The exact eigenvalue relationship 
can be written as
\begin{equation}
	j_m(n_1 \om R) h_m^{(1)\prime}(\om R)-n_1\,j'_m(n_1 \om R) h_m^{(1)}(\om R) =0.
	\label{eq:reson_SD}
\end{equation}

\begin{figure}
	\begin{tikzpicture}
	\draw( 0.00, 0.0) node {\includegraphics[scale=0.50]{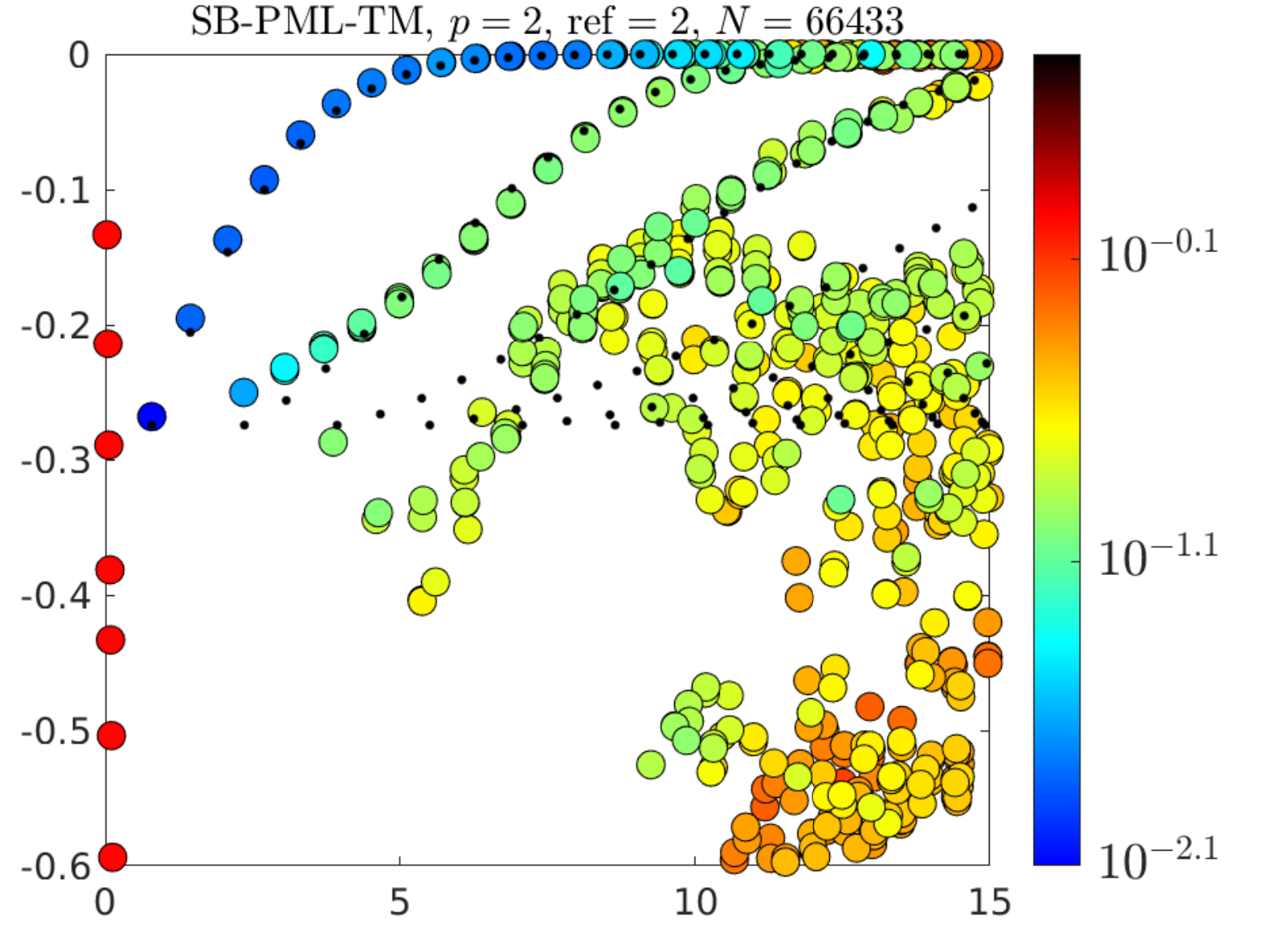} };
	\draw( 8.00, 0.0) node {\includegraphics[scale=0.50]{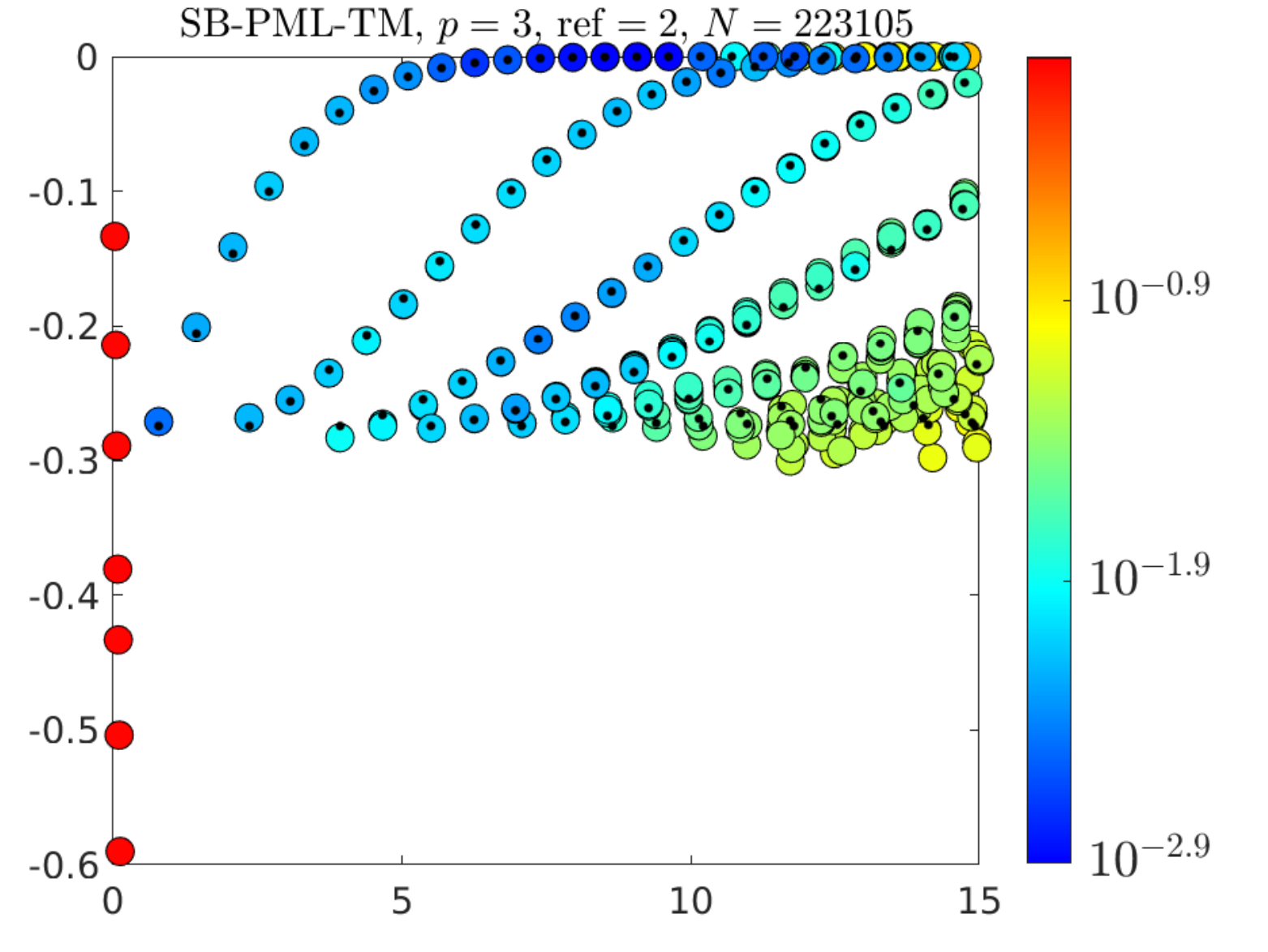} };
	\end{tikzpicture}
	\caption{\emph{Spectral window showing exact (dots) and FE eigenvalues (circles) the Single Ball problem described in Section \ref{sec:SB} for TM and TE polarizations. In colors we give $\tilde \pseudo_m$ corresponding to $\om_m^\fem$ from computations over ten points.} }
	\label{fig:filter_SB}
\end{figure}

\emph{Results for problems \ref{sec:bump}, and \ref{sec:SB}: } 
Similarly as pointed out in former discussions of this section,
the results gathered in Figures \ref{fig:filter_bump} and \ref{fig:filter_SB}, confirm 
the effectiveness from the application of the sorting scheme described in Remark \ref{def:approx_int_check}. 
The results in the figures illustrate a clear method to effectively sort the computed eigenpairs by using an approximation of the sorting indicator \eqref{eq:max_indicator}.


\section{Conclusions}
We have presented a sorting scheme, based on the Lippmann–Schwinger equation, for marking of potentially spurious scattering resonant pairs in $\mx R^d$ Helmholtz problems. For all computations and for TM and TE polarizations, numerical experiments on a broad range of problems illustrate that the sorting scheme provides valuable information on the location of potentially spurious resonances at low computational cost. This information can, for example, be used to determine the most important modes in quasimodal expansions.

\section*{Acknowledgments}
This work is founded by the Swedish Research Council under Grant No. 621-2012-3863.


\bibliographystyle{ieeetr} 

\end{document}